\documentclass[a4paper,fleqn]{cas-sc}

\usepackage[authoryear]{natbib}
\usepackage[utf8]{inputenc}
\usepackage{amsmath}
\usepackage[ruled,vlined]{algorithm2e} 
\usepackage{graphicx}
\usepackage{subcaption}
\usepackage{booktabs}
\usepackage{siunitx}
\usepackage{hyperref}

\def\tsc#1{\csdef{#1}{\textsc{\lowercase{#1}}\xspace}}
\tsc{WGM}
\tsc{QE}
\tsc{EP}
\tsc{PMS}
\tsc{BEC}
\tsc{DE}

\begin{document}
\let\WriteBookmarks\relax
\def\floatpagepagefraction{1}
\def\textpagefraction{.001}

\shorttitle{Offset-Free Robust Nonlinear Control Using Data-Driven Model}

\shortauthors{REBELLO, C. M. et~al.}

\title [mode = title]{Offset-Free Robust Nonlinear Control Using Data-Driven Model: A Nonlinear Multi-Model Computationally Efficient Approach}                      



%
\author[1]{Carine Menezes Rebello}[
                        orcid=0000-0002-0796-8116]



\ead{carine.m.rebello@ntnu.no}

\author[1]{Erbet Almeida Costa}[orcid=0000-0003-1397-9628]

\author[1]{Idelfonso B. R. Nogueira}[orcid=0000-0002-0963-6449]

\affiliation[1]{organization={Department of Chemical Engineering},
    addressline={Norwegian University of Science and Technology}, 
    city={Trondheim},
    postcode={793101}, 
    country={Norway}}

\cormark[1]
\cortext[cor1]{Corresponding author}



\begin{abstract}
Robust model predictive control (MPC) aims to preserve performance under model–plant mismatch, yet robust formulations for nonlinear MPC (NMPC) with data-driven surrogates remain limited. Integrating machine-learning surrogates into NMPC introduces risks such as structural misspecification and unreliable extrapolation, which can compromise constraint satisfaction and safety. This work proposes an offset-free robust NMPC scheme based on symbolic regression (SR). Using a compact NARX structure, we identify interpretable surrogate models that explicitly represent epistemic (structural) uncertainty at the operating-zone level, and we enforce robustness by embedding these models as hard constraints. Two robust controller configurations are investigated. In the single-zone variant (RNMPC\textsubscript{SZ}), synthetic data are generated within one operating zone to identify SR models that are embedded as constraints while the nominal predictor remains fixed. In the multi-zone variant (RNMPC\textsubscript{MZ}), zone-specific SR models from multiple operating zones are jointly enforced in the constraint set; at each set-point change, the nominal predictor is re-scheduled to the SR model of the newly active zone. In both cases, robustness is induced by the intersection of the admissible sets defined by the enforced SR models, without modifying the nominal cost or introducing tube/ancillary dynamics.
The approach was validated on a simulated pilot-scale electric submersible pump (ESP) system with pronounced nonlinearities and dynamically varying safety envelopes (downthrust/upthrust). 
RNMPC\textsubscript{SZ} and RNMPC\textsubscript{MZ} maintained disturbance tracking and rejection, and by intersecting models in the constraints, they increased margins and eliminated violations (especially near the upthrust), with a slight increase in settling time. Including up to four models per zone did not increase the time per iteration, maintaining real-time viability; RNMPC\textsubscript{MZ} presented the largest margins.
\end{abstract}


\begin{highlights}
\item Offset-free robust NMPC using symbolic regression models.
\item Applied to a pilot-scale electric submersible pump (ESP) system.
\item Ensures constraint satisfaction with low computational cost.
\end{highlights}


\maketitle

\section{Introduction}

The concept of robustness in control theory, articulated by \cite{GARCIA1989335,Morari1999} in the 1990s, refers to preserving a feedback system's performance despite deviations between the actual plant dynamics and those represented in the control model. This concern has long underpinned control theory research, with particular advances in linear systems. \cite{GARCIA1989335} noted that while substantial progress had been made for linear systems, robustness in nonlinear control remained largely unexplored at the time. Decades later, this gap is still pertinent. Although robust model predictive control (MPC) has matured significantly in its linear formulation, the systematic treatment of robustness for nonlinear MPC (NMPC), especially in the presence of model-plant mismatches, remains a challenging and active area of research \citep{Shardt2024, TUFA20161008,THANGAVEL20181074,Tian2012,rawlings2017model}.

This issue has become more acute with the increasing integration of machine learning (ML) models as surrogate models within MPC frameworks \citep{VAUPEL2020261,TATULEACODREAN20206031,MARTINSEN2022105024,LI2024108854,Marcellus2023}. These models offer a compelling alternative to first-principles models due to their ability to capture complex, nonlinear behaviour from data. However, embedding ML-based models within optimisation loops introduces new vulnerabilities. ML models can exhibit artificial minima in their prediction landscapes \citep{REBELLO2024126811,Rebello2024,NOGUEIRA2022243}. These features may arise from overfitting, poor extrapolation outside training domains, even within the domain in areas not well detailed by the training data, or a lack of physical constraints in the learning process \citep{LIMA2025100208}. When such models are used as predictors in NMPC, the optimiser may be misled by these artificial minima. This results in control actions that appear optimal for the surrogate model but are suboptimal or unsafe for the real system.

Despite this risk, many current NMPC implementations using ML surrogates do not explicitly address the robustness issue \citep{martins2024,MARTINS2021,TURAN2024,LIMA2022100052}. Typically, the ML model is used to predict the system's behaviour across the open-loop prediction horizon, and the control action is derived by minimising a cost function over this horizon. However, the validity of these predictions is rarely questioned, and the control action computed from them is assumed to be reliable \citep{martins2024,MARTINS2021}. This assumption is particularly problematic in mission-critical and constraint-sensitive systems. Suppose the surrogate model fails to accurately represent the system's behaviour due to, for instance, unmodeled disturbances, extrapolation beyond the training domain, or non-convex artefacts. In that case, the computed control action may violate system constraints or diverge from the accurate optimal trajectory. Such errors can result in operational failures, equipment damage, or safety incidents in sensitive applications \citep{BJORNBERG2006777,ASWANI20131216,hewing2019cautious}.

Robust MPC (RNMPC) methods vary in characterising and handling uncertainty, particularly in nonlinear settings where model fidelity and computational efficiency are often in tension. Polytopic uncertainty represents plant variations within the convex hull of a finite set of linear time-invariant (LTI) models \citep{KOTHARE19961361,Mayne1998}. Each vertex model defines a system matrix, e.g:
\begin{equation}
A(\theta) = \sum_{i=1}^{N} \theta_i A_i, \quad \sum \theta_i = 1, \quad \theta_i \geq 0).
\end{equation}

Robustness is typically achieved using parameter-dependent Lyapunov functions or solving linear matrix inequalities (LMIs), ensuring constraint satisfaction and stability over the full polytope. While this framework is tractable for LTI systems and widely adopted, it becomes computationally burdensome as the number of vertices increases. It can introduce conservativeness due to the convexity assumption \citep{Bemporad1999, MARTINS2016132}.

Similarly, the multi-plant framework represents uncertainty through a discrete, finite set of linear models, each corresponding to a different operating condition or regime. Unlike the polytopic case, there is no convex structure; the focus is on robust constraint satisfaction and performance for all models in the set individually. This approach is particularly attractive in the process industries, where systems often switch among identifiable modes. It enables the design of offset-free control without requiring steady-state target computation. Despite its practical appeal and reduced complexity, multi-plant MPC can be conservative, as it typically designs for the worst-case plant in the set \citep{Odloak2006,Odloak2004, MARTINS2016132}. The multi-plant MPC paradigm has been extensively developed and applied in linear MPC.

The set of models is described as:
\begin{align}
x_i(t+1) = A_i x_i(t) + B_i u(t),\\    
y_i(t) = C x_i(t),
\end{align}

for \( i = 1, 2, \dots, M \), where each pair \( (A_i, B_i) \) represents an independent linear time-invariant model in the uncertainty set.

Both polytopic and multi-plant frameworks offer theoretically sound and computationally tractable mechanisms for linear RMPC \citep{KOTHARE19961361,LEE2000463}. However, their generalisation to nonlinear systems remains limited in the literature. In particular, robust nonlinear MPC formulations often lack systematic methods to account for epistemic model uncertainty without incurring excessive complexity. This paper addresses this gap by extending the multi-plant strategy to the nonlinear regime using a data-driven approach based on symbolic regression. By constructing an ensemble of compact, interpretable surrogate models, our method captures epistemic uncertainty directly at the modelling stage. It embeds it into the control synthesis procedure, enabling robust constraint satisfaction in real-time applications.

On the other hand, multi-stage robust MPC approaches have emerged as promising solutions for robustness in nonlinear systems \citep{LUCIA20131306}. These methods represent uncertainty through branching scenario trees and compute optimal feedback policies across all sampled paths. While theoretically attractive, especially for handling non-convex or stochastic uncertainty, these approaches are often limited in practice due to their exponential growth in computational cost concerning the number of scenarios and prediction horizon. Despite efforts to mitigate this through sensitivity-based scenario selection \citep{KROG2024103270,MDOE2025108992}, multi-stage MPC still struggles to meet the real-time requirements of control systems applicable for mission-critical systems \citep{LUCIA20151015,LUCIA20131306,Maggioni2025,Paulson2020,Maggioni2025}.

Given these limitations, the present work proposes a computationally efficient, offset-free, robust nonlinear MPC scheme built upon data-driven symbolic regression models to address this challenge. These models offer explicit analytical expressions for system dynamics, enabling interpretable and efficient implementation. A nonlinear autoregressive with exogenous inputs (NARX) structure is used to organise the training data, capturing the plant’s temporal dependencies and exogenous drivers. Symbolic regression is then employed to identify compact, interpretable surrogate models that approximate the plant dynamics with minimal bias.

An advantage of symbolic regression is its ability to generate computationally inexpensive data-driven models while explicitly capturing epistemic uncertainty, that is, the uncertainty arising from limited knowledge or incomplete system representations \citep{LIMA2025110350,Rebello2024,VOLTOLINI2025132746}. By exploring different combinations of functional forms, symbolic regression yields a diverse set of plausible nonlinear models, each reflecting distinct structural hypotheses about the underlying process. This ensemble naturally defines a surrogate-based uncertainty set, suitable for robust control synthesis.

To handle this structural uncertainty, a multi-model MPC formulation is proposed that embeds the ensemble of symbolic models within an optimisation framework. The control law is derived by optimising performance while ensuring constraint satisfaction for all models in the set. Importantly, this control design is inspired by multi-model MPC formulations developed for linear systems, and it extends that principle to the nonlinear regime by guaranteeing that input and output constraints are satisfied across all surrogate models. This ensures robust feasibility and safety, even when the true system dynamics are unknown. In doing so, the proposed approach bridges a gap in RNMPC by tightly integrating data-driven modelling and robust synthesis in a computationally tractable manner.

This methodology was validated on a simulated pilot-scale electric submersible pump (ESP) system. The case study was deliberately chosen due to the challenging nature of its nonlinear operational constraints, which vary dynamically across different operating regions, a scenario where constraint enforcement is both critical and complex. To assess the approach, two configurations were investigated: a robust SR-based NMPC single-zone scheme (RNMPC\textsubscript{SZ}), in which SR models identified from one operating zone are embedded as hard constraints while the nominal model is kept fixed, and a robust SR-based NMPC multi-zone (RNMPC\textsubscript{MZ}) scheme, in which zone-specific SR models from multiple operating zones are jointly enforced in the constraint set and the nominal model is re-scheduled to the model of the newly active zone at each set-point change. In both cases, robustness is induced by the intersection of the admissible sets defined by the enforced SR models. The proposed RNMPC was then compared to a conventional NMPC based on a nonlinear mechanistic model (NMPC\textsubscript{MM}).

\section{Methodology}

\begin{figure}[h!]
	\centering
		\includegraphics[scale=.3]{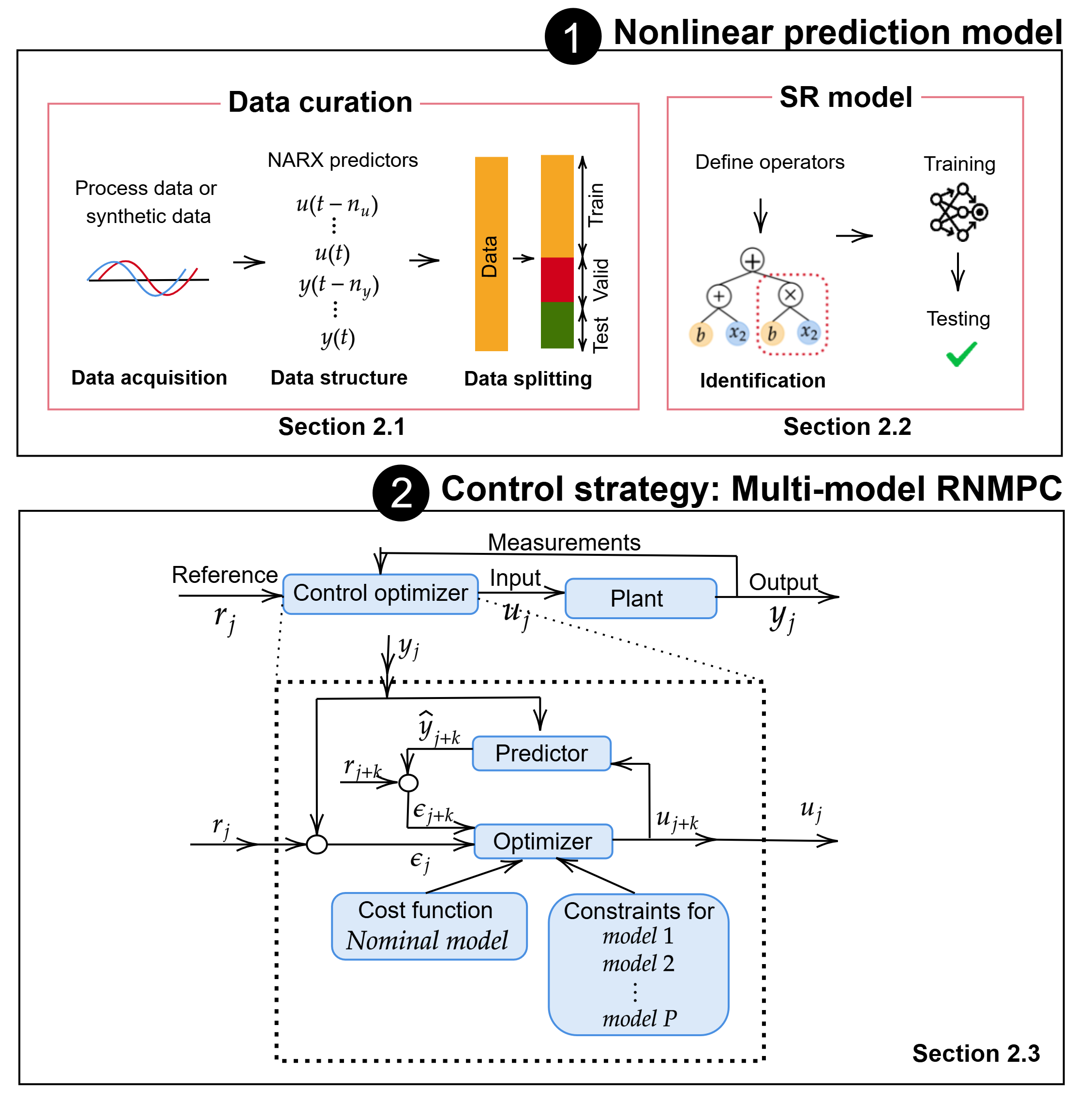}
	\caption{Methodology workflow for RNMPC via symbolic regression.}
	\label{fig:RNMPC_methodology}
 \end{figure}

The proposed methodology is structured into three main stages, as illustrated in Figure \ref{fig:RNMPC_methodology}. \textbf{Section~\ref{curation}} covers the \emph{data curation} phase, which includes acquisition of either process or synthetic data, structuring into a NARX-based multiple-input single-output (MISO) format, and subsequent splitting into training, validation, and testing subsets. This step ensures that the dynamic relationships between system inputs and outputs are properly captured and preprocessed for model identification.

\textbf{Section~\ref{regression}} presents the development of the \emph{symbolic regression models (SRMs)}. In this stage, a library of mathematical operators is defined to guide the structure of symbolic expressions. The surrogate models are then identified through an evolutionary or sparse regression process, trained, and validated using the previously curated datasets. This process results in a set of compact, interpretable models that approximate the system dynamics while capturing epistemic uncertainty.

\textbf{Section~\ref{control}} introduces the \emph{Control strategy: multi-model RNMPC}, wherein the SRMs are embedded into a robust NMPC framework. A multi-model optimisation problem is formulated to compute control actions \( u \) that satisfy performance and constraint requirements across all surrogate models in the ensemble. An offset-free formulation is employed to improve tracking performance in the presence of unmeasured disturbances \( w \), while maintaining real-time feasibility. The optimisation is executed in a receding horizon fashion using the symbolic models, which act as the predictive component of the controller.

Finally, \textbf{Section~\ref{ESP}} provides the validation of the proposed methodology based on a pilot-scale (ESP). The ESP system was selected for its nonlinear, dynamically constrained behaviour, making it an ideal benchmark for evaluating robust NMPC under realistic conditions.

\subsection{Data curation} \label{curation}

The first stage comprises generating a space-filling synthetic dataset: latin hypercube sampling (LHS) is employed to draw input vectors that excite the system throughout its admissible operating space, providing the data required to fit a surrogate model that will later act as the plant model. 

To incorporate data-driven surrogate models into the predictive control framework, it is essential to adopt a structured regression form capable of explicitly embedding the underlying system dynamics into the predictive formulation. In particular, the NARX is adopted due to its inherent suitability for dynamic modelling and predictive control tasks. The NARX structure explicitly encodes temporal correlations between the system's inputs and outputs by regressing future outputs on past measured inputs and outputs. This structure constrains the search space for symbolic regression, guiding the identification procedure toward dynamically consistent and causally interpretable mathematical expressions.

In the NARX formulation, the one-step-ahead system output prediction at time $t$ is defined by a nonlinear mapping $f(\cdot)$ of a suitably defined regression vector $\phi(t)$:

\begin{equation}
\hat{y}(t+1) = f\left(\phi(t)\right),
\label{eq:one_step}
\end{equation}

with the regression vector explicitly structured to encapsulate the system’s dynamic dependencies:

\begin{equation}
\phi(t) = \left[\, y^\top(t),\, y^\top(t-1),\, \dots,\, y^\top(t-n_y),\, u^\top(t),\, u^\top(t-1),\, \dots,\, u^\top(t-n_u)\, \right]^\top,
\label{eq:reg_vector}
\end{equation}

\noindent where $y(t)\in\mathbb{R}^{n_y}$ denotes the output vector and $u(t)\in\mathbb{R}^{n_u}$ denotes the input (or exogenous disturbance) vector at time step $t$. The model orders $n_y$ and $n_u$ represent the memory depth, determining the extent to which historical data influences the prediction.

This structured regression directly incorporates the dynamical constraints into the training dataset; as a result, the discovered symbolic surrogate models approximate the nonlinear system behaviour, maintaining the necessary structural integrity to be integrated reliably within a receding-horizon MPC formulation.

Within the predictive horizon $m_p$, the multistep output predictions are computed recursively as:

\begin{align}
\hat{y}(t+2|t) &= f\!\left(\phi(t+1|t)\right), \label{eq:pred2} \\
\hat{y}(t+3|t) &= f\!\left(\phi(t+2|t)\right), \label{eq:pred3} \\
&\vdots \nonumber \\
\hat{y}(t+N_p|t) &= f\!\left(\phi(t+m_p-1|t)\right), \label{eq:multi_step}
\end{align}

where $\hat{y}(t+k|t)$ represents the predicted system output at time step $(t+k)$, with $k=1,\dots,m_p$ indexing the prediction step within the horizon. The regression vector $\phi(t+k|t)$ is obtained by shifting past lags and inserting the most recent predicted output and planned future input.

For compactness, the entire horizon can be written as
\begin{align}
\hat{\mathbf{y}}(t\mid t)=
\begin{bmatrix}
\hat y(t{+}1\mid t)\\
\hat y(t{+}2\mid t)\\
\vdots\\
\hat y(t{+}m_p\mid t)
\end{bmatrix}
=
\begin{bmatrix}
f\!\big(\phi(t\mid t)\big)\\
f\!\big(\phi(t{+}1\mid t)\big)\\
\vdots\\
f\!\big(\phi(t{+}m_p{-}1\mid t)\big)
\end{bmatrix}. \label{eq:compact}
\end{align}

Structuring the symbolic regression step around the NARX format ensures that the surrogate models identified are inherently suited for the MPC law to be formulated in Section \ref{control} .

To prepare the data for symbolic regression model identification, the dataset structured according to the previously defined NARX representation is further proposed to be organised explicitly into a multiple-input single-output (MISO) regression format. In this approach, each training sample is represented by a regression vector containing past output values and both current and past exogenous inputs, paired with a single target output. The adoption of the MISO structure is motivated by findings in the literature which show that multi-output symbolic regression substantially increases the complexity of the search space and often results in less interpretable models \citep{WANG2025111289, MOYANO2021275}. In contrast, MISO formulations remain computationally tractable while still capturing interdependencies among variables through the regression vector, which makes them particularly suitable for reliable model identification in the NMPC framework. This formulation inherently encodes dynamic interactions within the data structure, guiding the symbolic regression algorithm toward identifying interpretable, and dynamically consistent surrogate models.

Mathematically, the MISO formulation derived from the general NARX structure Eq. \eqref{eq:reg_vector} is explicitly represented as follows:
\begin{equation}
y_i(t+1) = f_i\left(y_i(t), \dots, y_i(t - n_y), u(t), u(t - 1), \dots, u(t - n_u)\right),
\label{eq:miso_narx}
\end{equation}
where \( y_i(t+1) \) represents the single predicted output at time \( t+1 \), corresponding specifically to the \(i\)-th output variable of interest. Each surrogate model \( f_i(\cdot) \) thus explicitly incorporates the dynamic influence of historical outputs and exogenous input variables relevant to the specific output variable. Thus, t-1 represents the delays that need to be passed to the model for the prediction of the future instant. The number of delays can be tuned using different methods, including expert knowledge of the process and data or a dependency analysis between the predicted variable and the predictors, as shown on \citep{NOGUEIRA2022243,He1993}. By structuring data in this MISO format, separate symbolic regression models \( f_i(\cdot) \) can be individually identified for each controlled output variable, thereby eliminating unnecessary coupling among output predictions and facilitating targeted optimisation of distinct performance metrics.

For model training and evaluation, it is proposed to partition the dataset into two subsets: approximately 80\% of the samples are allocated for parameter estimation and hyperparameter tuning, while the remaining 20\% are reserved as an independent validation set for assessing model performance. The specific partition ratio, however, may vary and should be chosen according to the dataset size and application-specific constraints.

All input and output variables are normalised using min–max scaling to the interval \([0,1]\). The scaling parameters (minimum and maximum values) are computed exclusively from the training set to avoid information leakage. The identical linear transformation is subsequently applied to the test set, preserving data consistency across the subsets.

Stratified shuffling is employed during data partitioning to preserve the marginal distributions of input variables in both subsets. This ensures that the validation set adequately represents the same operational envelope covered by the training data, thereby providing robust, unbiased assessment conditions.

\subsection{Symbolic regression model} \label{regression}

Given the dataset structured in the previously defined NARX–MISO form, symbolic regression is proposed to be used to identify surrogate models by explicitly approximating the unknown nonlinear mapping function $f(\cdot)$ introduced in Equation \eqref{eq:one_step}. Symbolic regression is an approach that systematically searches for analytical expressions describing the relationship between inputs and outputs without assuming a fixed functional form. Specifically, symbolic regression aims to identify a nonlinear function $f(\cdot)$ that accurately relates the regression vector $\phi(t)$, containing historical input-output data, to the one-step-ahead predicted output $\hat{y}(t+1)$. The primary advantage of adopting this system identification approach lies in its ability to discover the simplest possible mathematical structures that accurately capture the system dynamics. By explicitly balancing model simplicity and predictive accuracy during the search process, symbolic regression yields compact and interpretable representations. This trade-off is particularly valuable in the proposed RNMPC framework, where computational efficiency and model transparency are essential for real-time feasibility and robust control synthesis.

Formally, symbolic regression seeks a mathematical expression of the form:
\begin{equation}
\hat{y}(t+1) = f\left(\phi(t)\right),
\end{equation}
where the function $f(\cdot)$ is represented by a symbolic expression composed of elementary mathematical operators (e.g., addition, multiplication, exponentiation), constants, and the elements of the regression vector defined in Equation \eqref{eq:reg_vector}. The symbolic regression algorithm systematically constructs candidate expressions, balancing model complexity and predictive accuracy, to find compact and interpretable representations of system dynamics \citep{Rebello2024,Tonda2024}.

\begin{figure}[h!]
\centering
\includegraphics[scale=.16]{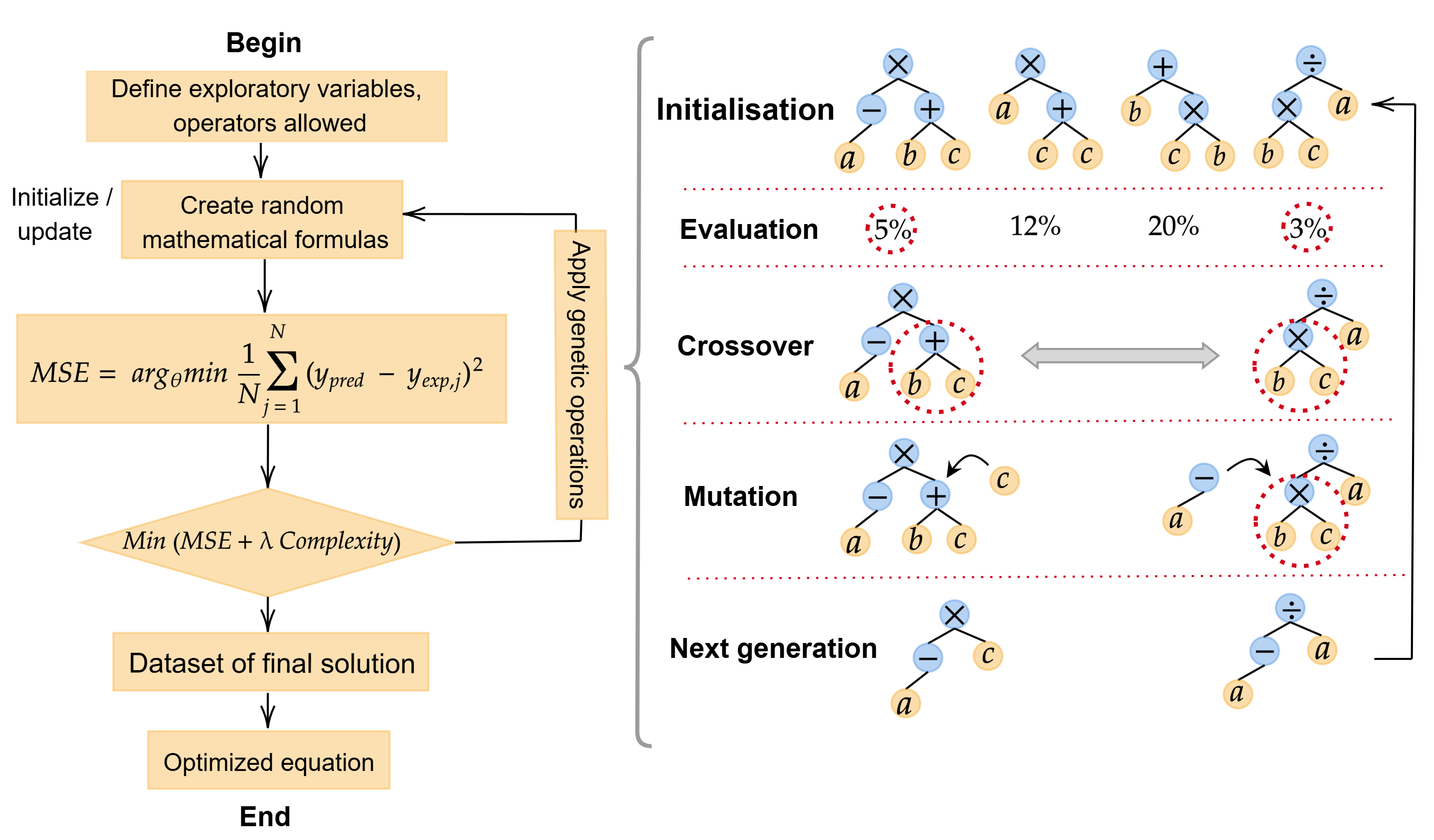}
\caption{Illustration of the symbolic regression process implemented by PySRRegressor, showing expression initialisation, evaluation, genetic operations, and model selection through evolutionary optimisation.}
\label{SR}
\end{figure}

In this study, symbolic regression uses evolutionary search techniques combined with symbolic expression trees. This evolutionary strategy explores the high-dimensional model space to evolve candidate equations iteratively. The iterative process of evolving and selecting optimal symbolic equations is illustrated in Figure \ref{SR}.

The left-hand side of Figure \ref{SR} presents the algorithmic workflow of symbolic regression. The process begins with the definition of the exploratory variables (input regressors) and the set of mathematical operators allowed during the search. From these, a population of random mathematical expressions is generated, which forms the initial set of candidate models. Each expression is then evaluated according to a loss function, where the mean squared error (MSE) is defined as
\begin{equation}
MSE = \arg \min \frac{1}{N} \sum_{i=1}^{N} \left( y_{\text{pred}} - y_{\text{exp}} \right)^2.
\label{eq:MSE}
\end{equation}
To enforce parsimony, a complexity penalty is added to the optimisation objective, resulting in:
\begin{equation}
\min \big( MSE + \lambda  \text{Complexity} \big).
\label{eq:MSE_complexity}
\end{equation}
Based on this criterion, the best-performing expressions are carried forward or recombined to form a new generation. When stopping criteria are satisfied, such as a maximum number of generations or convergence in accuracy, the process stores a dataset of the best-performing solutions and selects the final optimised equation that best balances accuracy and interpretability.

The right-hand side of Figure \ref{SR} illustrates the evolutionary process in terms of symbolic expression trees, which evolve across successive generations. The workflow comprises the following steps:

\begin{itemize}
\item \textbf{Initialisation:} A population of random mathematical expressions is generated from input variables and allowed mathematical operators. Examples of these initial expressions are shown in the first row of Figure \ref{SR}, including $(a - (b + c))$, $a \times (c + c)$, $b + (c \times b)$, and $((b \times c) \div a)$.

\item \textbf{Evaluation:} Each expression is evaluated based on a loss function, e.g. mean squared error, which compares its output with the target data. A complexity penalty is also applied to discourage overly complex expressions. In the figure, expressions are annotated with performance scores, such as 5\%, 12\%, 20\%, and 3\%, indicating their relative error or fitness levels.

\item \textbf{Genetic operations:} To improve the population over generations, two types of transformations are applied with distinct objectives:
\begin{itemize}
\item \textit{Crossover (exploration):} Aims to explore new regions of the search space by recombining building blocks of two parent expressions. Subexpressions (subtrees) are exchanged to generate new descendants. For example, from $(a \times (b + c))$ and $((b \times c) \div a)$, two blocks were selected to be removed and restructured: $(b+c)$ or $(b \times c)$, as illustrated in the crossover section of Figure \ref{SR}.
\item \textit{Mutation (diversification):} It aims to introduce variation and prevent premature convergence by randomly changing parts of an expression. This may involve substituting a variable (e.g., $(b+c)$ becomes $c$), changing an operator (e.g., $\times$ to $-$), or modifying a substructure. In the figure, this is exemplified by transforming $(a \times (b + c))$ into $(-a \times c)$ and restructuring $((b \times c) \div a)$ into $(-a \div a)$.
\end{itemize}

\item \textbf{Next generation:} Based on their evaluation scores, selected expressions are carried forward or combined to form a new generation. This iterative process, depicted in the bottom part of the figure, continues to refine the population, promoting expressions that better balance predictive accuracy and simplicity. For example, the resulting expressions $(a \times c)$ and $(a \div c)$ are preserved in the next generation due to their improved performance.

\item \textbf{Final solution:} As highlighted in the bottom blocks of the left-hand flowchart, the process stores a dataset of the best-performing solutions. From this set, the algorithm selects the final optimised equation that best balances accuracy and parsimony, marking the conclusion of the symbolic regression routine.
\end{itemize}

The PySRRegressor framework automates this process, allowing for customisation of the operator sets, loss functions, and expression constraints \citep{Tonda2024,cranmer2023}. The Figure \ref{SR} illustrates how the expressions are represented as trees, with operators as nodes and variables or constants as leaves.  This methodology is beneficial for identifying governing equations in physical systems, where data-driven modelling needs to preserve interpretability and remain aligned with domain knowledge.

After the evolutionary search, the algorithm returns a ranked Pareto front of candidate equations. The list is ordered by increasing structural complexity, measured, for example, by the total number of nodes or nonlinear operators. Predictive accuracy typically increases as complexity grows. To discourage unnecessarily complicated expressions, model selection is guided by a loss function that combines the prediction error (MSE, Eq. \eqref{eq:MSE}) with a complexity penalty (Eq. \eqref{eq:MSE_complexity}) \citep{Tonda2024,cranmer2023}. The complexity of a candidate expression is computed as the sum of the individual weights assigned to each operator, defined by the option \texttt{complexity of operators}, and scaled globally by the hyperparameter \texttt{complexity penalty}. By default, each operator contributes unit cost, but nonlinear operators may be assigned higher weights to reflect their additional complexity. This mechanism ensures that two models with similar accuracy will favour the simpler structure, enforcing parsimony and interpretability (see the \href{[https://astroautomata.com/PySR/options/}{PySR](https://astroautomata.com/PySR/options/}{PySR) documentation}). A detailed list of the default parameters of \texttt{PySRRegressor} is provided in Appendix \ref{app:pysr_defaults}.

Finally, the selected candidate models are validated against the test set, a subset of data that was used during the identification process. Evaluation against this unobserved data provides an unbiased estimate of generalisation performance and protects against overfitting of symbolic expressions.

\subsection{Control strategy: Multi-model RNMPC}\label{control}

\begin{figure}[h!]
	\centering
		\includegraphics[scale=.16]{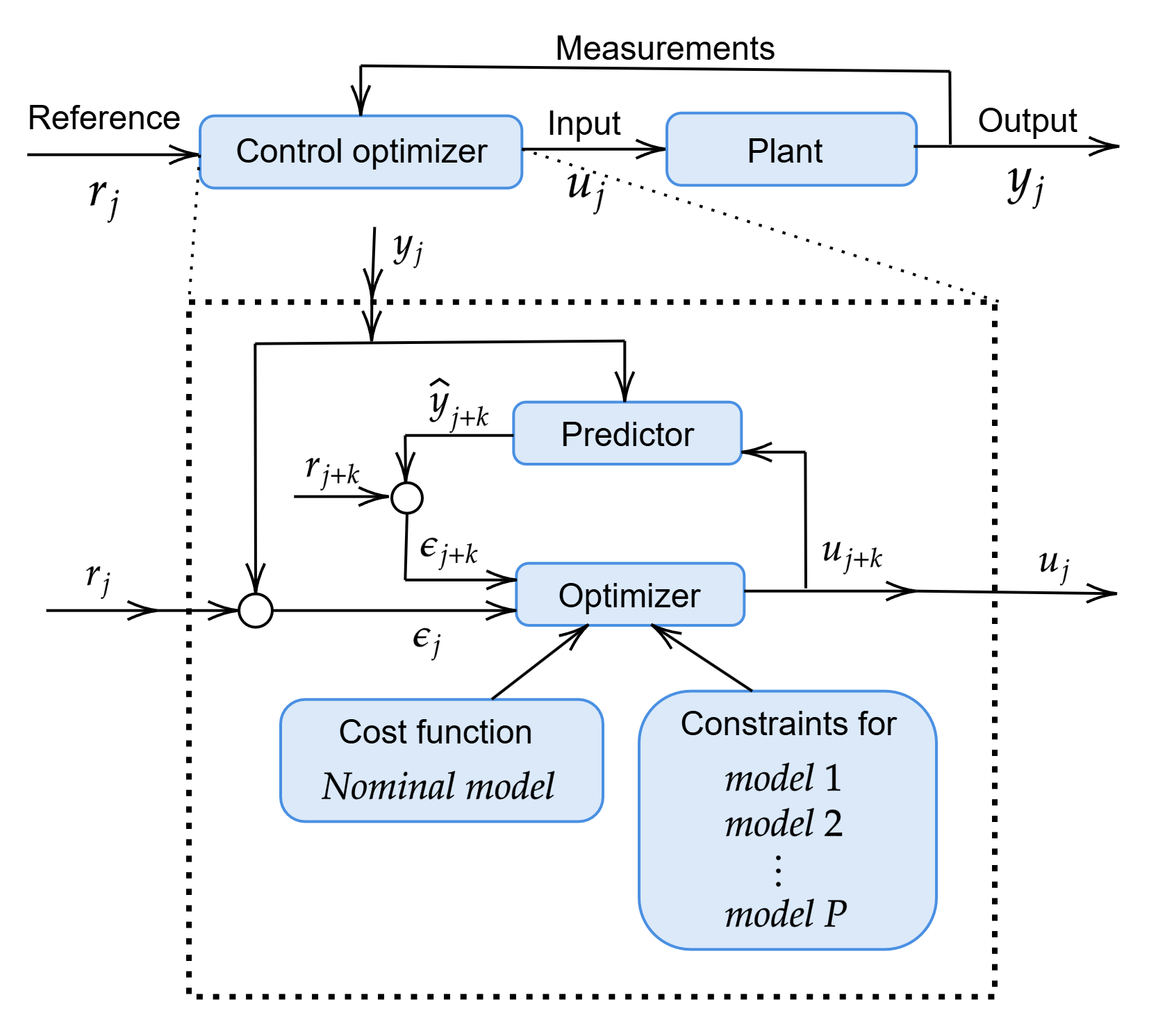}
	\caption{ Multi-model RNMPC control strategy incorporating symbolic regression constraints .}
	\label{fig:Control_strategy}
 \end{figure}

The RNMPC technique proposed in this work involves solving a constrained optimisation over a moving prediction horizon to compute the optimal control inputs sequentially while explicitly accounting for plant-model uncertainty. The control actions are updated at each discrete sampling interval according to the predicted system behaviour, as exemplified in Figure~\ref{fig:Control_strategy}.

Figure \ref{fig:Control_strategy} depicts the multi-model RMPC architecture used in this work. The outer loop shows the reference $r_j$, the computed input $u_j$ applied to the plant, and the measured output $y_j$. Inside the dashed box, the predictor generates forecasts $\hat y_{j+k}$ over the prediction horizon. The optimiser computes the sequence $\{u_{j+k}\}_{k=0}^{N_u-1}$ that minimises a nominal-model cost in terms of the tracking error $\varepsilon_{j+k}\!\triangleq\! r_{j+k}-\hat y_{j+k}$ (with $\varepsilon_j=r_j-y_j$), while  simultaneously enforcing hard constraints for all models $m=1,\dots,P$. Only the first control move is applied at each sampling instant, and the problem is resolved at the next step (receding-horizon principle), ensuring robust feasibility with respect to the considered model set.

The prediction model used within the RNMPC framework is constructed via symbolic regression. The analytical expressions that approximate system behaviour offer a compact and interpretable alternative to conventional black-box modelling strategies, such as neural networks \citep{JORDANOU2022101553,BONZANINI20205279,en15249335}, physics-informed neural networks \citep{ELORZACASAS2025109105,10418494}, support vector machines \cite{FENG20152048,s20113335,BAO2007691}, and ensemble methods \citep{LI2023119416, JEONG2020106875}, which are commonly used in the literature.
 Each model $f^p$ represents a possible system dynamics scenario derived from symbolic regression based on distinct training conditions or operating regimes. In addition to defining the system evolution, symbolic regression is employed to construct the expressions defining system constraints, including output bounds and actuator limitations, ensuring coherence between prediction and constraint enforcement.

The approach ensures robustness against structural mismatches and modelling errors by evaluating performance over an ensemble of plant models.

The proposed robust multi-plant MPC adapts the multi-model architecture introduced by \cite{LEE1997763} and subsequently extended by \cite{Odloak2004}, recasting it as a finite-horizon formulation suitable for the present application.

The proposed robust MPC formulation relies on a receding-horizon optimisation strategy that determines control actions based on a finite prediction window, based on the approach proposed by \cite{QIN2003733} and extended by \cite{Odloak2004}. At each discrete time step \(j\), given the current state information encoded within the NARX-MISO regression vector \(\phi(j)\), a finite sequence of control increments \(\{\Delta u_{j|j}, \Delta u_{j+1|j}, \dots, \Delta u_{j+m_c-1|j}\}\) is computed by solving a constrained optimisation problem. This optimisation minimises a quadratic cost function defined over a prediction horizon \(m_p\), while explicitly handling structural uncertainty by ensuring robust constraint satisfaction for all candidate surrogate models \(p \in \{1,\dots,P\}\).

Mathematically, the control law is computed as the solution to the following optimisation problem:
\begin{equation}
\min_{\{\Delta u_{j+k|j}\}_{k=0}^{m_c-1}}\Biggl\{\sum_{k=1}^{m_p}\bigl\|y_{j+k|j}^{p^\star}-r_{j+k}+e_j\bigr\|_Q^2 + \sum_{k=0}^{m_c-1}\bigl\|\Delta u_{j+k|j}\bigr\|_R^2\Biggr\}
\label{eq:robust_cost_narx}
\end{equation}
subject to constraints, for all surrogate models \(p = 1,\dots,P\):
\begin{subequations}\label{eq:constraints_robust}
\begin{align}
y_{j+k+1|j}^{p} &= f^{p}\bigl(\phi_{j+k|j}\bigr), &\quad k &= 0,\dots,m_p-1, \label{eq:model_dyn} \\
-\Delta u_{\text{max}} &\leq \Delta u_{j+k|j} \leq \Delta u_{\text{max}}, &\quad k &= 0,\dots,m_c-1, \label{eq:delta_u} \\
u_{\text{min}} &\leq u_{j+k|j} \leq u_{\text{max}}, &\quad k &= 0,\dots,m_c-1, \label{eq:u_limits} \\
y_{\text{min}} &\leq y_{j+k|j}^{p} \leq y_{\text{max}}, &\quad k &= 1,\dots,m_p, \label{eq:y_limits} \\
\Delta u_{j+k|j} &= 0, &\quad k &= m,\dots,m_c-1, \label{eq:blocking} \\
u_{j+k|j} &= u_{j+k-1|j}+\Delta u_{j+k|j}, &\quad u_{j-1|j} &= u_{j-1}, \label{eq:delta_u_def}
\end{align}
\end{subequations}

\noindent where \(y_{j+k|j}^{p^\star}\) denotes the predicted output at time \(j+k\) computed from the designated nominal surrogate model \(p^\star\), and \(r_{j+k}\) is the reference trajectory. Matrices \(Q\) and \(R\) are weighting matrices penalising output tracking errors and control move increments, respectively.

In Eq. ~\eqref{eq:robust_cost_narx}, the operator \textit{min}
selects the sequence of control increments that minimises a quadratic performance index evaluated on a designated \emph{reference model} \(p^\star\). The first summation term penalises deviations between the predicted outputs \(y^{p^\star}_{j+k|j}\) and the reference trajectory \(r_{j+k}\), while explicitly accounting for persistent bias \(e_j\) through the output penalty matrix \(Q\). This term ensures accurate tracking performance under the nominal model.

The second term penalises the magnitude of each control increment \(\Delta u_{j+k|j}\), using the weighting matrix \(R\), to enforce smooth actuator behaviour and prevent aggressive control actions. The matrices \(Q\) and \(R\) are selected to be positive semi-definite and positive definite, respectively, to guarantee appropriate scaling of the cost terms of the quadratic program. An optional terminal penalty term weighted by \(Q_{m_p}\) may be included at the final prediction step to further stabilise the closed-loop behaviour, without compromising the tractability of the optimisation.

Constraint \eqref{eq:model_dyn} incorporates the surrogate models \(f^{p}(\cdot)\), each identified via symbolic regression and structured according to the NARX-MISO representation. Here, the regression vector \(\phi_{j+k|j}\) is recursively updated using past outputs and inputs according to the NARX structure defined previously in Eq. \eqref{eq:reg_vector}. Constraints \eqref{eq:delta_u}–\eqref{eq:y_limits} enforce feasible operating ranges for input increments, inputs, and outputs across all surrogate models, ensuring robust constraint satisfaction despite model uncertainty. Constraint \eqref{eq:blocking} introduces move-blocking to limit computational complexity, holding control increments constant beyond a specified step \(m\). Finally, constraint \eqref{eq:delta_u_def} defines the relationship between incremental and absolute control inputs, anchoring the optimisation to the previously implemented input \(u_{j-1}\).

By simultaneously enforcing these constraints for each surrogate model, the proposed formulation ensures robust feasibility and performance across the defined model ensemble, explicitly addressing uncertainty and enhancing the applicability of real-time control. While the symbolic-regression model itself is unconstrained, operational bounds and actuator limitations are incorporated directly into the NMPC optimisation problem.

The term \(e_j\) in the first term of Eq. \eqref{eq:robust_cost_narx} is a steady-state error correction filter. This work follows the approach proposed by \citep{Plucenio2007,9664461,Plucenio2010} of applying a discrete filter to the error between the output that the model calculates $y^{m}[k]$, and the measured value in the plant $y^{p}[k]$ at time k. The filter is stable and leads to an offset-free control law through the use of convenient filter tuning. The filter derivation starts by defining the necessary correction factor $\eta[k]$ as a function of the error between the predicted variable and the measured variable as:
\begin{align}
\Delta \eta[k] &= (1 - \omega) \bigl(\hat{y}[k|k-1] - y^p[k]\bigr) + \omega \,\Delta \eta[k-1]
\label{DeltaFiltro}\\
\eta[k] &= \eta[k-1] + K\,\Delta \eta[k]\label{FiltroEta}, \\
\hat{y}[k|k-1] &= g\bigl(f(\mathbf{x}[k-1], \mathbf{u}[k-1])\bigr) + \eta[k-1]. \label{MeasuredVariablePlant}
\end{align}

The other variables involved are an integrator gain $K$, and a weight variable $\omega$ to ponder the robustness capability. $\hat{y}[k|k-1]$ is the corrected output prediction using the factor $\eta[k-1]$.

Now it is necessary to define the a priori error and the a posteriori error, which will be, respectively:

\begin{align}
    e_{pos}[k] &= y_m[k] - y_{\text{rm}}[k|k-1], \label{EpsylonK}\\
    e_{pri}[k] &= y_m[k] - \hat{y}[k-1]. \label{ErroK}
\end{align}

From those equations, it is possible to see that the a priori error is the difference between the measured value and the model's current prediction. In turn, after applying the correction, we have the a posteriori error.

Including Eq. \eqref{EpsylonK} in Eq. \eqref{ErroK} and parallelly including the a posteriori error in the correction factor Eqs. \eqref{DeltaFiltro} and \eqref{FiltroEta} all equations can be written as:

\begin{align}
e_{pri}[k] &= e_{pos}[k] - \eta[k-1], \label{ExpansioEk}\\
\Delta \eta[k] &= \omega \Delta \eta[k-1] + (1 - \omega) e[k], \label{DeltaEtaPrior}\\
\eta[k] &= \eta[k-1] + K \, \Delta \eta[k]. \label{EtakPrior}
\end{align}

The discrete Z transform can be used, in this context, conveniently to obtain the following equations in the discrete domain:

\begin{align}
e_{pri}(z) &= e_{pos}(z) - z^{-1} \eta(z),\\
\eta(z) &= \frac{K}{1 - z^{-1}} \Delta \eta(z),\\
\Delta \eta(z) &= \frac{1}{1 - \omega z^{-1}} e_{pri}(z).
\end{align}

With the appropriate algebraic manipulations, one can obtain the following transfer function in the discrete domain to relate the a priori and a posteriori error:
\begin{equation}
\frac{e_{pri}(z)}{e_{pos}(z)} = \frac{z^2 - (\omega + 1) z + \omega}{z^2 - (\omega + 1 - K (1 - \omega)) z + \omega}. \label{TransFuncEEpyslon}
\end{equation}

The filter can be chosen as a second-order function with the desired choice of the poles Eq. \eqref{TransFuncEEpyslon}. This implies that using $e_j=e_{pos}$, the error minimization between the plant and the controller will follow a second-order behaviour. Additionally, in this paper, the specific case of using $\omega=0$ and $K=1$ is used to have the practical effect of integration of the error and, consequently, the filter will act as an integrating action of the controller leading to $e_j(\infty)=e_{pos}(\infty)=0$ in the steady-state.

\subsection{Case study: electrical submersible pump}\label{ESP}

Electric Submersible Pumps are among the most effective artificial lift methods used in oil production, particularly in wells experiencing reduced reservoir pressure. These systems transform electrical energy into hydraulic force, allowing fluids to be efficiently transported from the wellbore to the surface. ESPs are favoured for their ability to accommodate various well geometries, including vertical, horizontal, and deviated configurations, and for supporting high flow rates, which can significantly enhance overall production. Their placement within the fluid stream also facilitates effective motor cooling, contributing to their operational reliability.

This work employs an ESP-based system as a case study, focusing on a vertical production well setup. The well comprises a 32-meter column with a multistage pump installed at a depth of 22.8 meters. The process fluid is a mineral lubricating oil with an API gravity of 10, and the system operates in a closed loop. Oil is continuously circulated from and back to a storage tank (T-101) using a 15-stage ESP driven by an 18-horsepower motor. This experimental facility is situated at the Technological Centre for Industrial Automation (CTAI) at the Federal University of Bahia (UFBA).

\begin{figure}[h!]
	\centering
		\includegraphics[scale=.2]{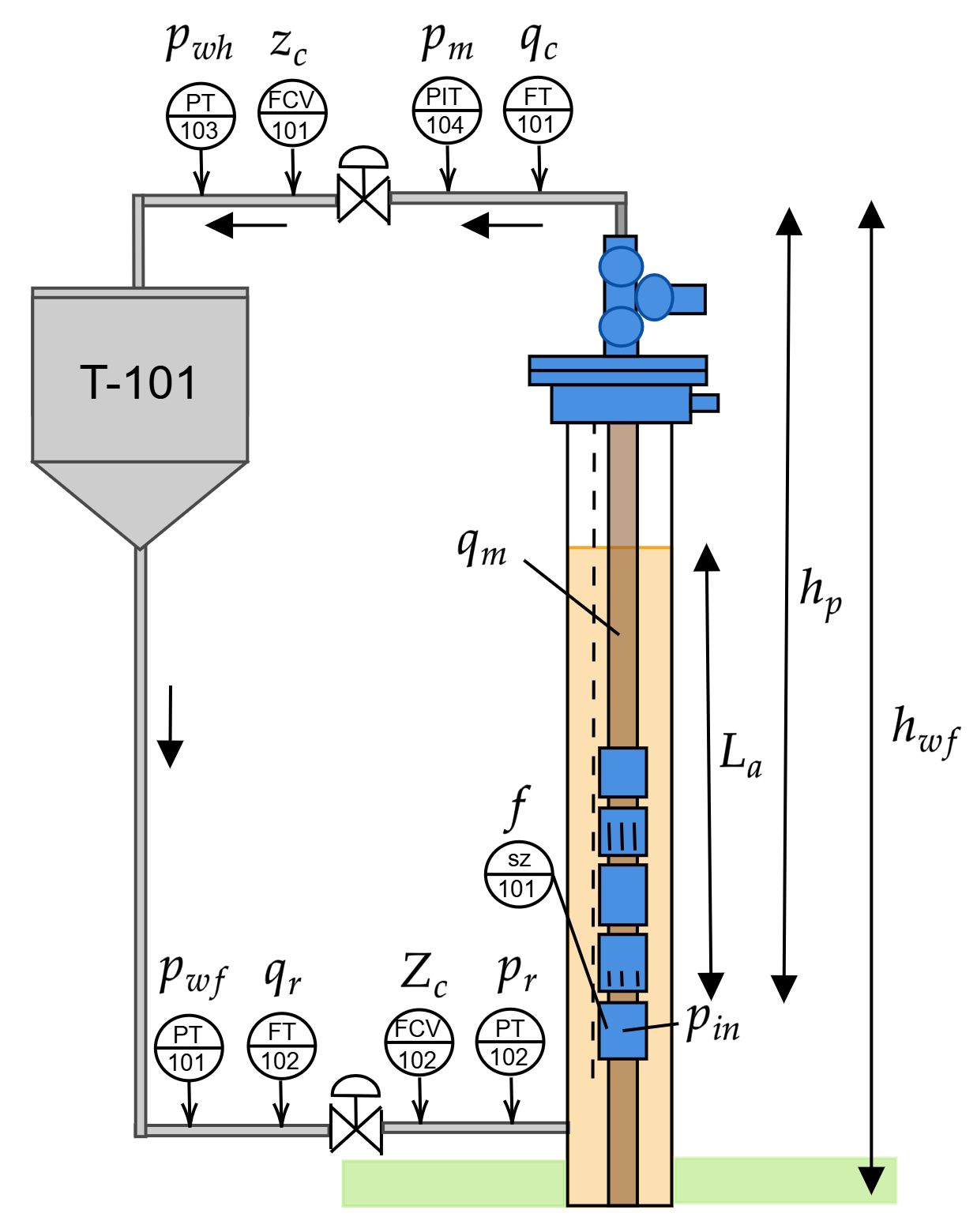}
	\caption{Illustrative layout of the pilot-scale ESP plant.}
	\label{fig:ESP}
 \end{figure}

Unlike full-scale industrial systems, the pilot plant utilises a simplified configuration with fewer pump stages and a lower-power motor. It is instrumented with sensors to monitor pressure, flow rate, and motor speed, and is supported by auxiliary equipment including valves, chokes, and a compressor unit. Figure~\ref{fig:ESP} provides a schematic representation of the ESP system and its instrumentation layout within the pilot installation.

This case study was selected explicitly due to the inherent dynamic constraints associated with ESP-operated systems. These constraints arise from the complex interactions between fluid properties, pump characteristics, and control elements such as valves and chokes. An aspect of ESP operation is maintaining the system within its designated operational envelope, defined by flow rate, pressure, and frequency limits. Deviations from this envelope can lead to various adverse effects, including excessive wear on mechanical components and pump degradation. Moreover, ESPs are often classified as mission-critical assets, meaning their failure can result in a complete production halt and significant revenue losses. Therefore, ensuring the system operates within safe and efficient boundaries and providing equipment integrity, process stability, continuous production, and economic risk minimisation is possible.

\subsection{ESP mechanistic model} \label{ESP_model}

The mathematical modelling of the ESP system dynamics is built upon fundamental conservation laws of mass and momentum. The assumptions adopted are: I) the flow is monophasic and incompressible, II) the system operates under isothermal conditions, III) both physical and thermal properties are constant over time. These simplifications help reduce model complexity while capturing the system's key dynamics. Table \ref{tab:esp_dynamics} summarises the formulation of the model used in this study, which was validated by \citep{COSTA2021108880}. 

\makeatletter
\newcommand{\espTableEqAnchors}{%
  \setcounter{equation}{27}
  \refstepcounter{equation}\label{eq:14}
  \refstepcounter{equation}\label{eq:15}
  \refstepcounter{equation}\label{eq:16}
  \refstepcounter{equation}\label{eq:17}
  \refstepcounter{equation}\label{eq:18}
  \refstepcounter{equation}\label{eq:19}
  \refstepcounter{equation}\label{eq:20}
  \refstepcounter{equation}\label{eq:21}
  \refstepcounter{equation}\label{eq:22}
  \refstepcounter{equation}\label{eq:23}
  \refstepcounter{equation}\label{eq:24}
  \refstepcounter{equation}\label{eq:25}
  \refstepcounter{equation}\label{eq:26}
  \refstepcounter{equation}\label{eq:27}
  \refstepcounter{equation}\label{eq:28}
}
\makeatother

\espTableEqAnchors

\begin{table}[ht]
\centering
\caption{Mathematical representation of ESP system dynamics and flow characteristics.}
\label{tab:esp_dynamics}
\renewcommand{\arraystretch}{1.7}
\begin{tabular}{p{0.4\textwidth} p{0.5\textwidth}}
\hline
\textbf{Description} & \textbf{Equation} \\[1ex]
\hline
Dynamics of intake pressure ($p_{in}$) & 
$\displaystyle \frac{d p_{in}}{dt} = \frac{g}{A}\left(q_r - q_m\right)$ \quad (Eq. \ref{eq:14}) \\[2ex]

Dynamics of choke pressure ($p_c$) & 
$\displaystyle \frac{d p_c}{dt} = \frac{1}{V}\left(q_m - q_c\right)$ \quad (Eq. \ref{eq:15}) \\[2ex]

Momentum balance for flow ($q_m$) & 
$\displaystyle \frac{d q_m}{dt} = \frac{1}{l}\Bigl(p_{wf} - p_c - \Delta P_p - \Delta P_h - \Delta P_f\Bigr)$ \quad (Eq. \ref{eq:16}) \\[2ex]

Hydrostatic pressure loss ($\Delta P_h$) & 
$\displaystyle \Delta P_h = \Delta P_{h1} + \Delta P_{h2}$ \quad (Eq. \ref{eq:17}) \\
 & $\displaystyle \Delta P_{h1} = \rho g\left(h_{wf} - h_p\right)$ \quad (Eq. \ref{eq:18}) \\
 & $\displaystyle \Delta P_{h2} = \rho g\left(h_p - h_c\right)$ \quad (Eq. \ref{eq:19}) \\[2ex]

Average density ($\bar{\rho}$) & 
$\displaystyle \bar{\rho} = \frac{\rho_1 V_1 + \rho_2 V_2}{V_1 + V_2}$ \quad (Eq. \ref{eq:20}) \\[2ex]

Pressure increment by ESP ($\Delta P_p$) & 
$\displaystyle \Delta P_p = H(f,q_b) g,\quad H(f,q_b)=H_0(q_b,f_0)\left(\frac{f}{f_0}\right)^2$ \quad (Eq. \ref{eq:21}) \\[2ex]

Pump polynomial for freshwater ($H_0$) & 
$\displaystyle H_0 = x_2\,q_m^2 + x_1\,q_m + x_0$ \quad (Eq. \ref{eq:22}) \\[2ex]

Frictional pressure loss ($\Delta P_f$) & 
$\displaystyle \Delta P_f = F_1 + F_2$ \quad (Eq. \ref{eq:23}) \\
 & $\displaystyle F_i = f_i \left(\frac{\rho L_i D_i}{4}\right) \left(\frac{q_i}{A_i}\right)^2$ \quad (Eq. \ref{eq:24}) \\[2ex]

Average system length ($\bar{l}$) & 
$\displaystyle \bar{l} = \frac{l_1 + l_2}{2}$ \quad (Eq. \ref{eq:25}) \\[2ex]

Flow rates through valves ($q_r,q_c$) & 
$\displaystyle q_r = K_r Z_r \sqrt{p_r - p_{wf}}$ \quad (Eq. \ref{eq:26}) \\
 & $\displaystyle q_c = K_c Z_c \sqrt{p_c - p_{wf}}$ \quad (Eq. \ref{eq:28}) \\
\hline
\end{tabular}
\end{table}

The system behaviour is described using a set of differential-algebraic equations that govern process variables such as the pump intake pressure ($p_{in}$), choke pressure ($p_{c}$), and the mean flow rate through the production column ($q_m$) (Eq. \eqref{eq:14} - \eqref{eq:16}). The intake pressure change rate ($\frac{d p_{in}}{dt}$) is determined by the difference between the reservoir inflow ($q_r$) and the column flow rate ($q_m$), adjusted for gravitational effects ($g$) and pressure contributions from the reservoir ($p_r$ and $p_{wf}$), modulated by valve parameters ($K_r$, $Z_r$) and cross-sectional area ($A$) (Eq. \eqref{eq:14}).

Similarly, the choke pressure ($p_c$) evolves depending on the imbalance between the inflow from the column ($q_m$) and the outflow through the production choke ($q_c$), modulated by choke valve dynamics ($K_c$, $Z_c$), and the fluid volume ($V$) (Eq. \eqref{eq:15}). The momentum equation describes the transient flow behavior of the column flow rate ($q_m$) and accounts for the total pressure drop along the wellbore, including contributions from the pump ($\Delta P_p$), hydrostatic effects ($\Delta P_h$), and frictional losses ($\Delta P_f$), distributed over the column length ($l$) (Eq. \eqref{eq:16}).

The hydrostatic pressure difference ($\Delta P_h$) is separated into two components ($\Delta P_{h1}$ and $\Delta P_{h2}$), each corresponding to a vertical segment of the well and calculated as a product of fluid density ($\rho$), gravitational acceleration ($g$), and height differences ($h_{wf}$, $h_p$, $h_c$) (Eqs. \eqref{eq:17}–\eqref{eq:19}). The average fluid density ($\bar{\rho}$) between control volumes is computed using a volume-weighted average of segment densities ($\rho_1$, $\rho_2$) and volumes ($V_1$, $V_2$) (Eq. \eqref{eq:20}).

The pressure gain provided by the ESP ($\Delta P_p$) is determined using a frequency-dependent head function ($H(f,q_b)$), which scales the reference pump curve ($H_0(q_b, f_0)$) using affinity laws involving the actual ($f$) and reference ($f_0$) frequencies (Eq. \eqref{eq:21}). The head at the reference condition ($H_0$) is modelled as a second-order polynomial of the flow rate ($q_m$), with coefficients $x_0$, $x_1$, and $x_2$ empirically determined (Eq. \eqref{eq:22}).

Frictional pressure loss ($\Delta P_f$) is evaluated by summing losses from individual pipe sections ($F_1$, $F_2$), where each term is derived from a standard Darcy–Weisbach-type formulation involving friction factors ($f_i$), fluid density ($\rho$), pipe length ($L_i$), diameter ($D_i$), area ($A_i$), and flow rate ($q_i$) (Eqs. \eqref{eq:23}–\eqref{eq:24}). The system's average effective length ($\bar{l}$) is computed as the mean of the two principal segments ($l_1$, $l_2$) (Eq. \eqref{eq:25}).

Valve flow rates for both the reservoir ($q_r$) and choke ($q_c$) are represented using standard orifice equations, where flow is proportional to the square root of the pressure drop across the valve and scaled by the valve constant ($K$) and opening ($Z$) (Eqs. \eqref{eq:26}–\eqref{eq:28}).

The sampling time was set to 30 s based on expert knowledge of the plant and previous work by \citep{Costa2025DYCOPS,deAbreu2025}. This sampling time can be considered high for typical ESP systems, relative to the response times of some variables; however, it must be high enough to accommodate network and communication latency, as well as processing and implementation issues.

These equations, summarised in Table \ref{tab:esp_dynamics}, collectively form the dynamic model used to simulate and analyse the ESP system. Numerical simulations are performed using stiff solvers such as MATLAB’s ODE15s, which ensures accurate time integration of the stiff dynamics. This model provides a physically consistent and computationally efficient framework for studying system responses under varying operational scenarios. All physical parameters and model-specific constants employed in the mathematical formulation are detailed in Table~\ref{ESP_parameters}.

\begin{table}[h!]
\centering
\caption{Parameters for the ESP Model}
\renewcommand{\arraystretch}{1.4}
\begin{tabular}{p{0.38\textwidth} c c}
\hline
\textbf{Parameter} & \textbf{Value} & \textbf{Unit} \\ \hline
Gravity constant, \( g \) & 9.81 & m/s\(^2\) \\ 
Fluid density, \( \rho \) & 855 & kg/m\(^3\) \\ 
Fluid viscosity, \( \mu \) & 0.00888667 & Pa·s \\ 
Bulk modulus, \( \beta_f \) & \( 1.8 \times 10^9 \) & Pa \\
Cross-section area of the annulus, \( A \) & 0.033595 & m\(^2\) \\ 
Average cross-section area, \( A \) & 0.0108 & m\(^2\) \\ 
Pipe radius below ESP, \( r_1 \) & 0.110 & m \\ 
Pipe radius above ESP, \( r_2 \) & 0.0375 & m \\ 
Length from the pump intake to choke, \( l_2 \) & 22.7 & m \\ 
Length from the reservoir to choke, \( l_1 \) & 9.3 & m \\ 
Average length, \( L \) & 16 & m \\ 
Pipe volume below ESP, \( V_1 \) & 0.3286 & m\(^3\) \\ 
Pipe volume above ESP, \( V_2 \) & 0.103 & m\(^3\) \\ 
Head’s polynomial coefficient, \( x_2 \) & -75.397 & s\(^2\)/m\(^5\) \\ 
Head’s polynomial coefficient, \( x_1 \) & 7.8962 & s/m\(^3\) \\ 
Head’s polynomial coefficient, \( x_0 \) & 180.7986 & m \\ 
Reservoir valve constant, \( K_r \) & 0.7221 & m\(^3\)/s\(^0.5\) \\ 
Choke valve constant, \( K_c \) & 0.2283 & m\(^3\)/s\(^0.5\) \\ 
Minimum flow rate, \( q_{\text{min}} \) & \( 1 \times 10^{-3} \) & m\(^3\)/s \\ 
Maximum flow rate, \( q_{\text{max}} \) & 0.1 & m\(^3\)/s \\ 
ESP characteristics ref freq, \( f_0 \) & 60 & Hz \\ 
Height from the pump intake to choke, \( h_p \) & 22.7 & m \\ 
Height from the reservoir to choke, \( h_{\text{bh}} \) & 32 & m \\ 
\hline
\end{tabular}
\label{ESP_parameters}
\end{table}

\subsection{Implementation of the control strategy}

The proposed control strategy relies on accurately modelling the causal relationship between manipulated and controlled variables. Symbolic-regression models were identified in a NARX framework, mapping past values of the manipulated variables to future values of the controlled variables; this preserves causality and enables reliable multi-step-ahead predictions. 

\begin{figure}[h!]
	\centering
		\includegraphics[scale=.5]{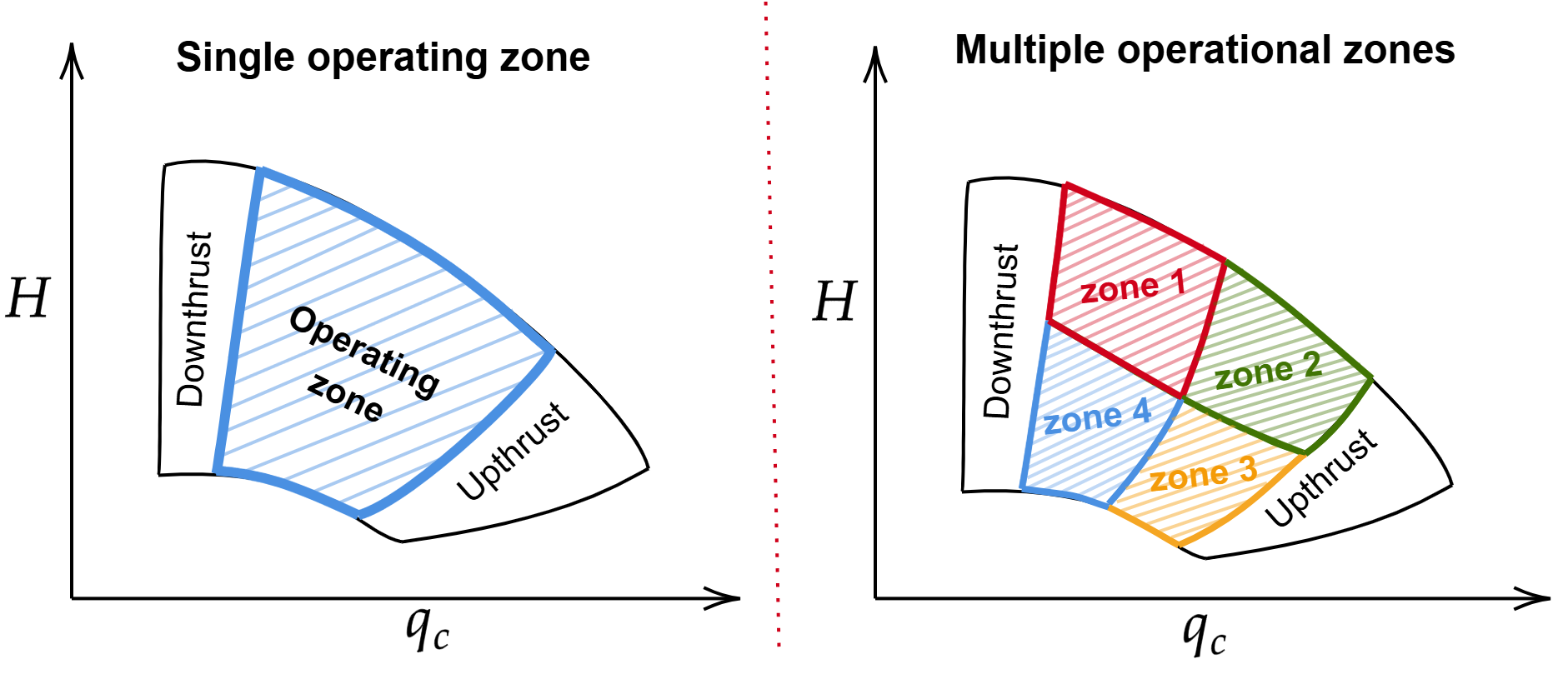}
	\caption{Illustrative of single-zone and multi-zone data-acquisition strategies for SR–NARX identification on the ESP operating envelope.}
	\label{fig:control_strategy_SZ_MZ}
 \end{figure}

Two complementary data-acquisition strategies were adopted for the identification of symbolic regression models (Figure \ref{fig:control_strategy_SZ_MZ}):

\begin{enumerate}
    \item \textbf{Singles operating zones data acquisition (SZ):} a single dataset was generated to cover the entire admissible operating envelope of the ESP system, treating it as a single operating region. This approach exposes the symbolic regression algorithm to the full dynamic range, enabling the discovery of expressions valid across the whole envelope.

    \item \textbf{Multiple operating zones data acquisition (MZ):} the operating envelope was divided into four subregions, and an independent dataset was acquired for each. This zonal partition allows the regression models to focus on local nonlinearities and regime-dependent behaviours that may be masked in the global approach. The zone 1 (high
    $f$, small $Z_c$ ) favours maximum pressure, small production and higher energy cost, zone 2 (high $f$, high $Z_c$) sustains high flow and closer to operational limits, zone 3 (low $f$, high $Z_c$) reduces energy consumption at the expense of reduced throughput, zone 4 (low $f$, small $Z_c$) ensures stable operation within mechanical constraints, with lower production.

\end{enumerate}

These two strategies yield two controllers: RNMPC\textsubscript{SZ} and RNMPC\textsubscript{MZ}.

The first one is the proposed robust NMPC scheme, which integrates an ensemble of symbolic regression models for a single operational zone. The nominal SR model is selected via closed-loop (MPC-in-the-loop) testing, evaluating setpoint tracking, disturbance rejection, and constraint satisfaction, and chosen as the variant with the best overall performance; the remaining SR variants stay in the ensemble and are enforced through robust constraints. The second one integrates an ensemble of symbolic regression models for the four operational zones. The nominal SR model is setpoint-scheduled: as the setpoint changes and drives the operating point across regions of the envelope, the controller switches the nominal to the SR model of the active zone, while the remaining zone models stay in the ensemble and are enforced via robust constraints.

The control scheme adopts a 2 × 2 configuration comprising:

\begin{itemize}
  \item \textbf{Manipulated variables:}
  \begin{itemize}
    \item \textit{Motor frequency} (\(f\)): directly affects the operation of the pump.
    \item \textit{Choke valve opening} (\(Z_c\)): adjusts downstream pressure and flow.
  \end{itemize}
  
  \item \textbf{Controlled variables:}
  \begin{itemize}
    \item \textit{Pump head} (\(H\)): represents the pressure increase generated by the pump.
    \item \textit{Production flow rate} (\(q_c\)): corresponds to the fluid production delivered.
  \end{itemize}
\end{itemize}

Specifically, the manipulated variables,$f$ and $Z_c$, are responsible for actuating changes in the process, while the controlled variables, $H$ and $q_c$, represent the key performance indicators to be regulated.

An ensemble of structurally distinct symbolic models, arising from the different operator combinations, is explicitly appended as additional constraints in the robust MPC optimisation problem to capture this uncertainty.
The controller is coded in MATLAB R2023a. Every 30 seconds, the interior-point SQP solver fmincon, supplied with analytical gradients, computes the optimal move and immediately applies it to the ESP mechanistic model, closing the loop at the same rate. 

It is essential to highlight that although detailed mechanistic equations are not required to learn the SR models, minimal process knowledge is used to define operating zones and safety envelopes (downthrust/upthrust), and to interpret controller performance. For validation, it was adopted a three-step protocol: (1) evaluation on held-out test data (MSE/MAE); (2) one-step-ahead dynamic validation with delay realignment at each sample, at time $t$, the model receives measured inputs/outputs and predictions $t{+}1$, to assess temporal propagation and drift; and (3) tested different scenarios in closed-loop use of the SR model in the MPC predictor, assessing tracking, disturbance rejection, robustness.

\section{Results and discussions}

\subsection{Data curation and symbolic regression model}

The first stage of the methodology involved generating and structuring a comprehensive dataset that covered the entire admissible operating envelope of the ESP system, treating it as a single operating zone. 
This dataset provides the foundation for identifying nonlinear SRMs to be embedded in the predictive controller, ensuring that both transient and steady-state dynamics across the entire operational range are represented.

In a complementary stage, the methodology considered a partition of the admissible operating envelope into four distinct operating zones. 

For both the single-zone and multi-zone data-acquisition strategies, the LHS was adopted to ensure consistent dataset generation.

\begin{figure}[h!]
    \centering
    \begin{subfigure}[b]{0.42\textwidth}
        \centering
        \includegraphics[width=\textwidth]{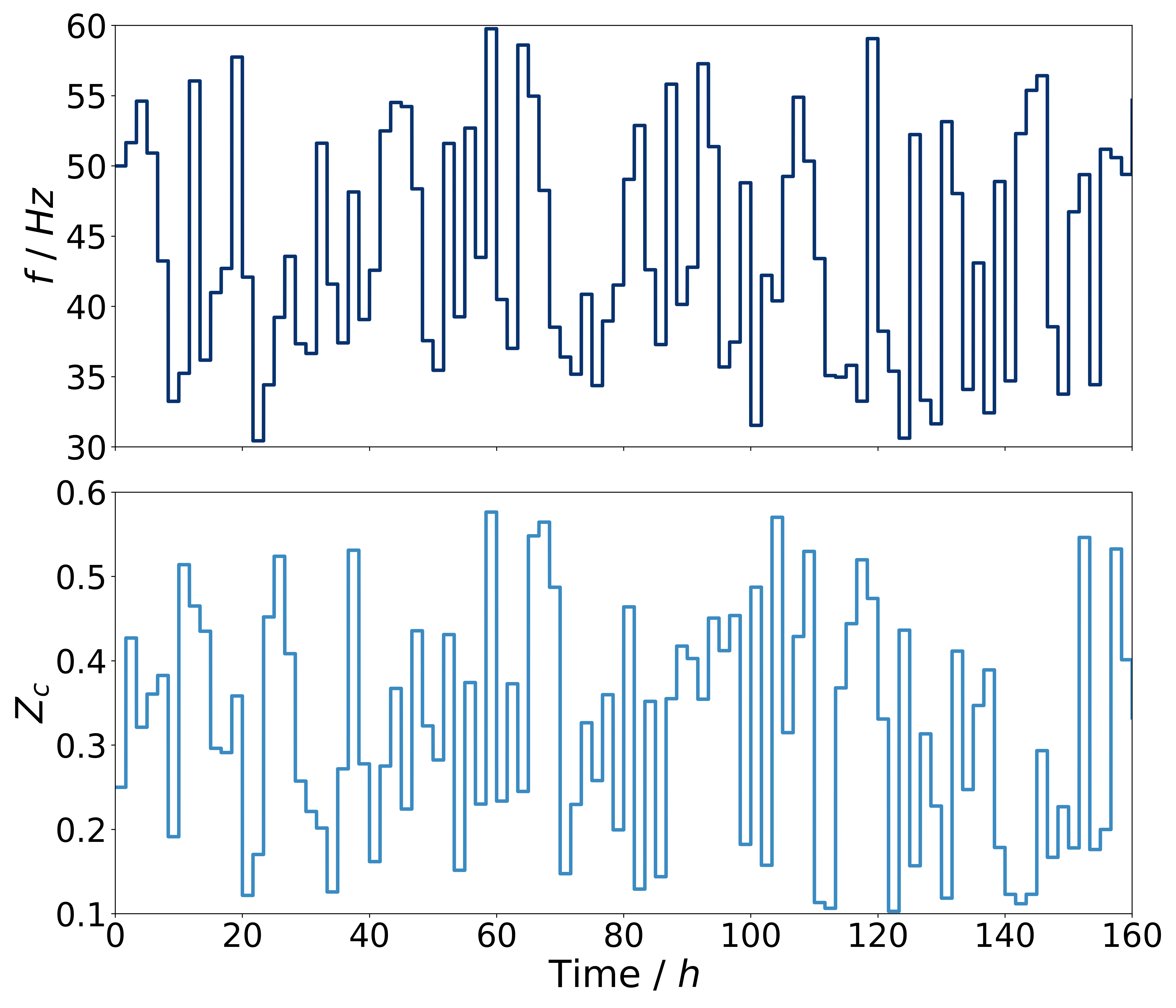}
        \caption{Input dynamic signal.}
        \label{fig:dinamica}
    \end{subfigure}
    \hspace{-0.2cm} 
    \begin{subfigure}[b]{0.50\textwidth}
        \centering
        \includegraphics[width=\textwidth]{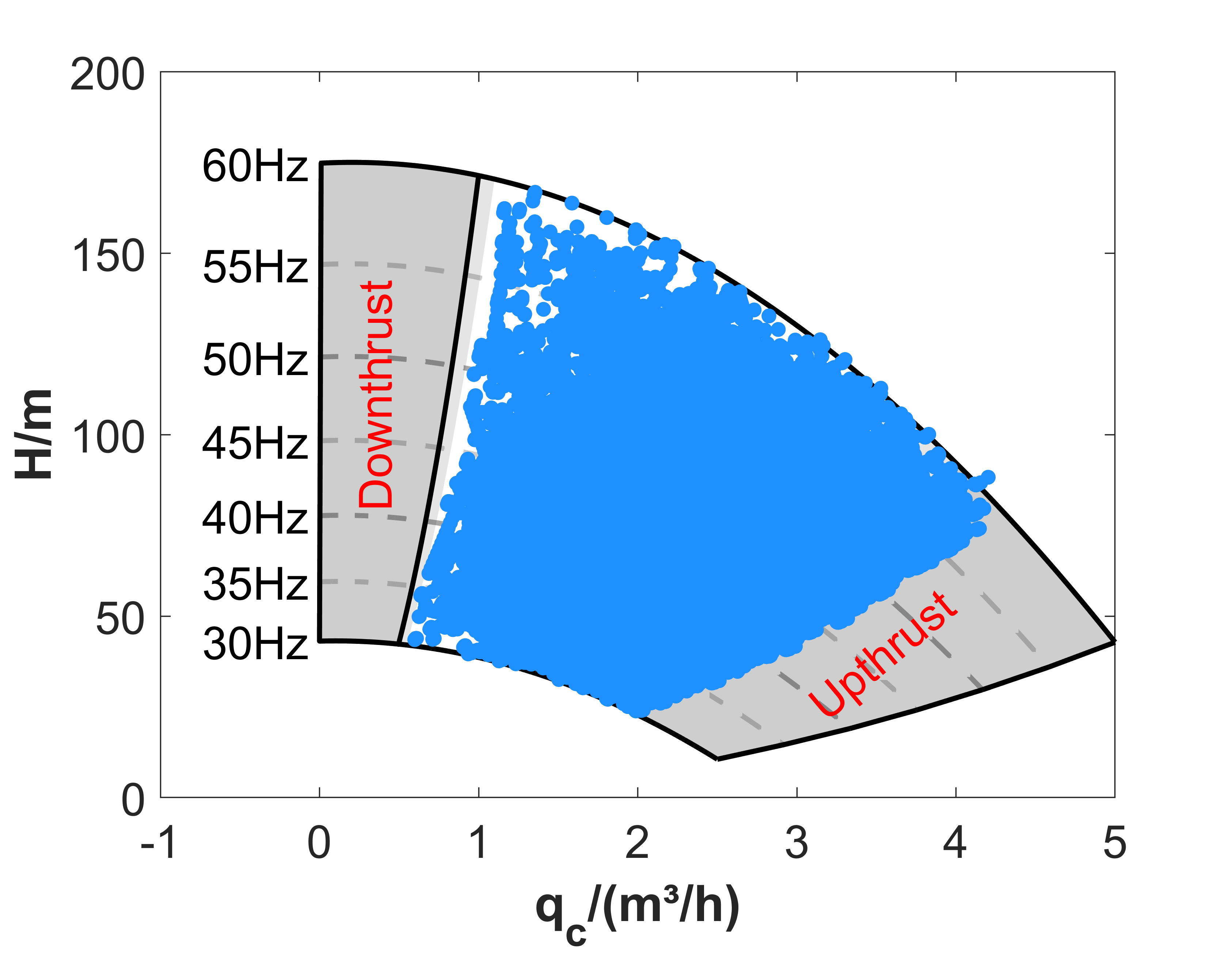}
        \caption{Outputs plotted on the ESP envelope.}
        \label{fig:envelope_lhs}
    \end{subfigure}
    \begin{subfigure}[b]{0.49\textwidth}
        \centering
        \includegraphics[width=\textwidth]{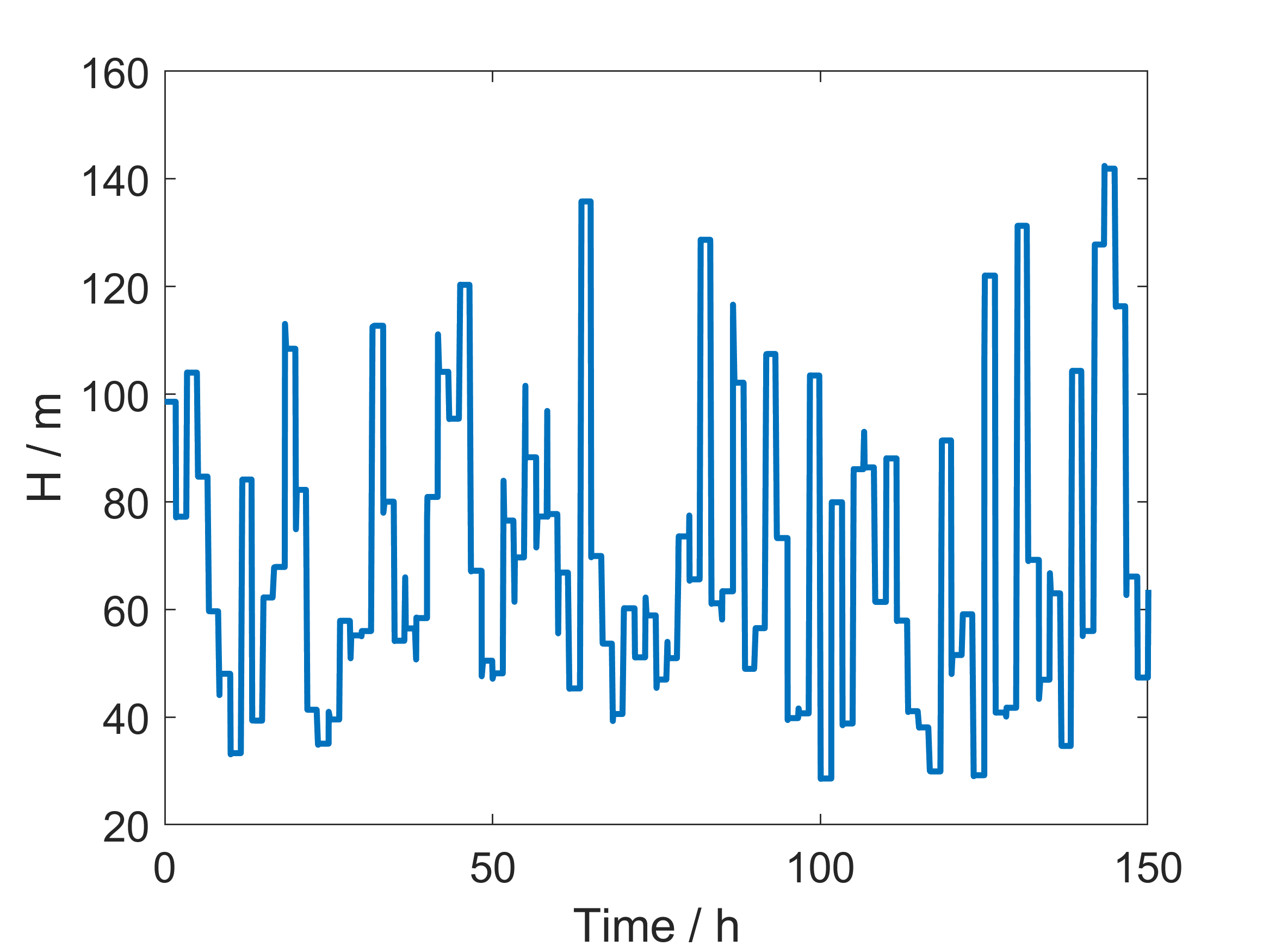}
        \caption{Dynamic $H$ response.}
        \label{fig:Head}
    \end{subfigure}
    \begin{subfigure}[b]{0.49\textwidth}
        \centering
        \includegraphics[width=\textwidth]{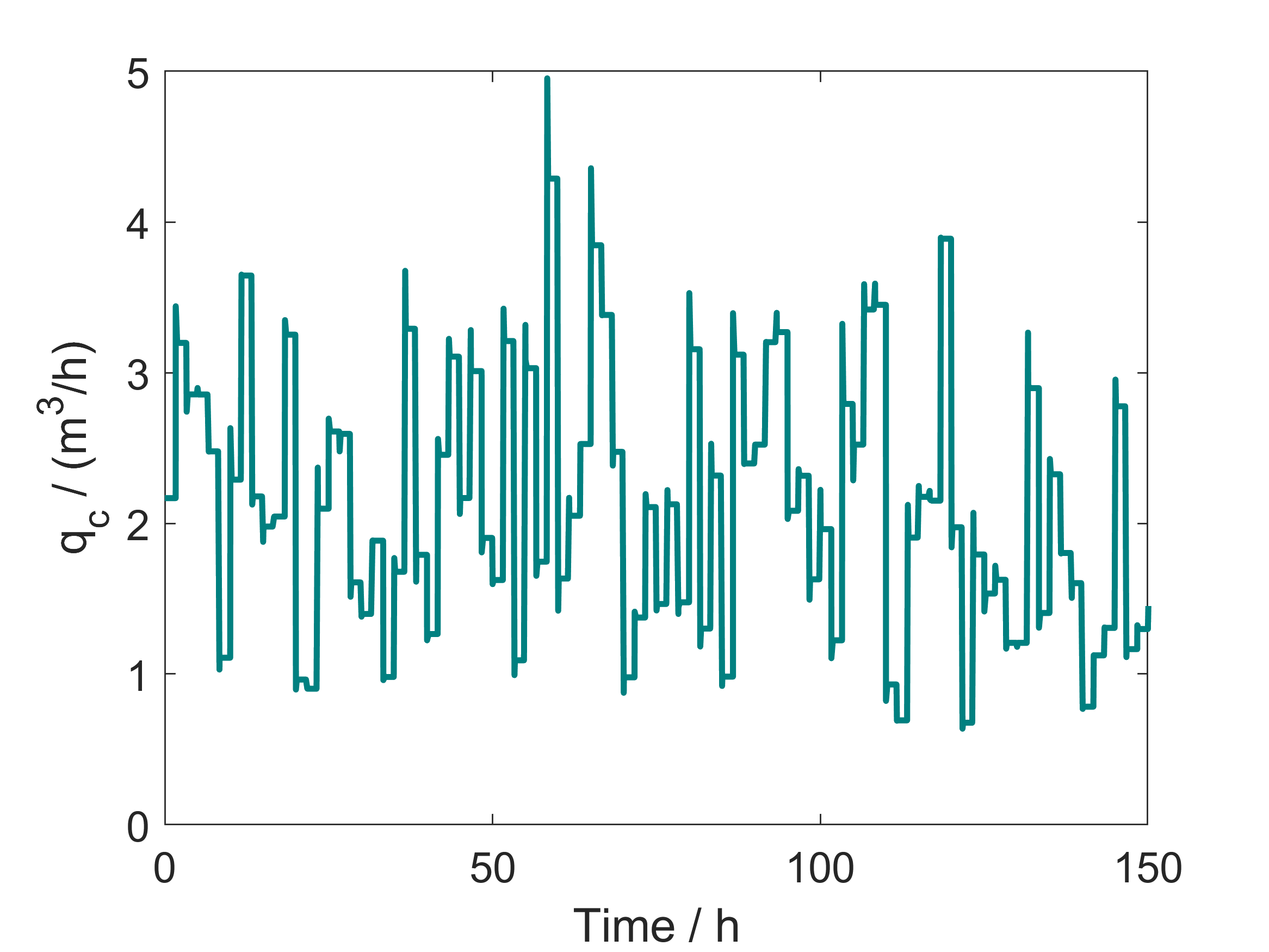}
        \caption{Dynamic $q_c$ response.}
        \label{fig:vazao}
    \end{subfigure}
    \caption{(a) Synthetic perturbation trajectories, (b) corresponding LHS coverage of the operational envelope, (c) dynamic response of the pump head $H$ to the applied LHS perturbations, and 
(d) dynamic response of the flow rate $q_c$ under the same  LHS excitation.}
    \label{fig:dinamica_plots}
\end{figure}

For a single zone approach, LHS was employed to create 30000 input trajectories, motor frequency \(f\) (30–60 Hz) and choke opening \(Z_c\) (0.1–0.6), for perturbing the mechanistic ESP model. Each simulation ran for 150 samples of 30 s each to reach steady state and produce a synthetic output dataset. Figure~\ref{fig:dinamica} shows the first 160 h as an example; these trajectories traverse the full operational envelope with stratified, non‐repetitive intervals, enabling accurate capture of both transient and steady‐state behaviour across diverse conditions, as it is possible to see in Figure \ref{fig:envelope_lhs}, which shows all generated data.

The corresponding dynamic responses are depicted in Figures \ref{fig:Head} and \ref{fig:vazao}, where the head $H$ and flow rate $q_c$ exhibit rich variations induced by the LHS perturbations. Occasional transient jumps in both traces arise from water-hammer caused by abrupt changes in choke opening; the same phenomenon was observed in the experimental plant and reported by \cite{COSTA2021108880}. Moreover, we deliberately included long steady-state holds because the plant typically requires $\approx 2\,\mathrm{h}$ to reach steady state across all variables \citep{COSTA2021108880}, which motivated longer stationary periods during data acquisition.

In the multi-zone strategy, the operating envelope was partitioned into four distinct regions, as summarised in Table~\ref{tab:zones}. 

\begin{table}[H]
\centering
\caption{Operating ranges considered in the multi-zone data acquisition strategy.}
\label{tab:zones}
\begin{tabular}{c c c c c}
\hline
\textbf{Zone} & \(\; f_{\min} \,[\text{Hz}]\;\) & \(\; f_{\max} \,[\text{Hz}]\;\) & \(\; Z_{c,\min}\;\) & \(\; Z_{c,\max}\;\) \\ 
\hline
1 & 45 & 60 & 0.10 & 0.30 \\
2 & 45 & 60 & 0.30 & 0.58 \\
3 & 30 & 45 & 0.30 & 0.58 \\
4 & 30 & 45 & 0.10 & 0.30 \\
\hline
\end{tabular}
\end{table}

\begin{figure}[h!]
    \centering
    \begin{subfigure}[b]{0.45\textwidth}
        \centering
        \includegraphics[width=\textwidth]{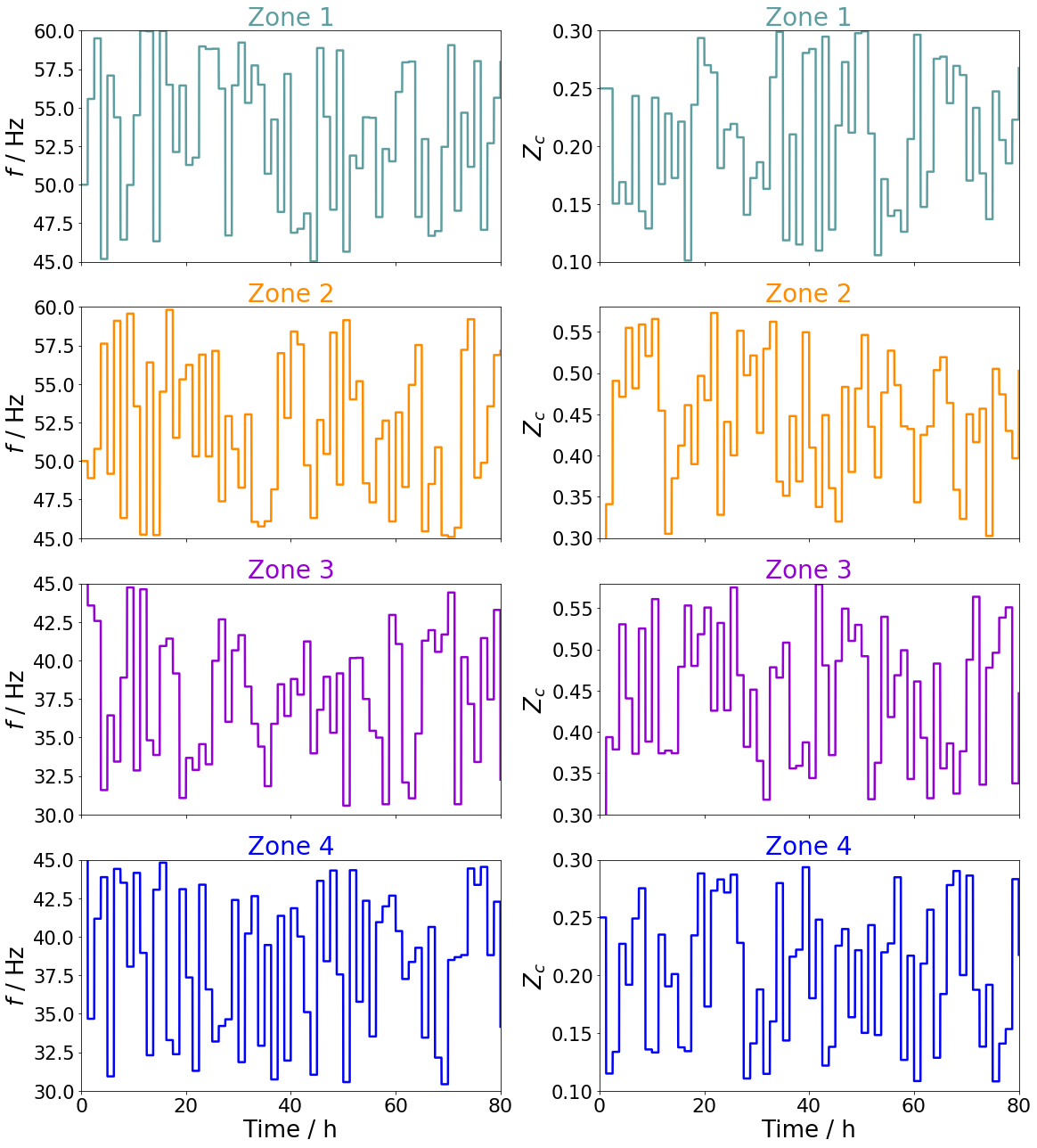}
        \caption{Input dynamic signal.}
        \label{fig:entradas_dinamicas_zonas}
    \end{subfigure}
    \hspace{-0.2cm} 
    \begin{subfigure}[b]{0.45\textwidth}
        \centering
        \includegraphics[width=\textwidth]{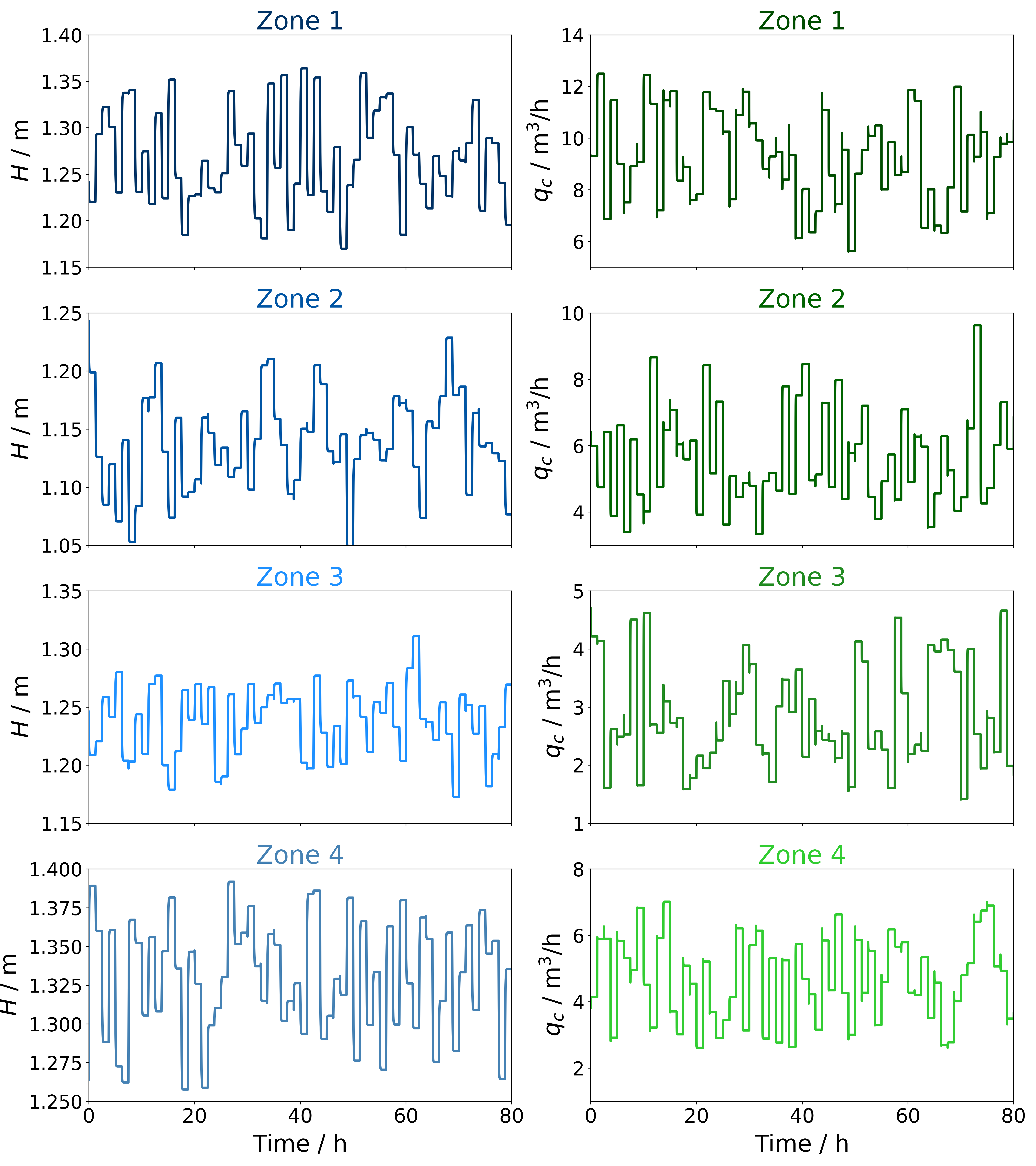}
        \caption{Outputs plotted on the ESP envelope.}
        \label{fig:saida_dinamicas_zonas}
    \end{subfigure}
    \begin{subfigure}[b]{0.6\textwidth}
        \centering
        \includegraphics[width=\textwidth]{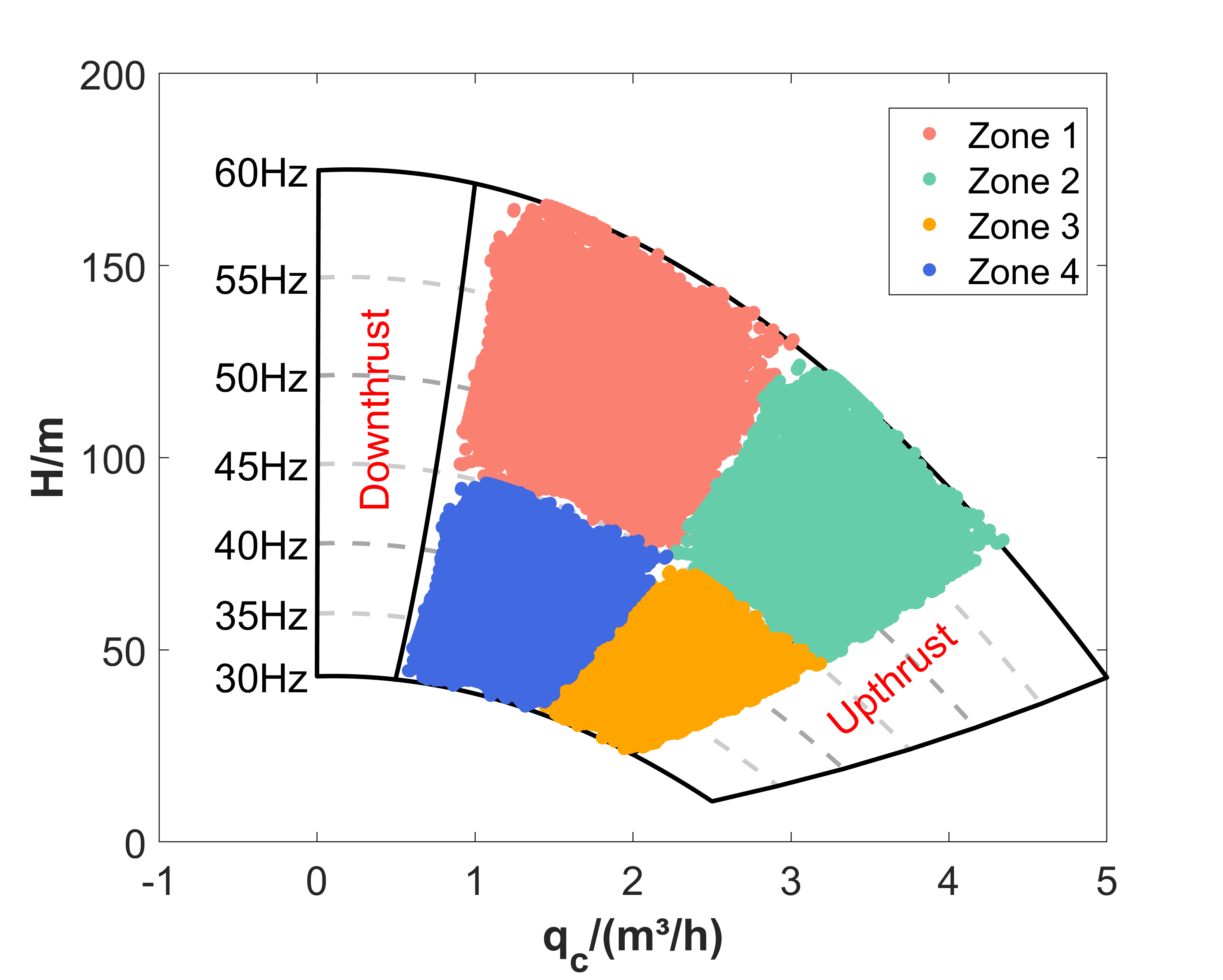}
        \caption{Outputs plotted on the ESP envelope.}
        \label{fig:zonas_operacao}
    \end{subfigure}
    \caption{(Synthetic multi-zone dataset generated by LHS perturbations: 
    (a) excitation signals in  $f$ and $Z_c$, 
    (b) dynamic response of the $H$, and $q_c$,c) ESP operating envelope is divided into four regions according to Table~\ref{tab:zones}, used to guide the multi-zone modeling strategy.}
    \label{fig:dinamica_plots}
\end{figure}

Figure~\ref{fig:zonas_operacao} illustrates the multi-zone data acquisition strategy. 
For each region of the operating envelope, an independent dataset was generated using LHS within the corresponding ranges of $f$ and  $Z_c$. 
Figure \ref{fig:entradas_dinamicas_zonas} shows the synthetic excitation signals, while panel~(b) presents the resulting dynamic responses of pump head $H$ and flow rate $q_c$. 
Figure \ref{fig:saida_dinamicas_zonas} depicts the ESP operating envelope divided into four zones, as defined in Table~\ref{tab:zones}, highlighting the trade-offs between maximum flow, energy consumption, and mechanical constraints. 
This zonal partition enables the symbolic regression models to capture the local, nonlinear behavior specific to each regime, which would otherwise be masked in a global approach.

The next step was to structure the synthetic time series in an NARX format with $n_u = 1$ (input lag) on each manipulated variable and $n_y = 1$ (output lag), applied to both the single and the multi-zone datasets. In this setup, each training sample comprises $\{f(t-1), Z_c(t-1), y(t-1)\}$ and is used to predict $y(t)$, embedding the system’s immediate memory for one-step-ahead forecasting. Two independent symbolic regression models were identified under this formulation: one for the $H$ and another for the $q_c$. During prediction within the NMPC horizon, these models are simulated iteratively in parallel: at each step, the candidate input sequence from the optimiser is applied to both models, generating the predicted values of $H$ and $q_c$, which are then recursively used as regressors for the subsequent steps. This procedure ensures that the coupled behaviour of the two outputs is reconstructed within the prediction horizon.

\begin{figure}[h!]
	\centering
		\includegraphics[scale=.6]{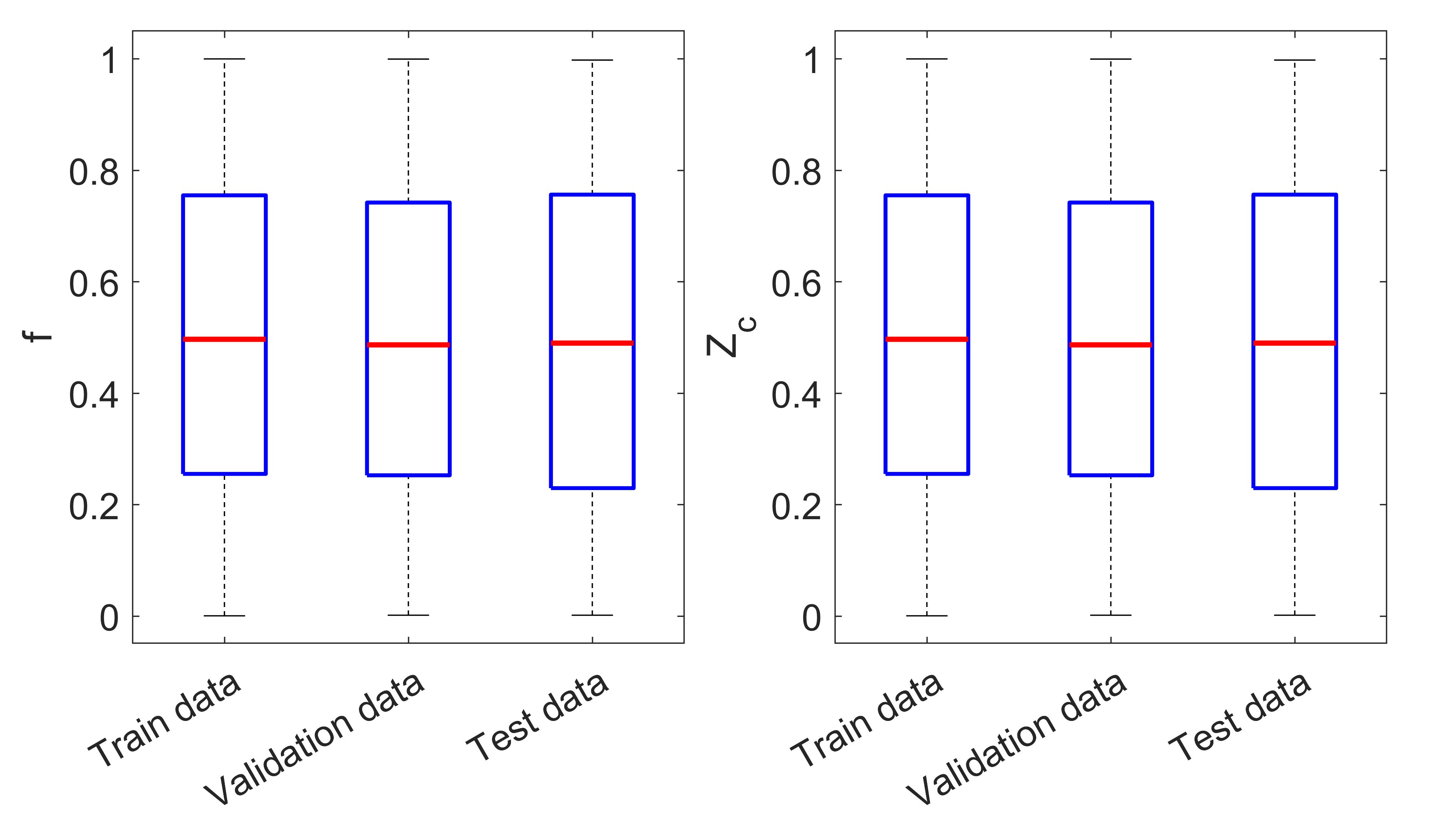}
	\caption{Boxplot of input–output variable distributions across training and test sets.}
	\label{fig:boxplot}
 \end{figure}

Finally, both dataset was partitioned into training (20,000 samples), validation (10\,000 samples), and test (10\,000 samples) sets, and all variables were normalised to zero mean and unit variance. The dataset was maintained in chronological order to preserve the temporal structure important for dynamic modelling. The boxplots in Figure~\ref{fig:boxplot} display aligned medians, interquartile ranges, and whiskers across the three subsets, indicating no significant outliers and a uniform distribution of values for both the single and each zone in the multi-zone datasets approach. This consistency prevents sampling bias, reduces the risk of overfitting, and ensures that model evaluation and controller testing reflect realistic operational scenarios.

The symbolic‐regression search is executed for 1000 generations, minimising the MSE.

Candidate expressions are constructed from the operator set summarised in Table \ref{tab:pysr_operators} to explore a rich yet tractable search space. The four binary arithmetic operators ($+,-,\times,\div$) furnish the basic algebraic scaffold, whereas the unary primitives comprise trigonometric ($\sin,\cos,\tan$), hyperbolic ($\sinh,\cosh,\tanh$), exponential, logarithmic, reciprocal ($ x ^{-1}$), low-order polynomial terms, and the signum function ($\mathrm{sign}$). This configuration, applied in both the single and the multi-zone approaches, strikes a balance between the expressiveness of the search space and control over model complexity and fitting accuracy. It is important to mention that fitting functions to data may require the use of previously unknown nonlinear relationships. Suppose you have prior knowledge about the relationship between the data and its nonlinearity, as well as what the final function should look like. In that case, you can use this information to define the initial set of proposed functions. On the other hand, if this information isn't available, you need to assume functions and test them against the data, eliminating functions or compositions of functions that don't fit the data well. In this paper's approach, it is assumed that there is an unknown relationship between the input and output data, and this is the reason for obtaining a more complex sequence of functions.

\begin{table}[H]
\centering
\caption{Mathematical operators made available to PySRRegressor during the evolutionary search.}
\label{tab:pysr_operators}
\begin{tabular}{lll}
\toprule
\textbf{Operator type} & \textbf{Symbol} & \textbf{Description} \\
\midrule
\multirow{4}{*}{Binary} 
    & $+$ & Addition \\
    & $-$ & Subtraction \\
    & $*$ & Multiplication \\
    & $/$ & Division \\
\midrule
\multirow{12}{*}{Unary} 
    & $\cos(x)$   & Cosine \\
    & $\sin(x)$   & Sine \\
    & $\tan(x)$   & Tangent \\
    & $\cosh(x)$  & Hyperbolic cosine \\
    & $\sinh(x)$  & Hyperbolic sine \\
    & $\exp(x)$   & Exponential \\
    & $\log(x)$   & Natural logarithm \\
    & $\mathrm{inv}(x)=1/x$ & Reciprocal \\
    & $p1(x)=x^{2}$ & Quadratic term \\
    & $p2(x)=x^{3}$ & Cubic term \\
    & $p3(x)=x^{4}$ & Quartic term \\
    & $\mathrm{sign}(x)$ & Sign function: $\{-1,0,1\}$ \\
\bottomrule
\end{tabular}
\end{table}

For the single-zone approach, the best-fitting symbolic regression equations identified to represent the system's dynamic behaviour are summarised in Table~\ref{tab:symb_models}. These models were selected based on their ability to capture the accurate input–output relationships between the manipulated variables and the controlled outputs using the NARX structure. Each expression encodes the system’s nonlinearities and memory effects, offering mathematical analytical formulations.

\begin{table}[h!]
\centering
\caption{Identified SR Models for ESP system (single-zone approach).}
\label{tab:symb_models}
\renewcommand{\arraystretch}{2.0}
\resizebox{\textwidth}{!}{%
\begin{tabular}{c|p{0.79\textwidth}}
\hline
\textbf{Equation} & \textbf{Expression} \\
\hline
\multicolumn{2}{c}{$H(t)$} \\
\hline
\textbf{Nominal model} & $ H(t) = \left| H(t-1) + \left( \sin \left( \left( f(t) - f(t-1) \right) \cdot e^{\left( - \sin^2 \left( Z_c(t-1) \right) \right)} \right) - \sin \left( 0.32 \cdot \left( Z_c(t) - Z_c(t-1) \right) \right) \right) \cdot 0.83 \right| $ \\
\textbf{Model I} & $ H(t) = \left| H(t-1) + \left( \sin \left( \left( f(t) - f(t-1) \right) \cdot e^{ \left( - \sin^2 \left( Z_c(t-1) \right) \right)} \right) - \left( 0.32 \cdot \left( Z_c(t) - Z_c(t-1) \right) \right) \right) \cdot 0.83 \right| $ \\
\textbf{Model II} & $ H(t) = \left| H(t-1) + \big(f(t) - \big(f(t-1) + f(t)\,e^{-Z_c(t)^2}\,(Z_c(t)-Z_c(t-1))\big)\big)\,e^{-Z_c(t-1)} \right| $ \\
\hline
\multicolumn{2}{c}{$ q_c(t)$} \\
\hline
\textbf{Nominal model} & $ q_c(t) = \sin \left( \left( Z_c(t) \right)^{0.58} \cdot \cos \left( \sqrt{ \sin \left( \cos \left( \sqrt{f(t)} + 0.44 \right) \right) }^3 \right) \right) $ \\
\textbf{Model I} & $ q_c(t) = q_c(t-1) + \left( \left( \sin \left( Z_c(t) \right) \cdot \sin \left( f(t) - (-0.33) \right) \right) - q_c(t-1) \right) \cdot \left| \text{sign} \left( Z_c(t) - Z_c(t-1) \right) \right| $ \\
\textbf{Model II} & $ q_c(t) = q_c(t) + \dfrac{\left( -f(t-1) + f(t) \right)\left( \sinh \left( \tan \left( q_c(t) \right) \right) + 0.40 \right)^{Z_c(t)}} {\cosh \left( \tan \left( \cos \left( f(t)^{Z_c(t-1)} \right) \right) \right)} $ \\
\hline
\end{tabular}}
\end{table}

\begin{figure}[h!]
    \centering
    \begin{subfigure}[b]{0.42\textwidth}
        \centering
        \includegraphics[width=0.8\linewidth]{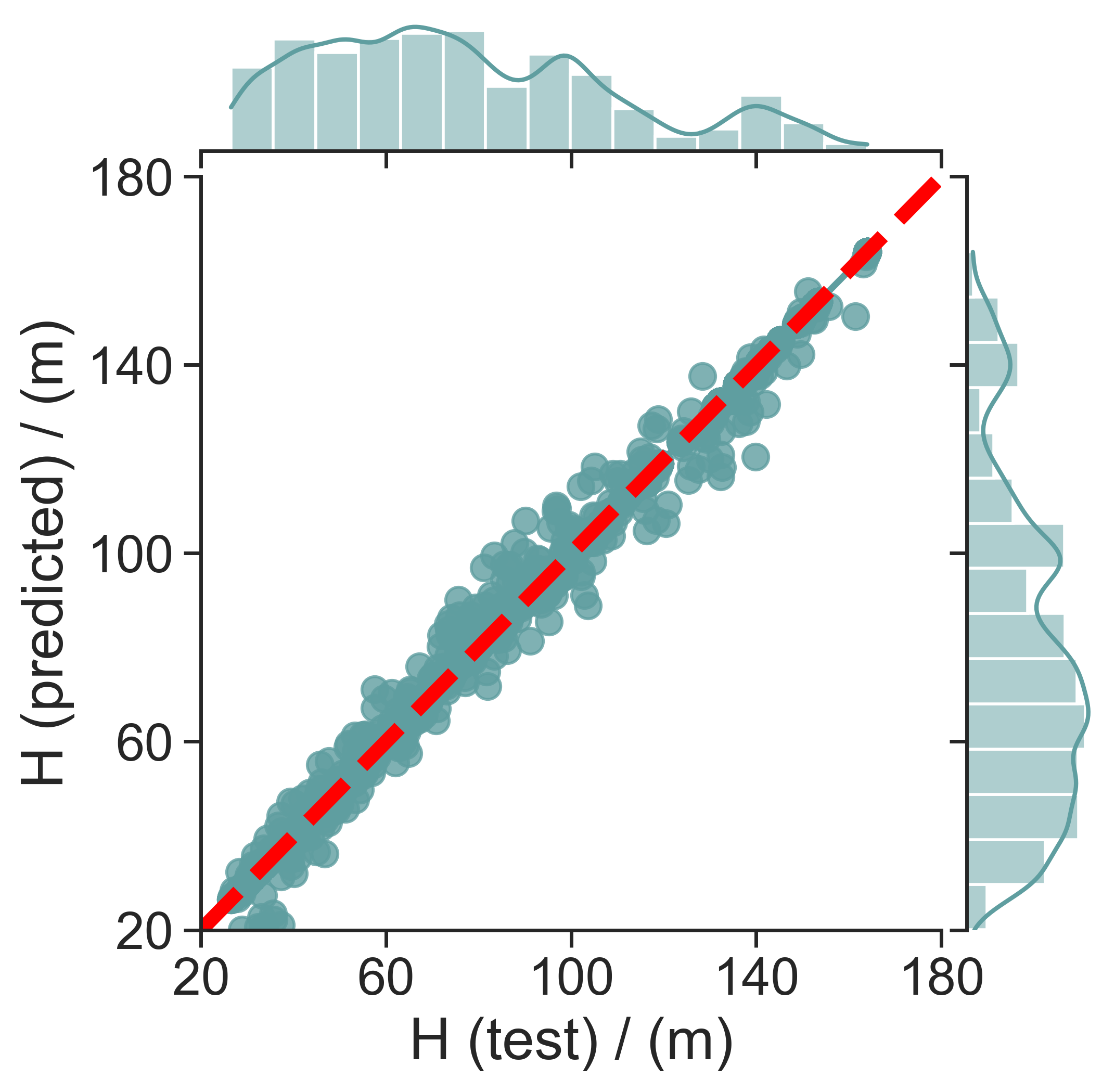}
        \caption{$H$ }
        \label{fig:narx_h_teste_1}
    \end{subfigure}
    \hspace{0.015\textwidth}
    \begin{subfigure}[b]{0.42\textwidth}
        \centering
        \includegraphics[width=0.8\linewidth]{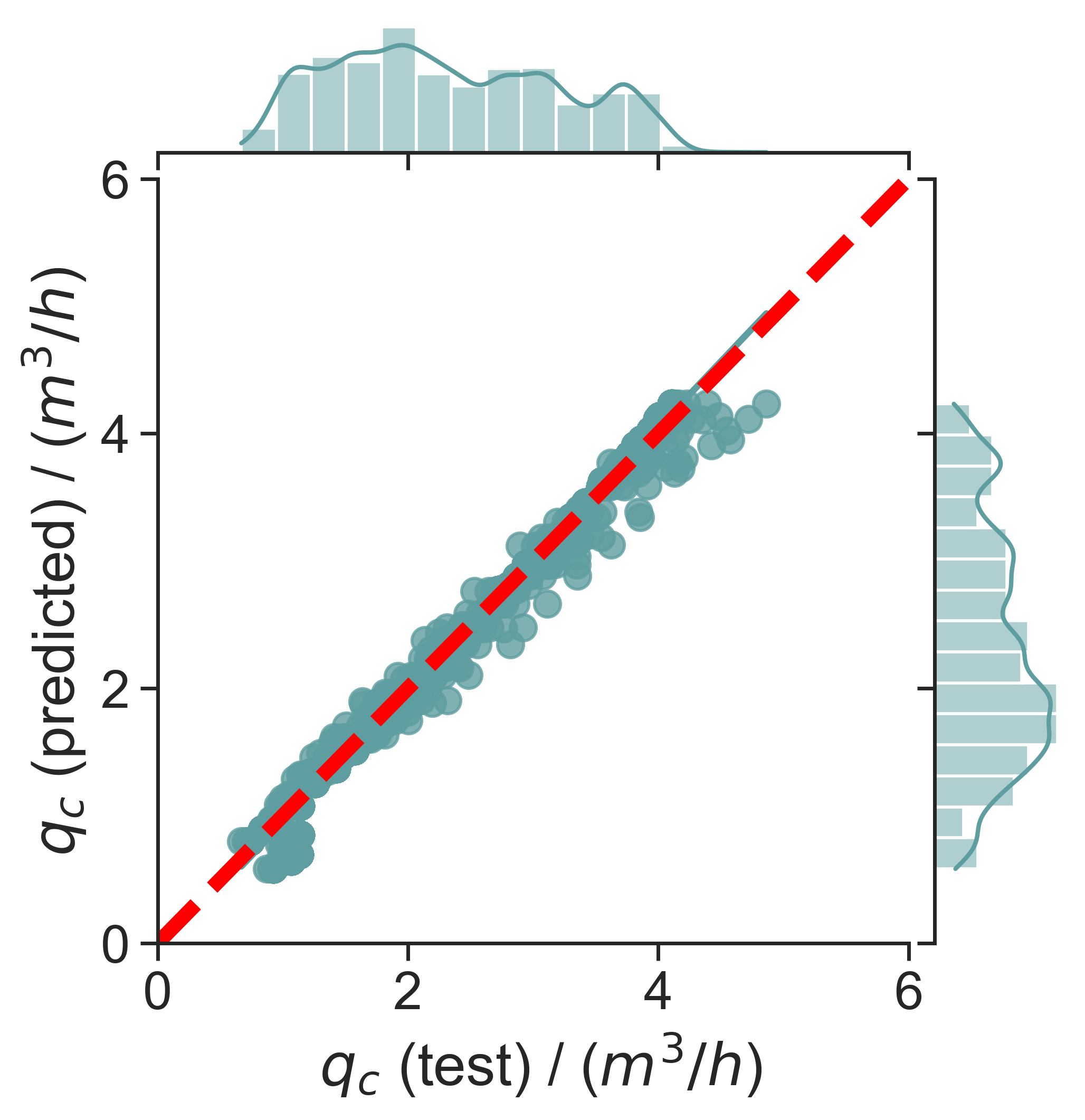}
        \caption{$q_c$ }
        \label{fig:narx_qm_teste_1}
    \end{subfigure}

    \caption{Parity plots of SR models on test data.}
    \label{fig:comparacao_testes_1}
\end{figure}

For the single-zone approach, the symbolic regression models were tested using an independent dataset not seen during training to evaluate the predictive accuracy and generalisation capability. Figure \ref{fig:comparacao_testes_1} presents the parity plots for both $H(t)$ and $q_m(t)$, demonstrating the alignment between predicted and actual values. The points are well distributed around the identity line, indicating that the symbolic models accurately capture the underlying system dynamics. 

In addition, Figure \ref{fig:comparacao_testes_2} also displays the time-domain trajectories of the nominal models over the test set, reinforcing their ability to reproduce the transient and steady-state behaviours of the system under realistic scenarios. 

\begin{figure}[h!]
    \centering 
    \begin{subfigure}[b]{0.45\textwidth}
        \centering
        \includegraphics[width=\textwidth]{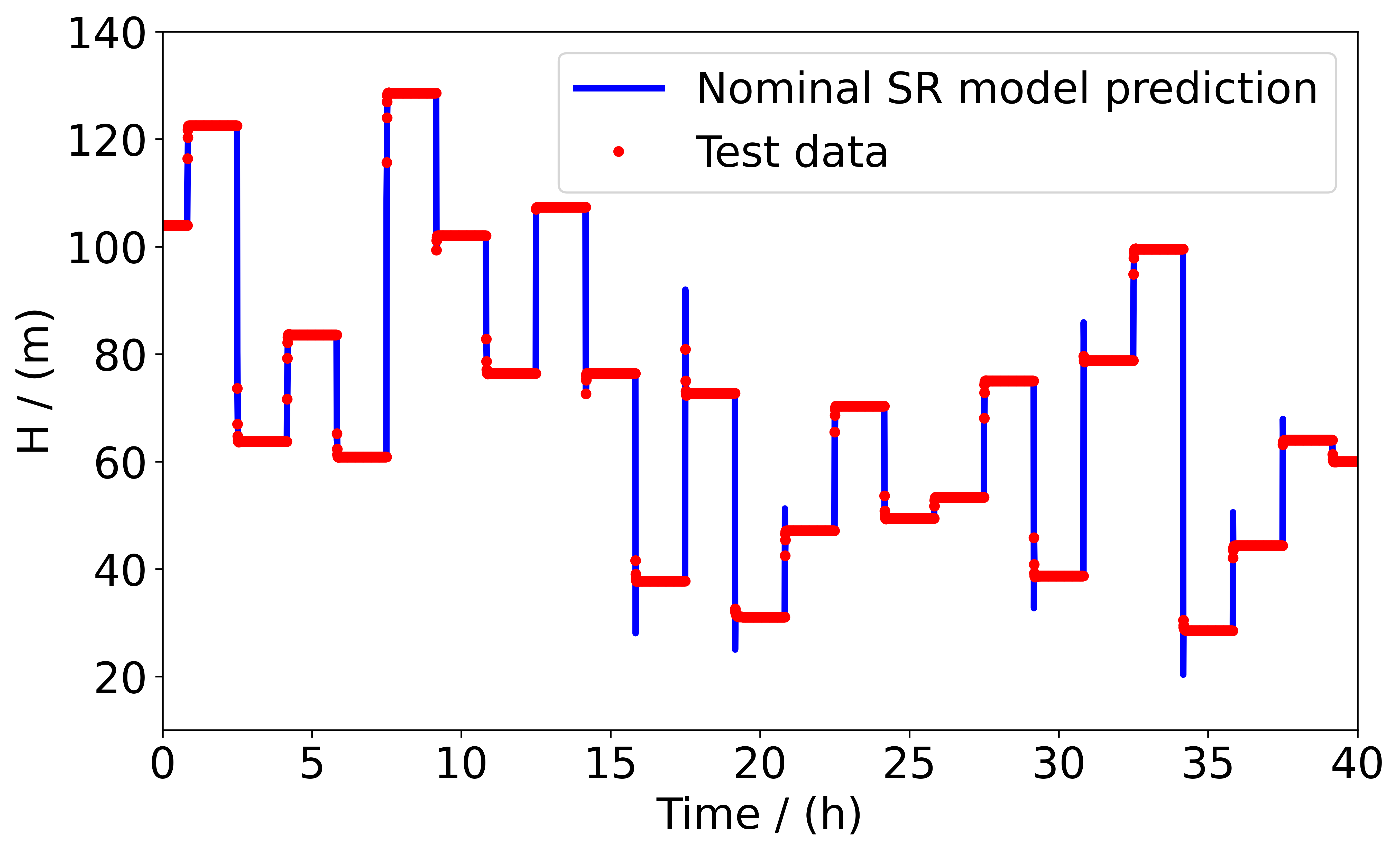}
        \caption{$H$}
        \label{fig:narx_h_teste}
    \end{subfigure}
    \hspace{0.015\textwidth}
    \hfill
    \begin{subfigure}[b]{0.45\textwidth}
        \centering
        \includegraphics[width=\textwidth]{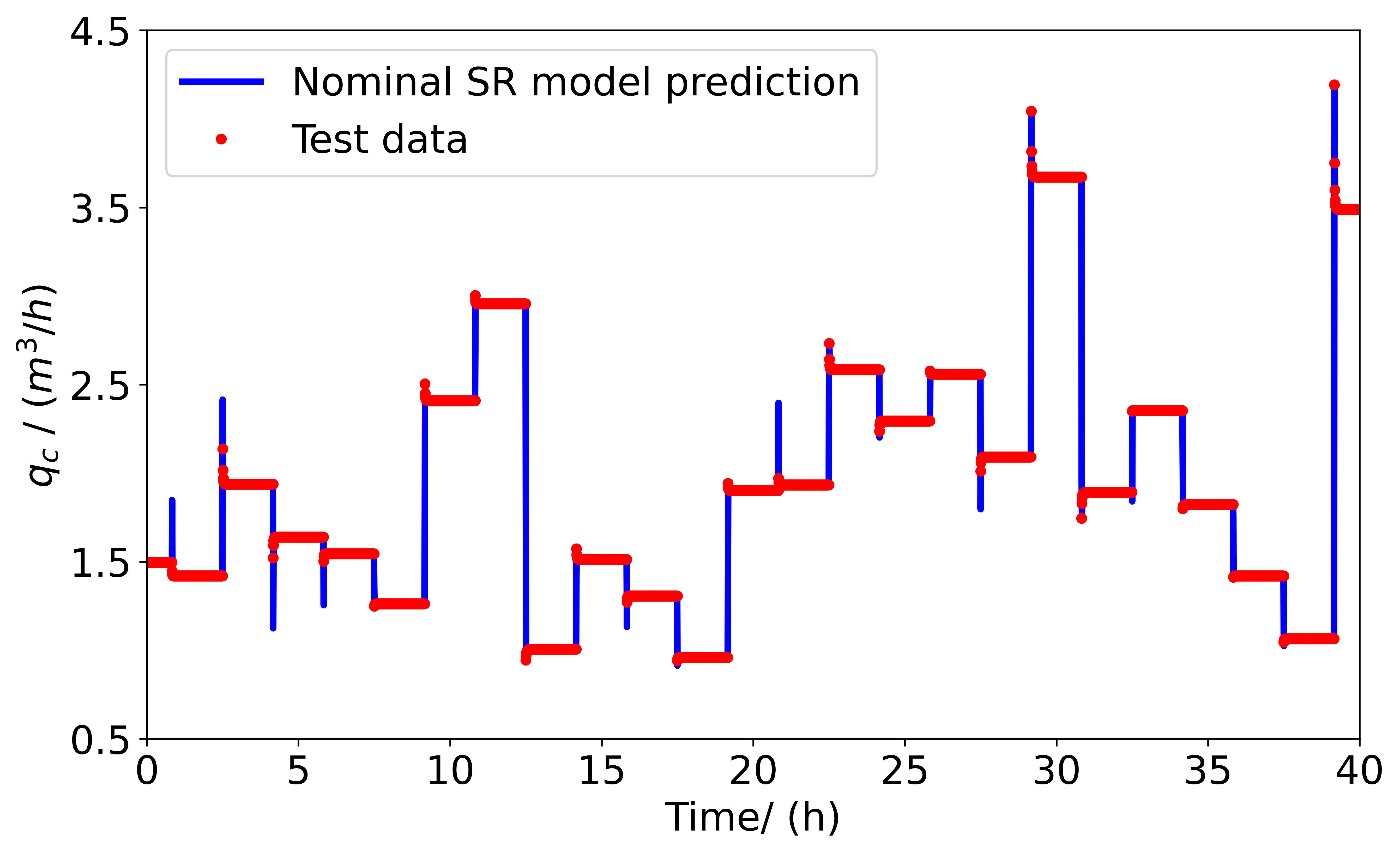}
        \caption{$q_c$ }
        \label{fig:narx_qm_teste}
    \end{subfigure}
\caption{Time-series dynamics of the nominal SR model predictions.}
    \label{fig:comparacao_testes_2}
\end{figure}

In the single-zone case, the quantitative performance of the symbolic models is reported in Table~\ref{tab:metrics_single_grouped} through the MSE and mean absolute error (MAE). 
For $H(t)$, both the nominal and model~II and III variants achieved the highest accuracy, with MSE values on the order of $10^{-5}$. The $q_c$ models exhibited close error levels across all formulations. 
To avoid redundancy, only the parity and dynamic responses of the nominal models are presented here, while the complete set of results can be found in Appendix~\ref{appendix}.

\begin{table}[h!]
\centering
\caption{MSE and MAE for SR models (single-zone), grouped by model.}
\label{tab:metrics_single_grouped}
\renewcommand{\arraystretch}{1.2}
\begin{tabular}{l cc cc cc}
\toprule
 & \multicolumn{2}{c}{\textbf{Model I (Nominal)}} & \multicolumn{2}{c}{\textbf{Model II}} & \multicolumn{2}{c}{\textbf{Model III}} \\
\cmidrule(lr){2-3}\cmidrule(lr){4-5}\cmidrule(lr){6-7}
 & \textbf{MSE} & \textbf{MAE} & \textbf{MSE} & \textbf{MAE} & \textbf{MSE} & \textbf{MAE} \\
\midrule
$H(t)$      & $4.06\times10^{-5}$ & $8.45\times10^{-4}$ & $4.04\times10^{-5}$ & $8.45\times10^{-4}$ & $3.96\times10^{-5}$ & $8.24\times10^{-4}$ \\
$q_c(t)$    & $2.45\times10^{-4}$ & $9.79\times10^{-4}$ & $2.41\times10^{-4}$ & $7.09\times10^{-4}$ & $3.28\times10^{-5}$ & $6.94\times10^{-4}$ \\
\bottomrule
\end{tabular}
\end{table}

In the multi-zone approach, the operating envelope was partitioned into four regions, and independent symbolic regression models were identified for each of them. Table~\ref{tab:symb_models_zones} presents the symbolic regression models identified for that approach. 
For each operating region defined in Table~\ref{tab:zones}, independent surrogate models were obtained for the $H(t)$ and the $q_c(t)$.

\begin{table}[h!]
\centering
\caption{Identified SR models by operating zone (multi-zone approach).}
\label{tab:symb_models_zones}
\renewcommand{\arraystretch}{1.9}
\begin{tabular}{c|p{0.78\textwidth}}
\hline
\textbf{Equation} & \textbf{Expression} \\
\hline
\multicolumn{2}{c}{$H(t)$} \\
\hline
\textbf{Zone 1} &
$\displaystyle
H(t) = H(t-1) + \sin\!\Big( \big(f(t) - f(t-1)\big) \cdot 
\Big( \cos\!\big(\sqrt{Z_c(t-1)}\big) + \tan\!\big(\operatorname{sign}(f(t) - f(t-1))\big) \cdot \big(Z_c(t-1) - Z_c(t)\big) \Big) \Big).
$
\\
\textbf{Zone 2} &
$\displaystyle
H(t) = \Big|\, 
0.27\, Z_c(t-1) 
- 0.27 \, Z_c(t) 
+ H(t-1) 
+ 0.83 \, \big(-f(t-1) + f(t)\big) \, e^{- \, Z_c(t-1)^2} 
\,\Big|.
$
\\
\textbf{Zone 3} &
$\displaystyle
H(t) = 
\Big( H(t-1) - \big(Z_c(t) - Z_c(t-1)\big) \, H(t-1) \Big) 
- \Big( \big(f(t-1) - f(t)\big) \, \sin^{2}\!\Big( \Big(\tfrac{1.37}{1.8}\Big)^{Z_c(t)} \Big) \Big).
$
\\
\textbf{Zone 4} &
$\displaystyle
H(t) = H(t-1) \;-\; 
\frac{f(t-1) - f(t)}{\sinh^{2}\!\Big( 
\cosh\!\Big( \big(Z_c(t-1) - f(t)\big) \, e^{f(t-1)} + H(t-1) - Z_c(t) \Big) 
\Big)}.
$
\\
\hline
\multicolumn{2}{c}{$q_c(t)$} \\
\hline
\textbf{Zone 1} &
$\displaystyle
q_c(t) = 
\left( 
\frac{q_c(t-1)}{\sinh\!\big(\sqrt{f(t-1)}\big)} 
- \sin\!\Big( \big(Z_c(t-1) - Z_c(t)\big) \cdot 
\cos\!\big(\sin(\sqrt{\sqrt{q_c(t-1)}})\big) \Big) 
\right) 
\cdot \sinh\!\big(\sqrt{f(t)}\big).
$
\\
\textbf{Zone 2} &
$\displaystyle
q_c(t) = \Bigg| \; 
\Big| \, q_c(t-1) + 
\tan\!\Big(\sinh\!\big(f(t) - f(t-1)\big)\Big) \cdot 
\big(0.29^{\,1.13}\big) \Big| 
+ \big(Z_c(t) - Z_c(t-1)\big) \cdot \sqrt{0.29} 
\;\Bigg|.
$
\\
\textbf{Zone 3} &
$\displaystyle
q_c(t) = 
\sin\!\Bigg( 
\frac{\sinh\!\big(q_c(t-1)\big) \cdot \sin\!\big(\sin(Z_c(t))\big)}
     {\sin\!\big(\sin(Z_c(t-1))\big)} 
\cdot \exp\!\left(\frac{f(t) - f(t-1)}{0.81}\right) 
\Bigg).
$
\\
\textbf{Zone 4} &
$\displaystyle
q_c(t) = 
\exp\!\Bigg( 
- \Bigg[ 
\Big( Z_c(t) + \sin\!\big(Z_c(t-1)\big) + \cos\!\big(e^{\,Z_c(t-1)^3}\big) \Big)^{f(t)} 
- f(t) 
\Bigg]^{2} 
\Bigg) 
\cdot \Big( \sinh\!\big(Z_c(t)\big) + 0.05 \Big).
$
\\
\hline
\end{tabular}
\end{table}

\begin{figure}[h!]
    \centering
    \begin{subfigure}[b]{0.22\textwidth}
        \centering
        \includegraphics[width=\linewidth]{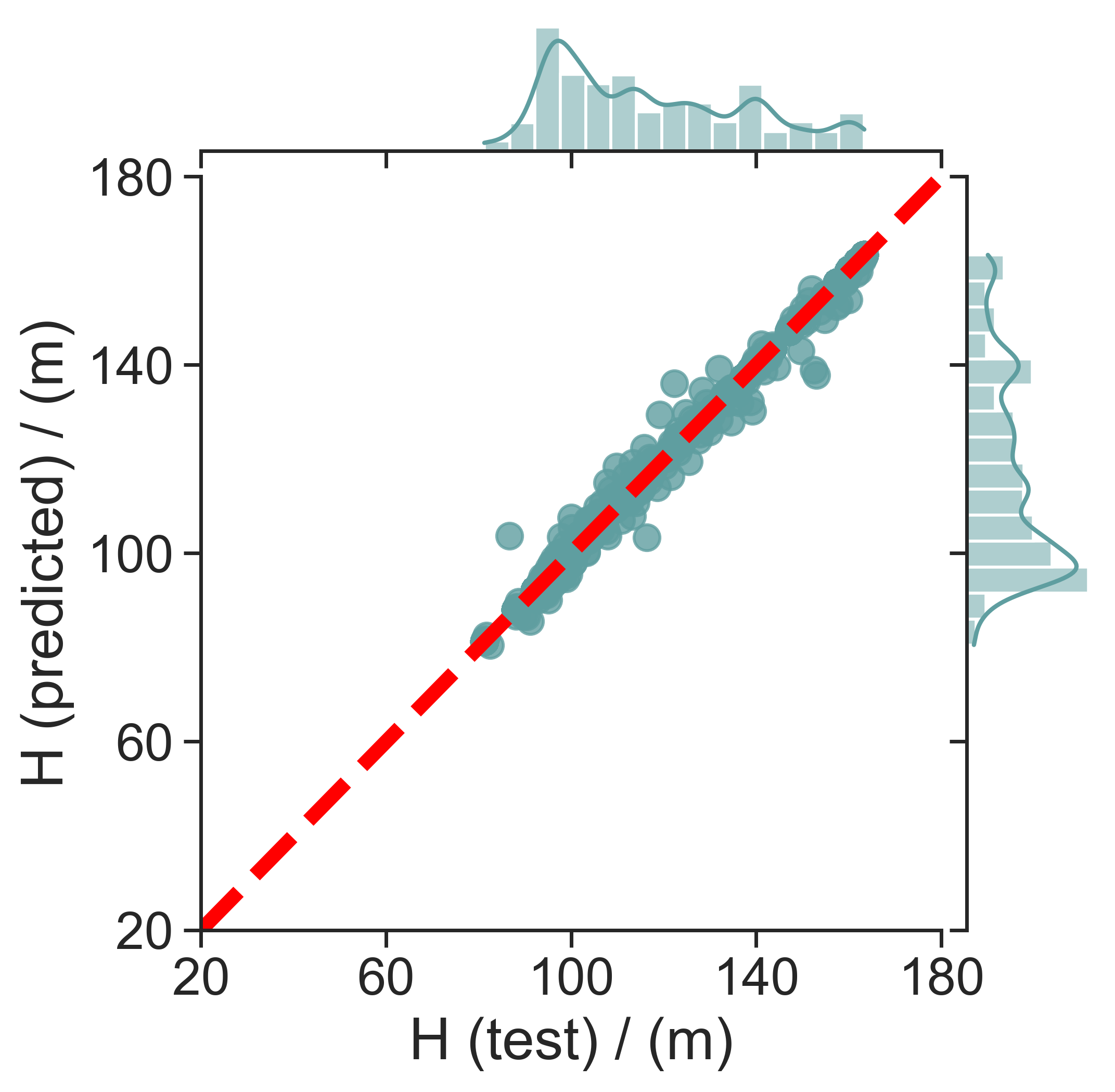}
        \caption{$H$ model I.}
    \end{subfigure}
    \hspace{0.005\textwidth}
    \begin{subfigure}[b]{0.22\textwidth}
        \centering
        \includegraphics[width=\linewidth]{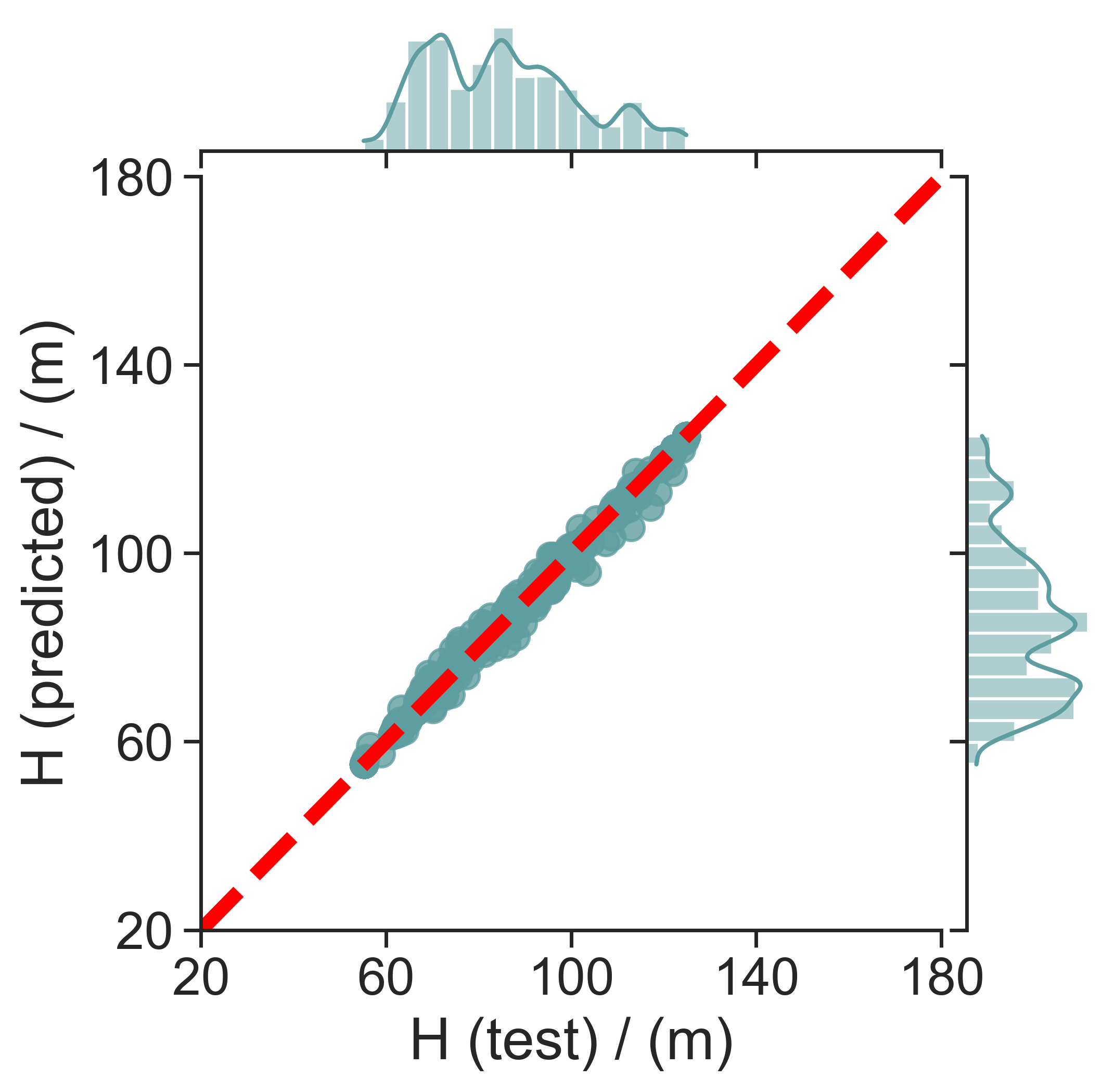}
        \caption{$H$ model II.}
    \end{subfigure}
    \hspace{0.005\textwidth}
    \begin{subfigure}[b]{0.22\textwidth}
        \centering
        \includegraphics[width=\linewidth]{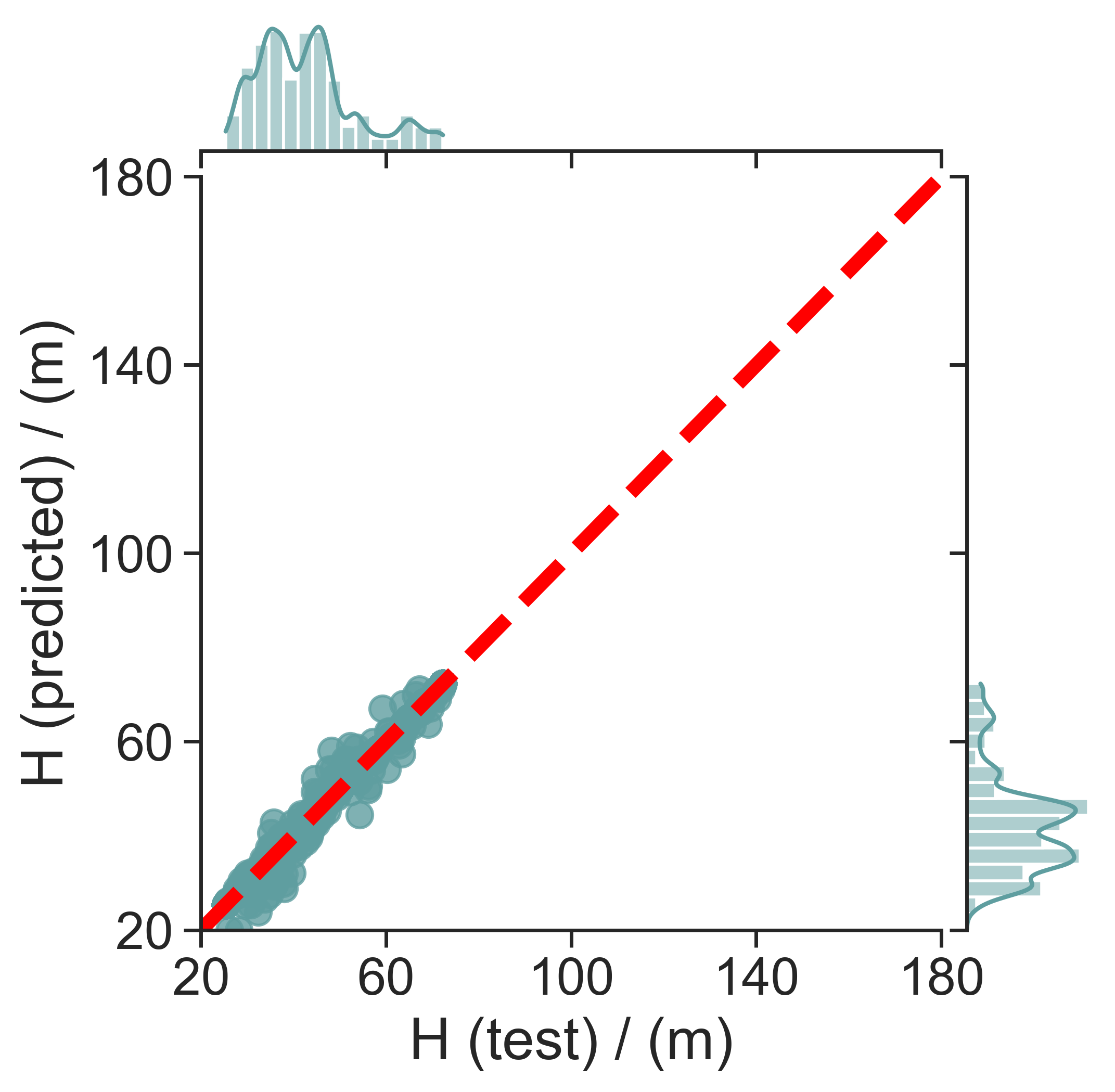}
        \caption{$H$ model I.}
    \end{subfigure}
    \hspace{0.005\textwidth}
    \begin{subfigure}[b]{0.22\textwidth}
        \centering
        \includegraphics[width=\linewidth]{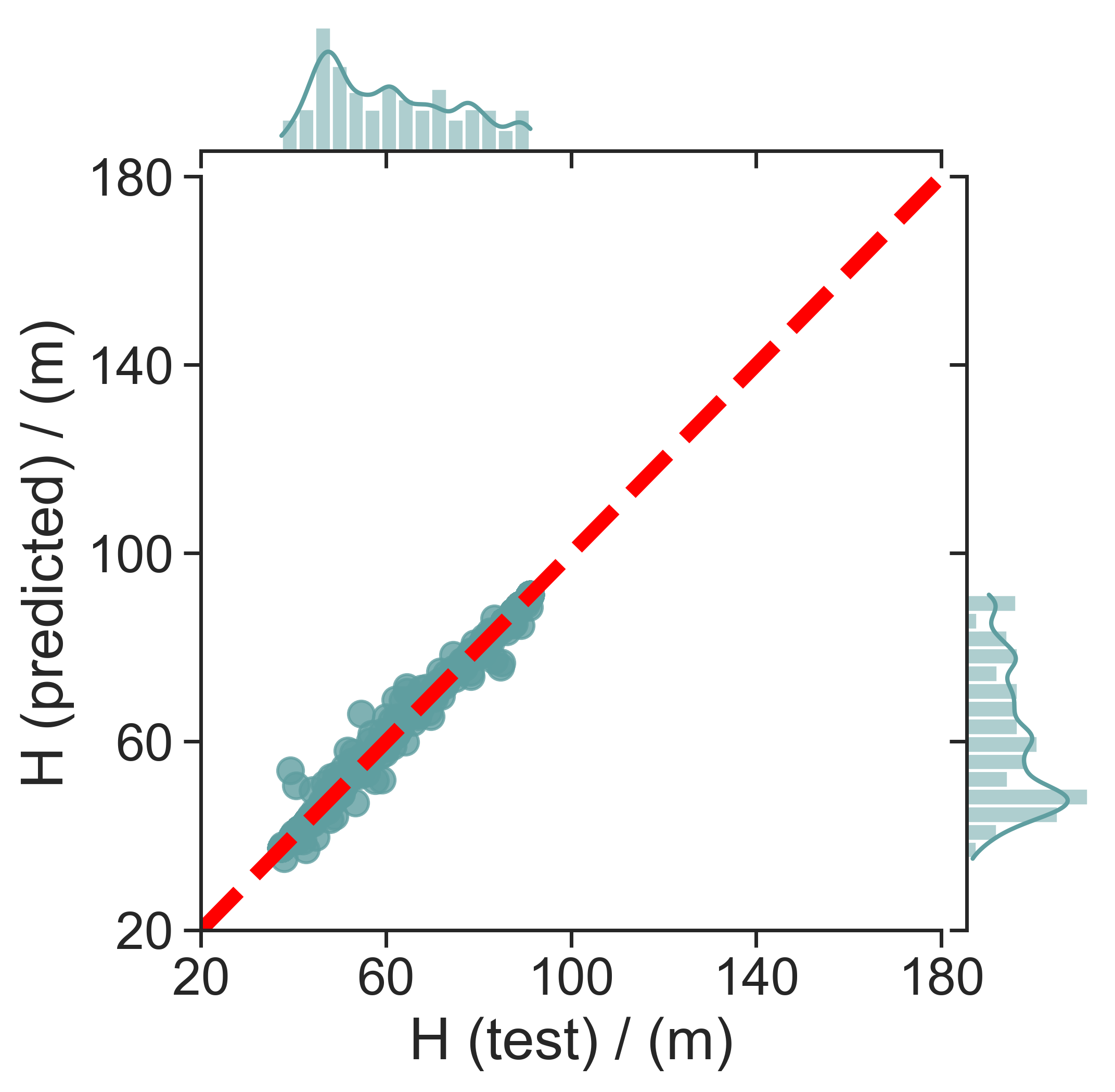}
        \caption{$H$ model II.}
    \end{subfigure}

    \vspace{0.3cm}
    \begin{subfigure}[b]{0.22\textwidth}
        \centering
        \includegraphics[width=\linewidth]{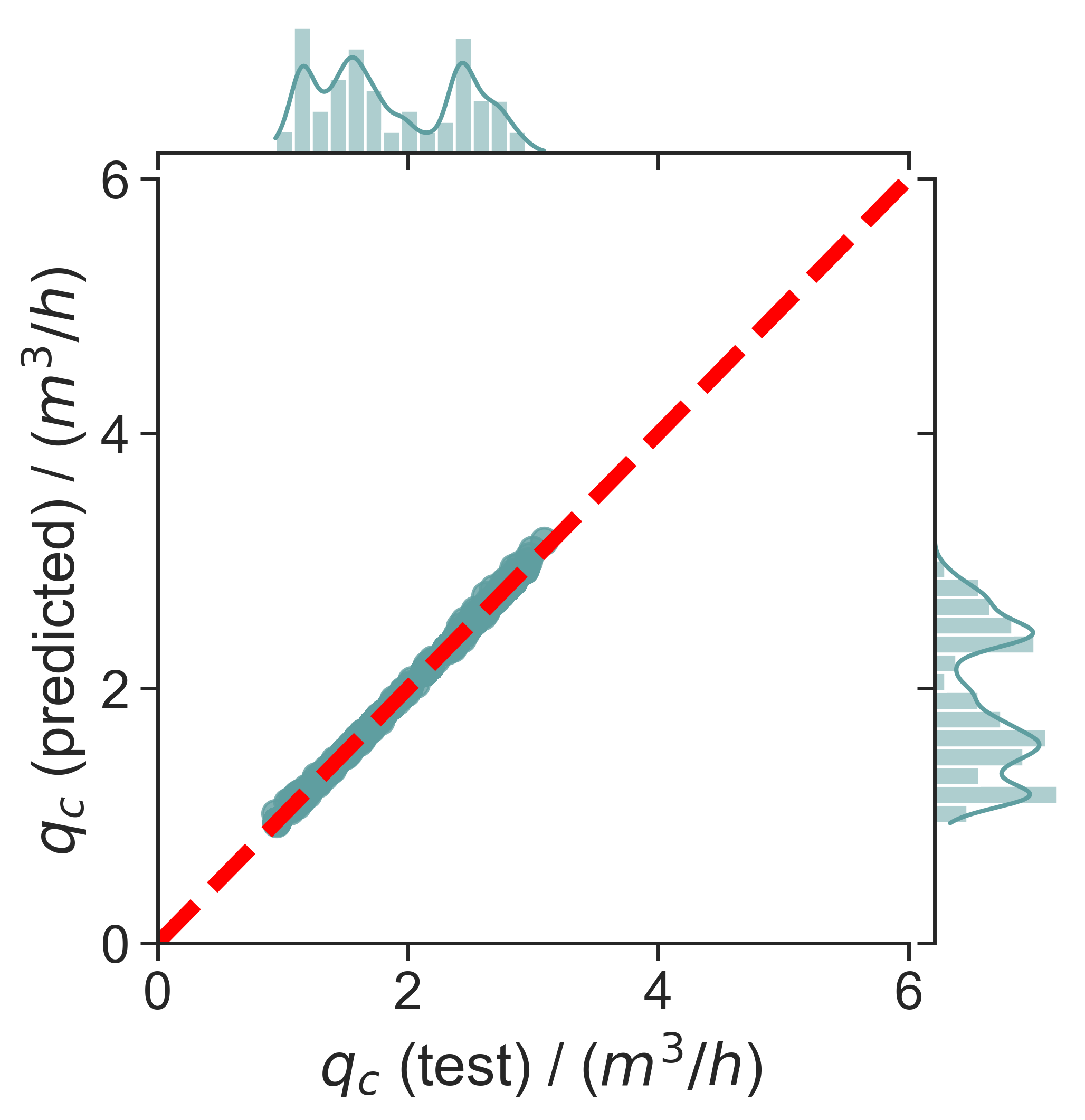}
        \caption{$q_c$ model III.}
    \end{subfigure}
    \hspace{0.005\textwidth}
    \begin{subfigure}[b]{0.22\textwidth}
        \centering
        \includegraphics[width=\linewidth]{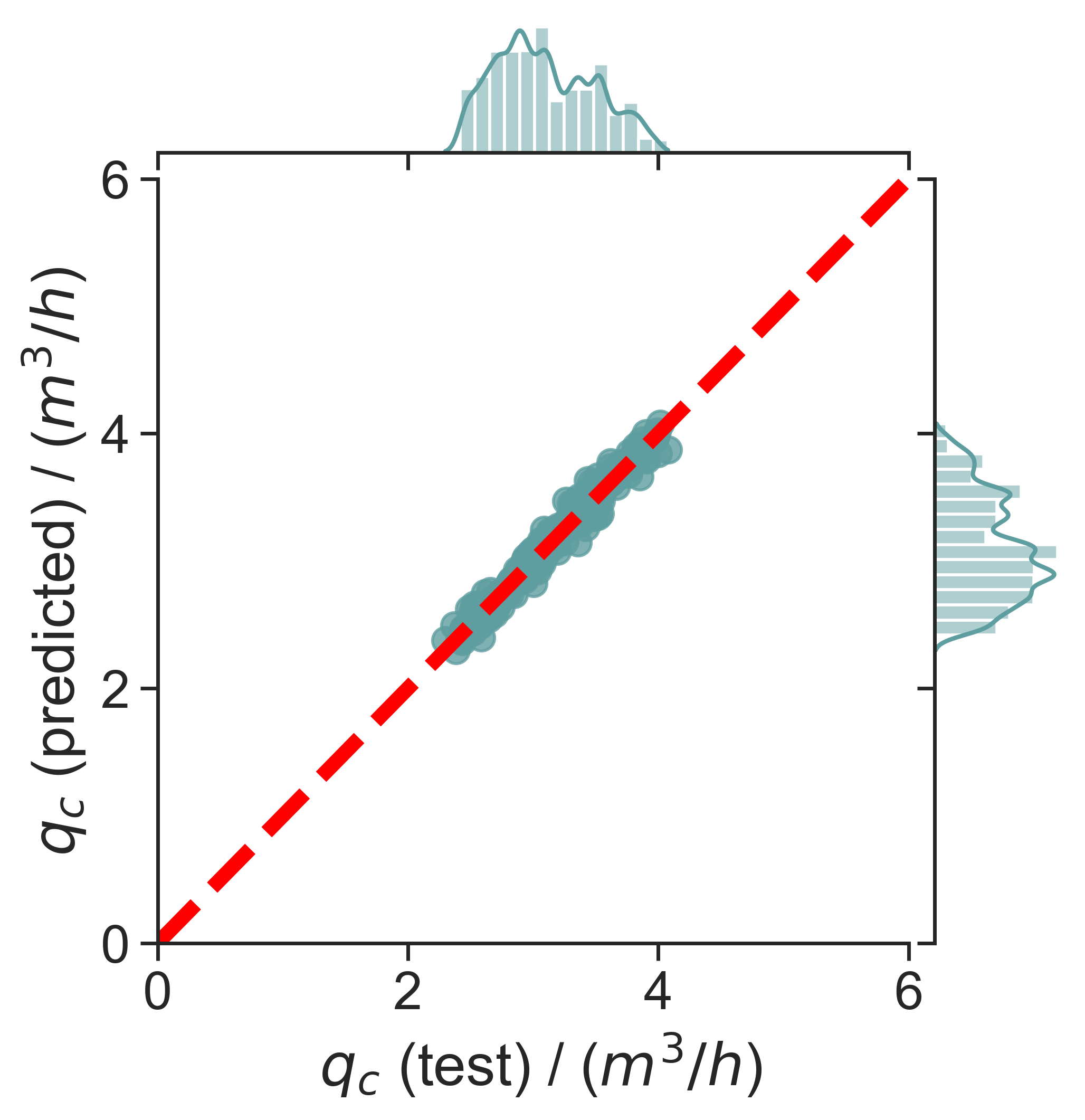}
        \caption{$q_c$ model IV.}
    \end{subfigure}
    \hspace{0.005\textwidth}
    \begin{subfigure}[b]{0.22\textwidth}
        \centering
        \includegraphics[width=\linewidth]{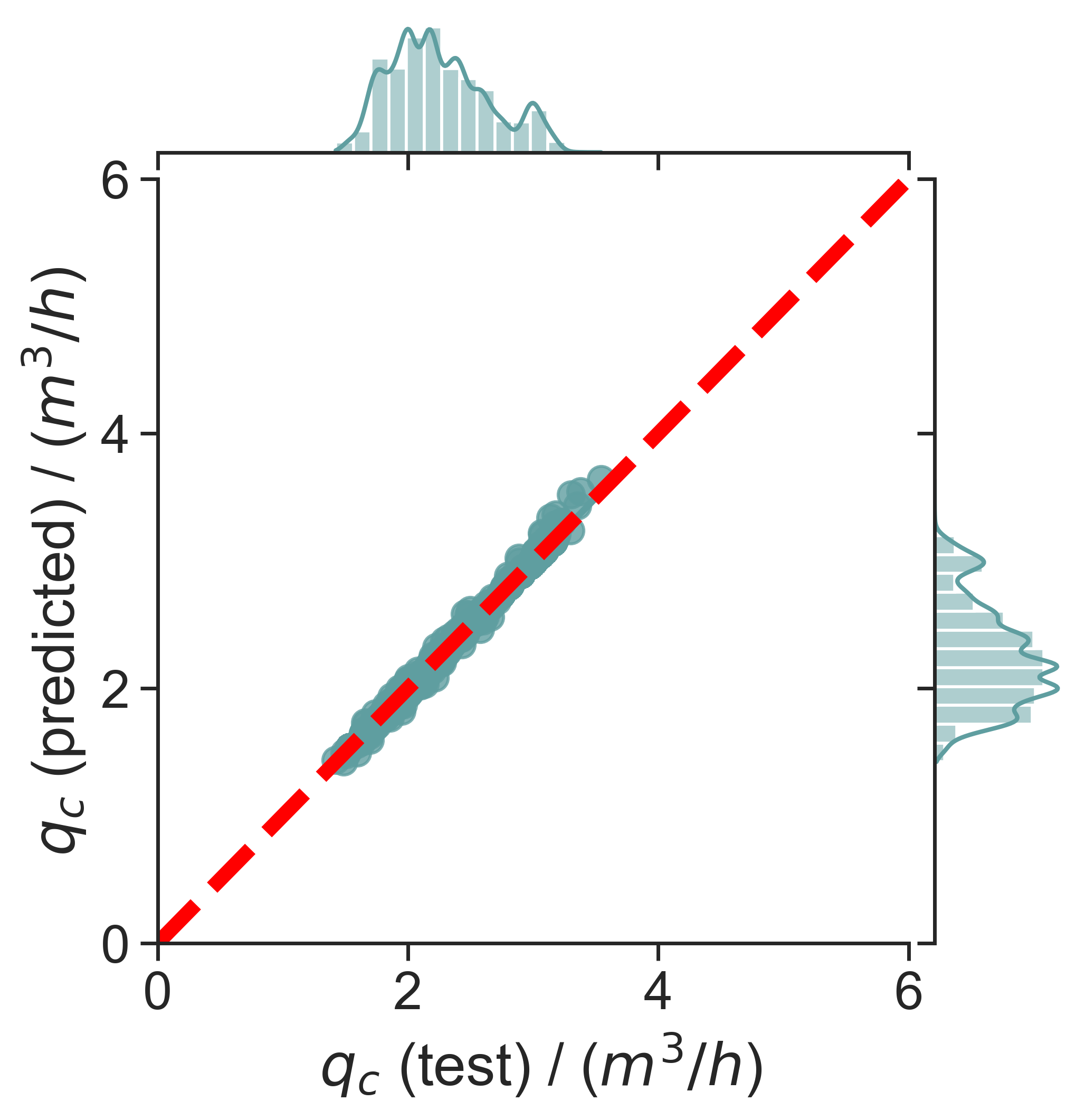}
        \caption{$q_c$ model III.}
    \end{subfigure}
    \hspace{0.005\textwidth}
    \begin{subfigure}[b]{0.22\textwidth}
        \centering
        \includegraphics[width=\linewidth]{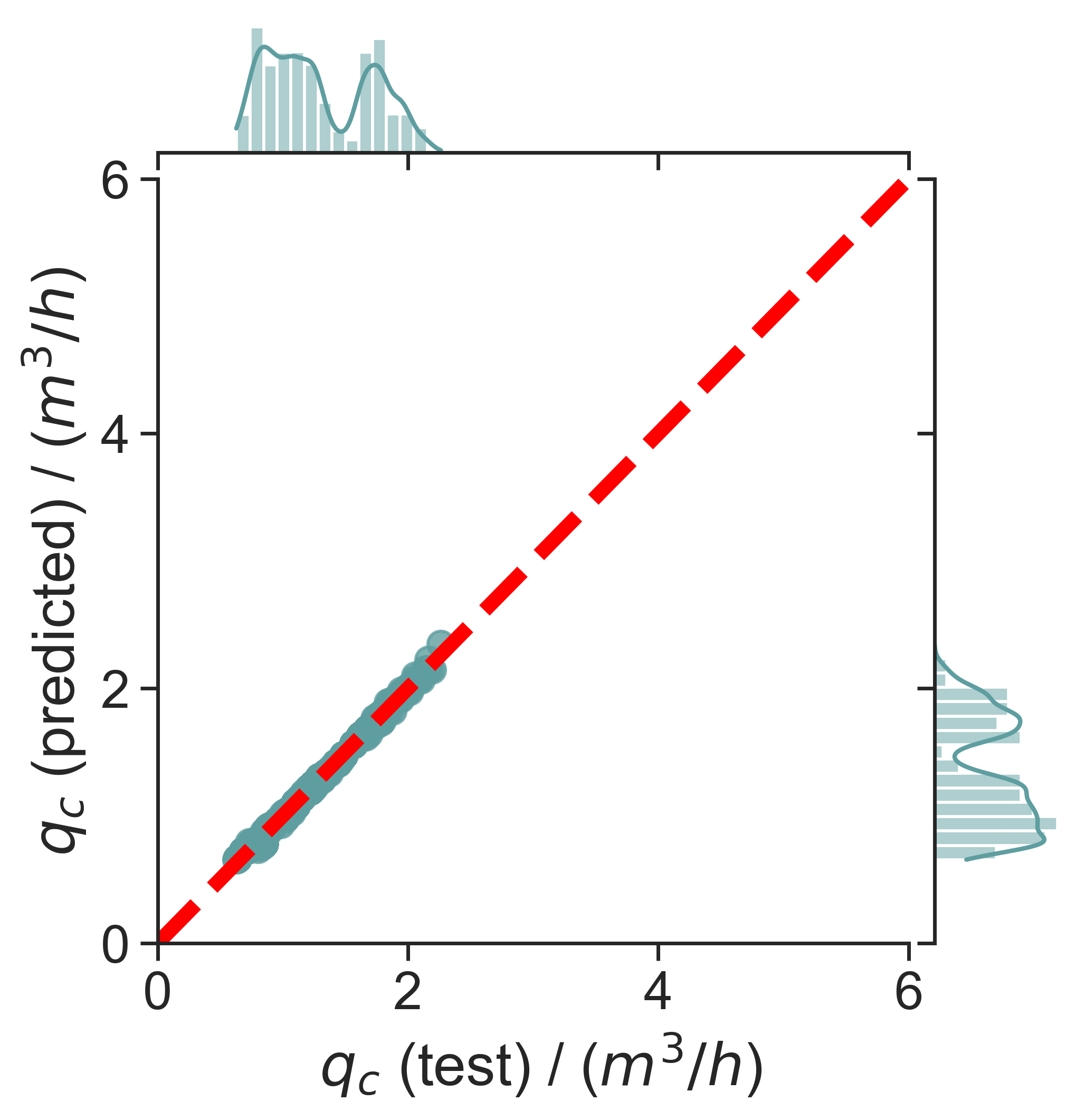}
        \caption{$q_c$ model IV.}
    \end{subfigure}

    \caption{Parity plots of SR models on test data for different variants of $H$ and $q_c$.}
    \label{fig:Parity_plots_multi_zone}
\end{figure}

Figure~\ref{fig:Parity_plots_multi_zone} shows parity plots for the zone-specific SR models of pump head $H$ and flow rate $q_c$ across zones~1–4. The scatter points tightly around the 45° identity line, indicating negligible bias and near-unity gain; dispersion is uniformly small for $H$ and only slightly larger for $q_c$. The absence of systematic curvature or stratification suggests that the local models capture the nonlinear input–output mapping across the full range, which is consistent with the low MAE/MSE reported in Table~\ref{tab:metrics_zones}.

It evaluated the dynamics of the nominal SR models per operating region. Predictions were generated in a multi-step free-run (NARX) using measured $f(t)$ and $Z_c(t)$ as exogenous inputs and feeding back $\{\hat H,\hat q_c\}$ as regressors. As seen in Figure \ref{fig:comparacao_testes_2}, the predicted trajectories closely track the measurements without systematic drift, capturing faster variations in all zones. The residual transients are consistent with the MAE/MSE reported in Table~\ref{tab:metrics_zones}, supporting the fidelity of the zone-specific models in the original domain.

\captionsetup[subfigure]{font=small,labelfont=bf,aboveskip=2pt,belowskip=2pt,justification=centering}

\begin{figure}[h!]
    \centering

    \begin{subfigure}[b]{0.42\textwidth}
        \centering
        \includegraphics[width=\linewidth]{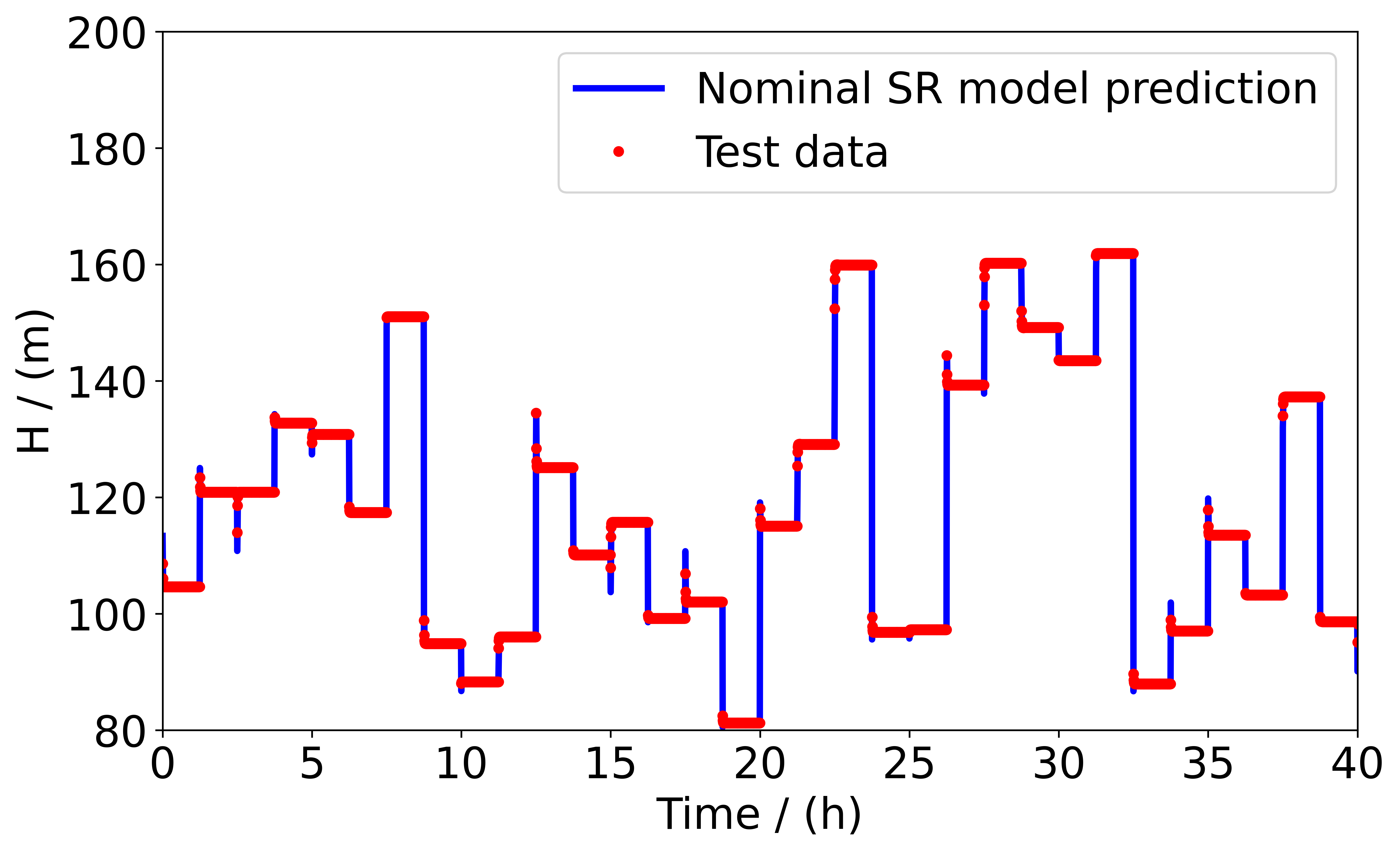}
        \caption{Zone 1 — $H$}
    \end{subfigure}\hspace{0.02\textwidth}%
    \begin{subfigure}[b]{0.42\textwidth}
        \centering
        \includegraphics[width=\linewidth]{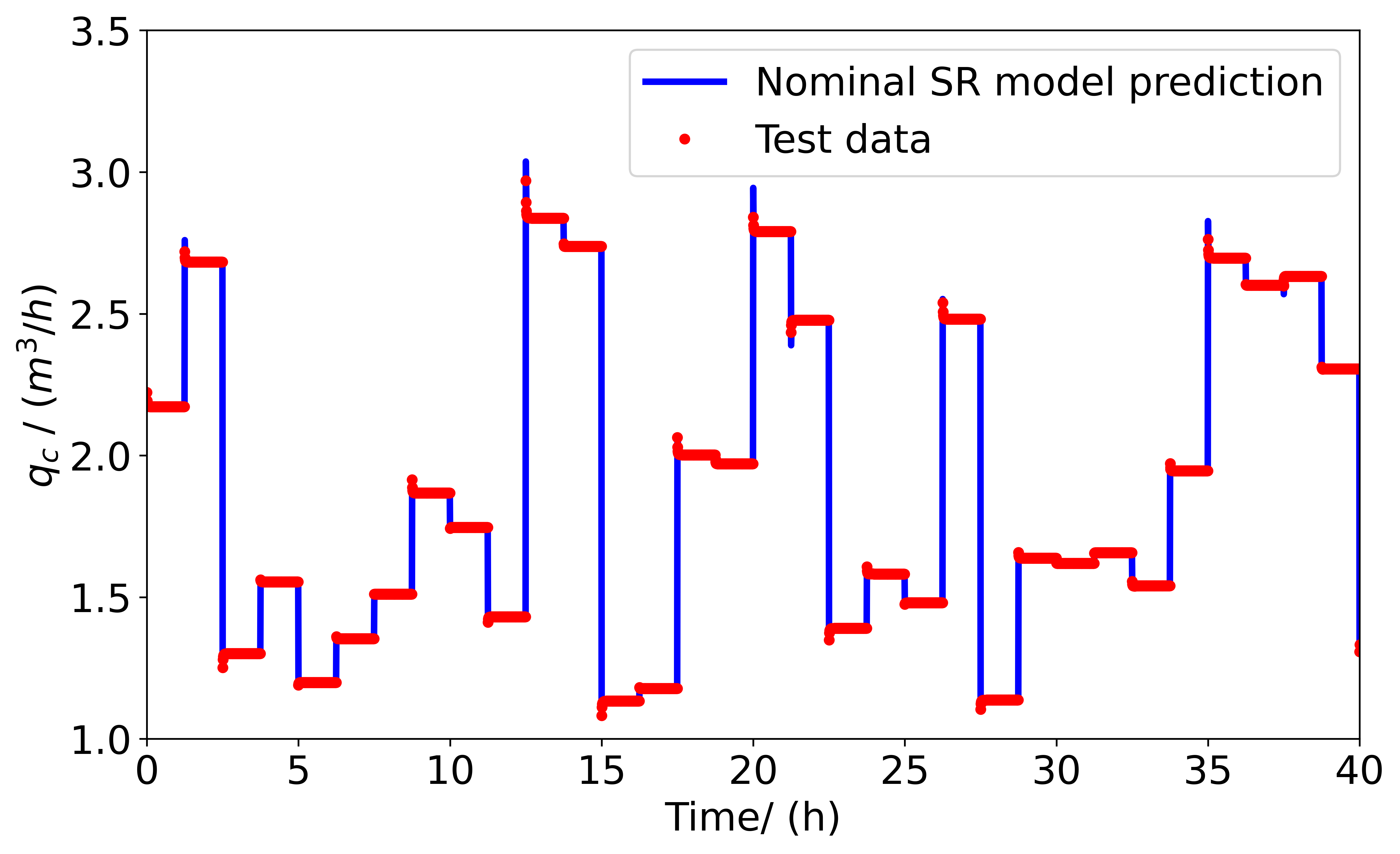}
        \caption{Zone 1 — $q_c$}
    \end{subfigure}

    \vspace{0.05cm}

    \begin{subfigure}[b]{0.42\textwidth}
        \centering
        \includegraphics[width=\linewidth]{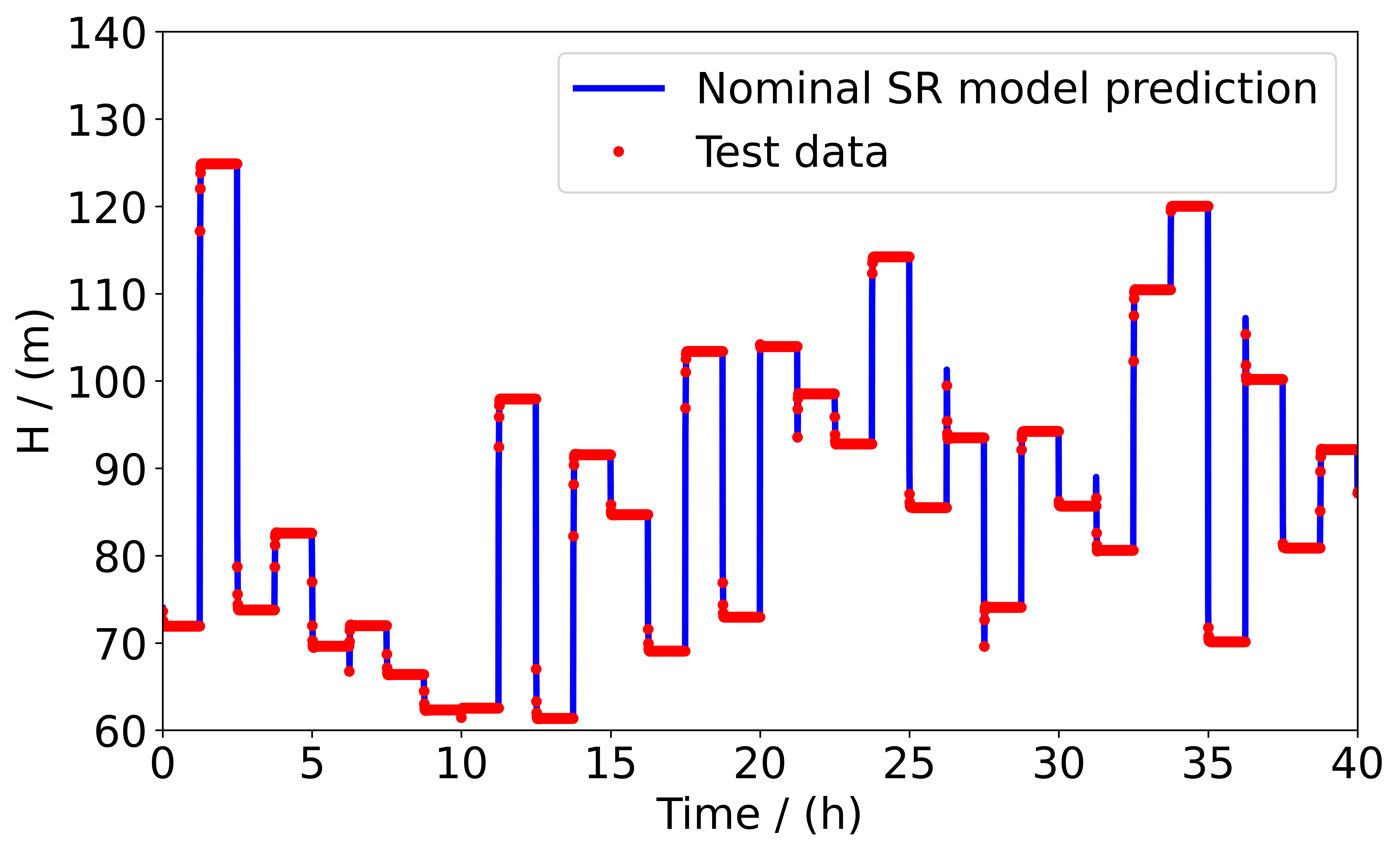}
        \caption{Zone 2 — $H$}
    \end{subfigure}\hspace{0.02\textwidth}%
    \begin{subfigure}[b]{0.42\textwidth}
        \centering
        \includegraphics[width=\linewidth]{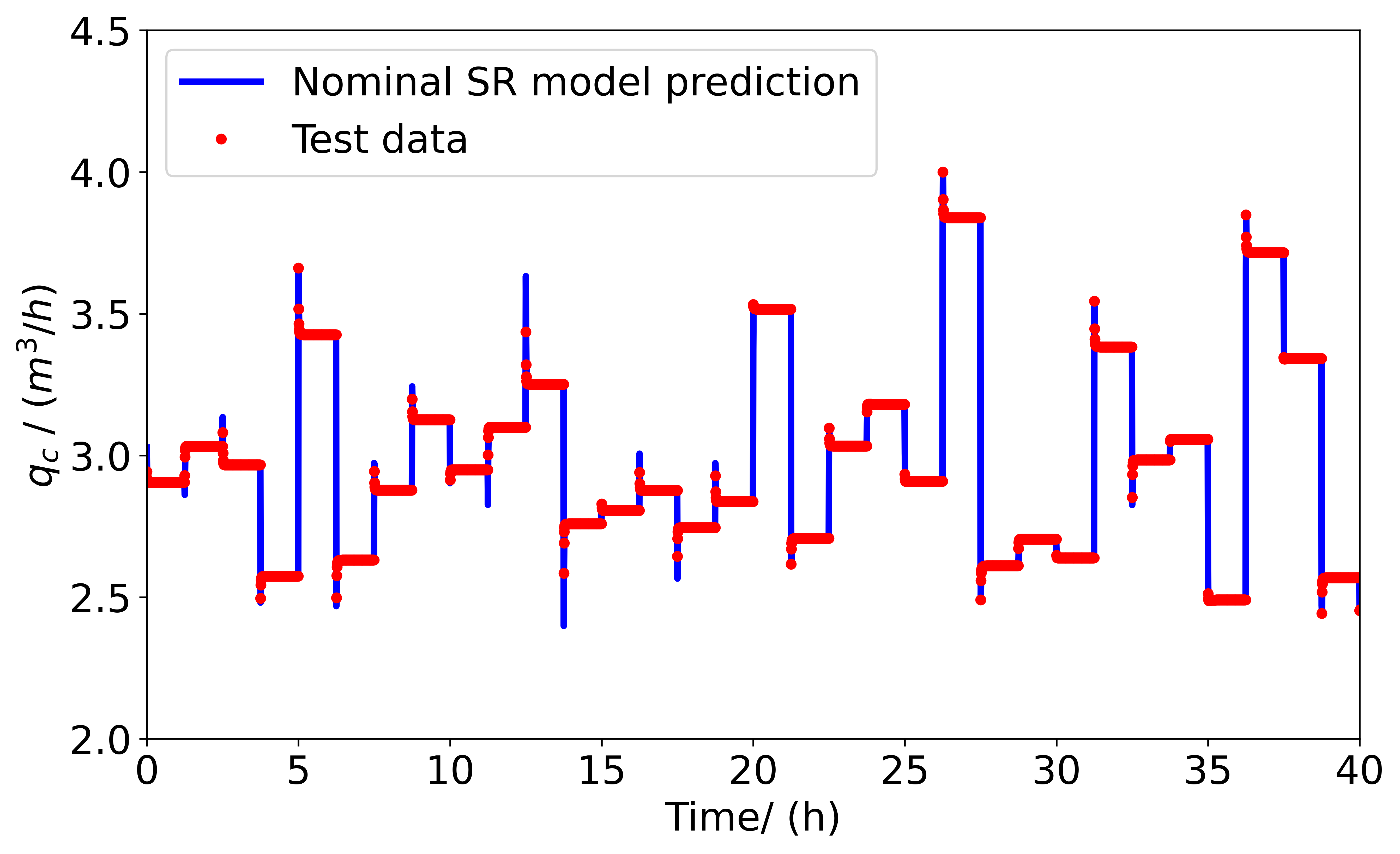}
        \caption{Zone 2 — $q_c$}
    \end{subfigure}

    \vspace{0.05cm}

    \begin{subfigure}[b]{0.42\textwidth}
        \centering
        \includegraphics[width=\linewidth]{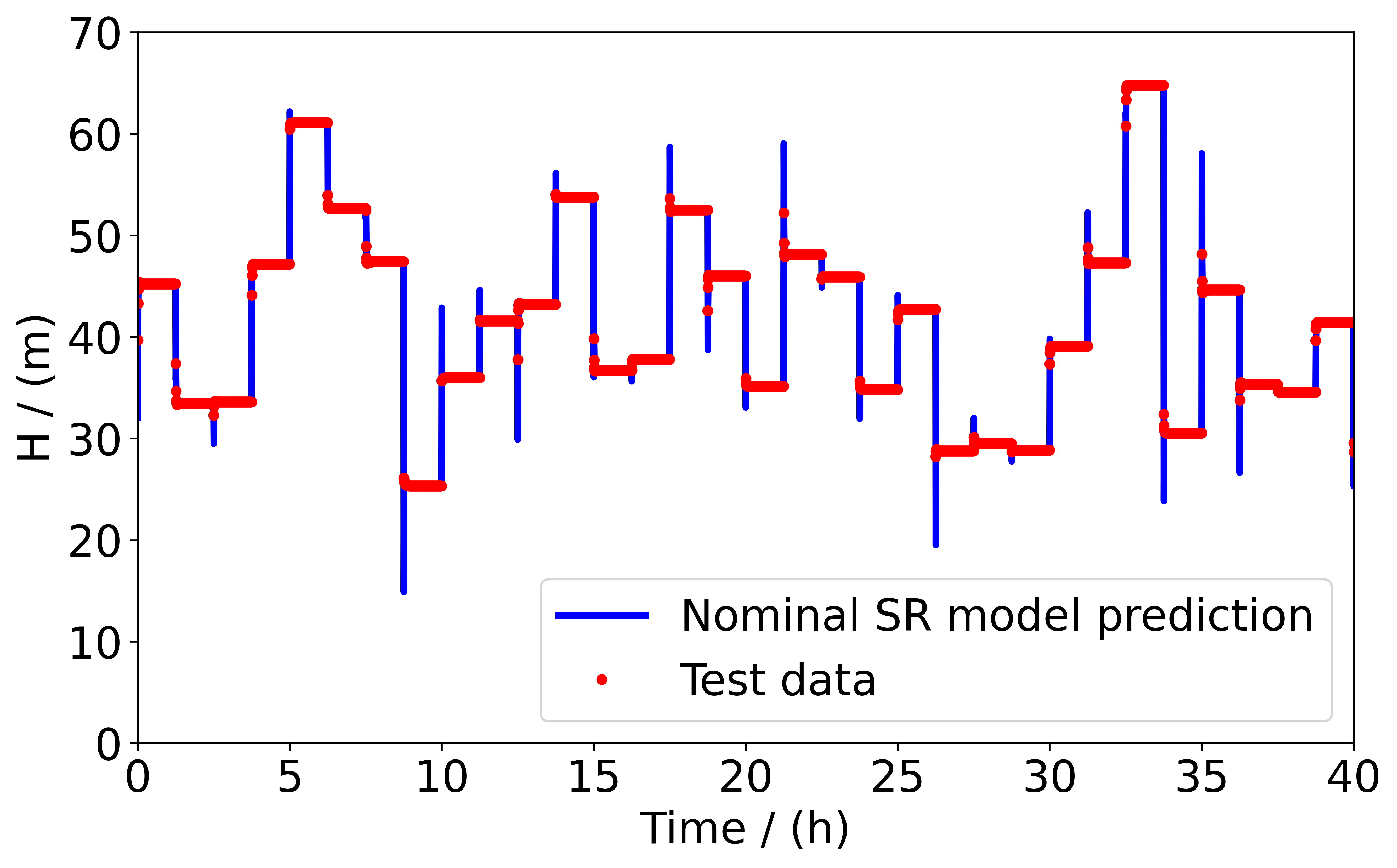}
        \caption{Zone 3 — $H$}
    \end{subfigure}\hspace{0.02\textwidth}%
    \begin{subfigure}[b]{0.42\textwidth}
        \centering
        \includegraphics[width=\linewidth]{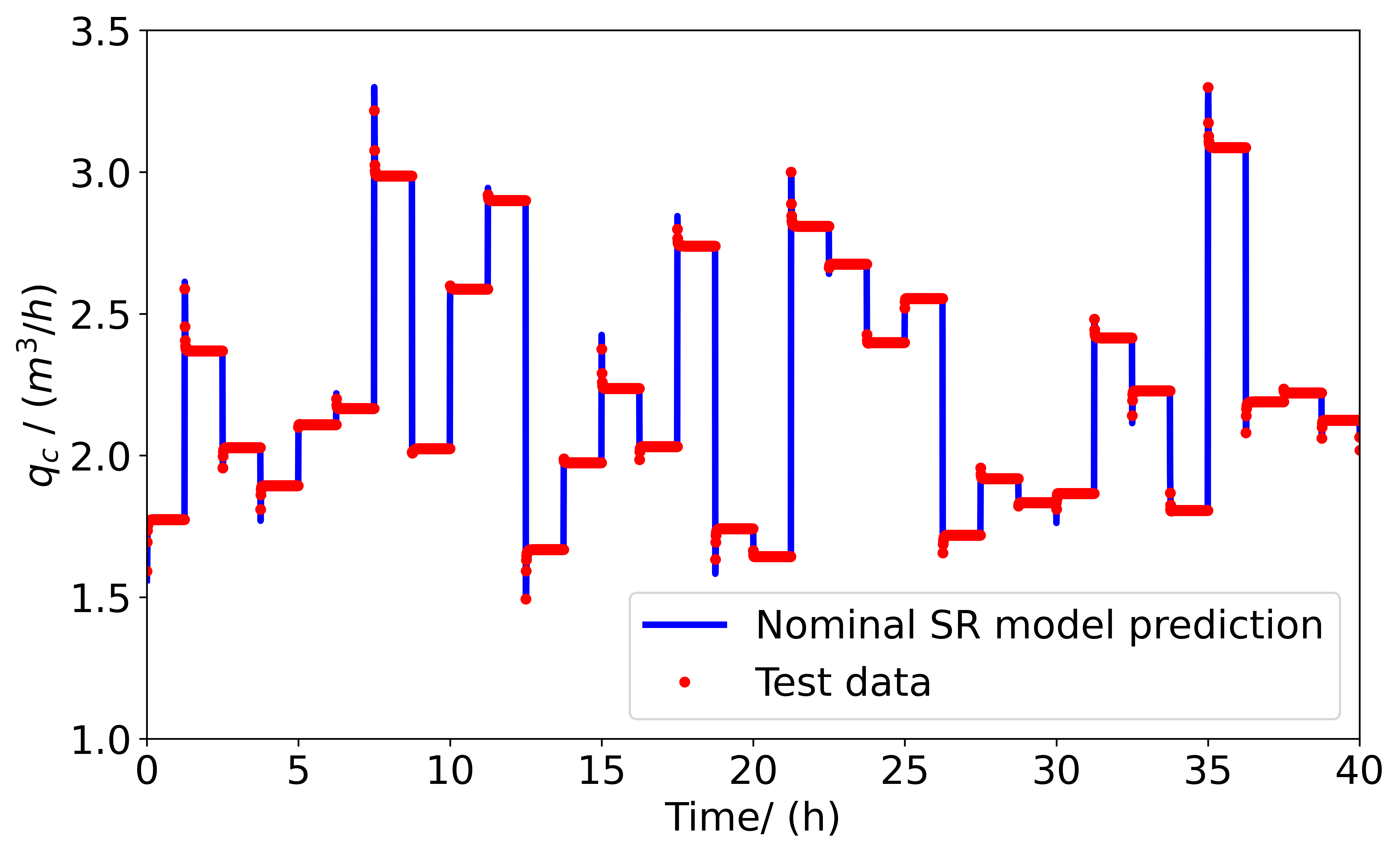}
        \caption{Zone 3 — $q_c$}
    \end{subfigure}

    \vspace{0.05cm}

    \begin{subfigure}[b]{0.42\textwidth}
        \centering
        \includegraphics[width=\linewidth]{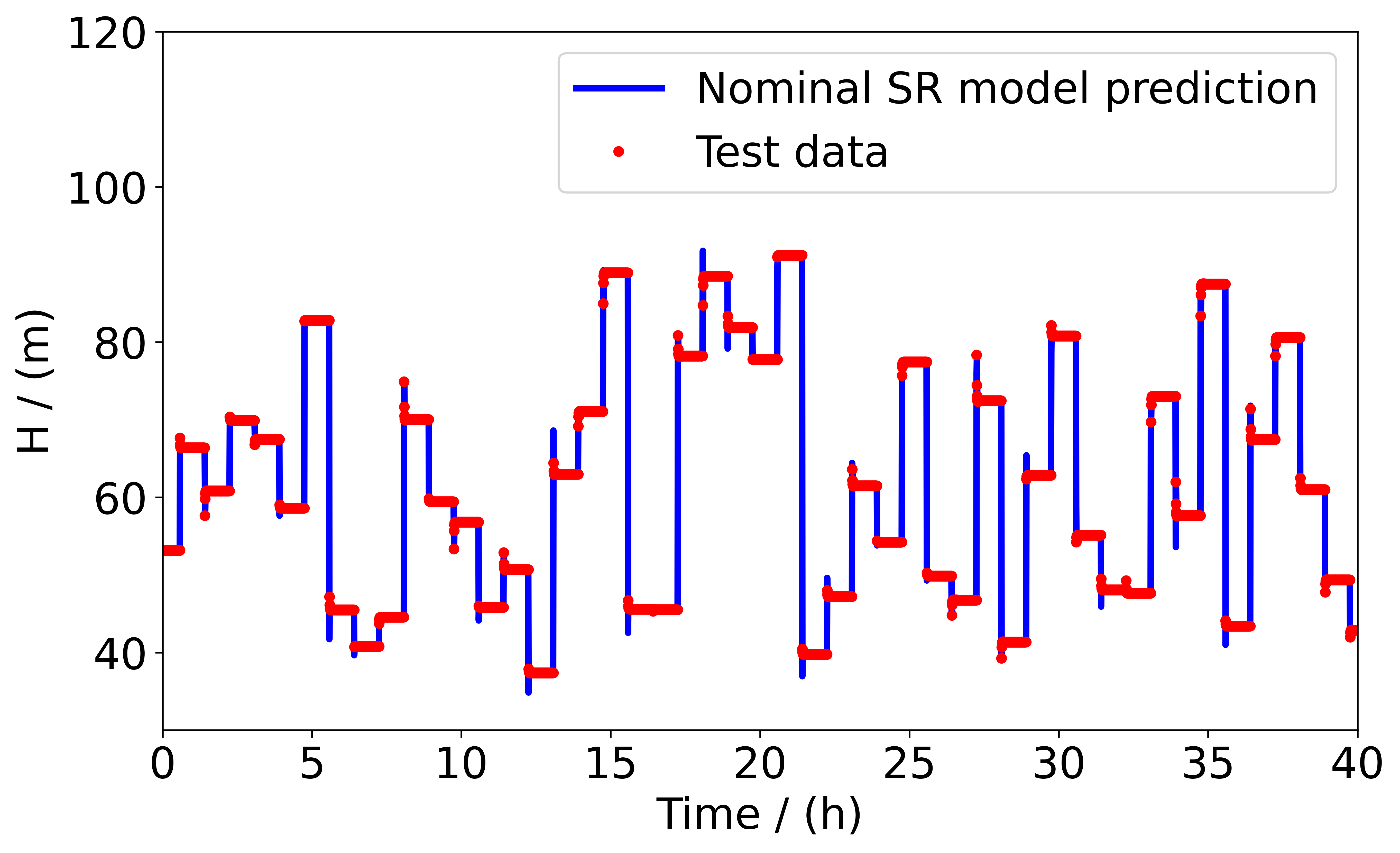}
        \caption{Zone 4 — $H$}
    \end{subfigure}\hspace{0.02\textwidth}%
    \begin{subfigure}[b]{0.42\textwidth}
        \centering
        \includegraphics[width=\linewidth]{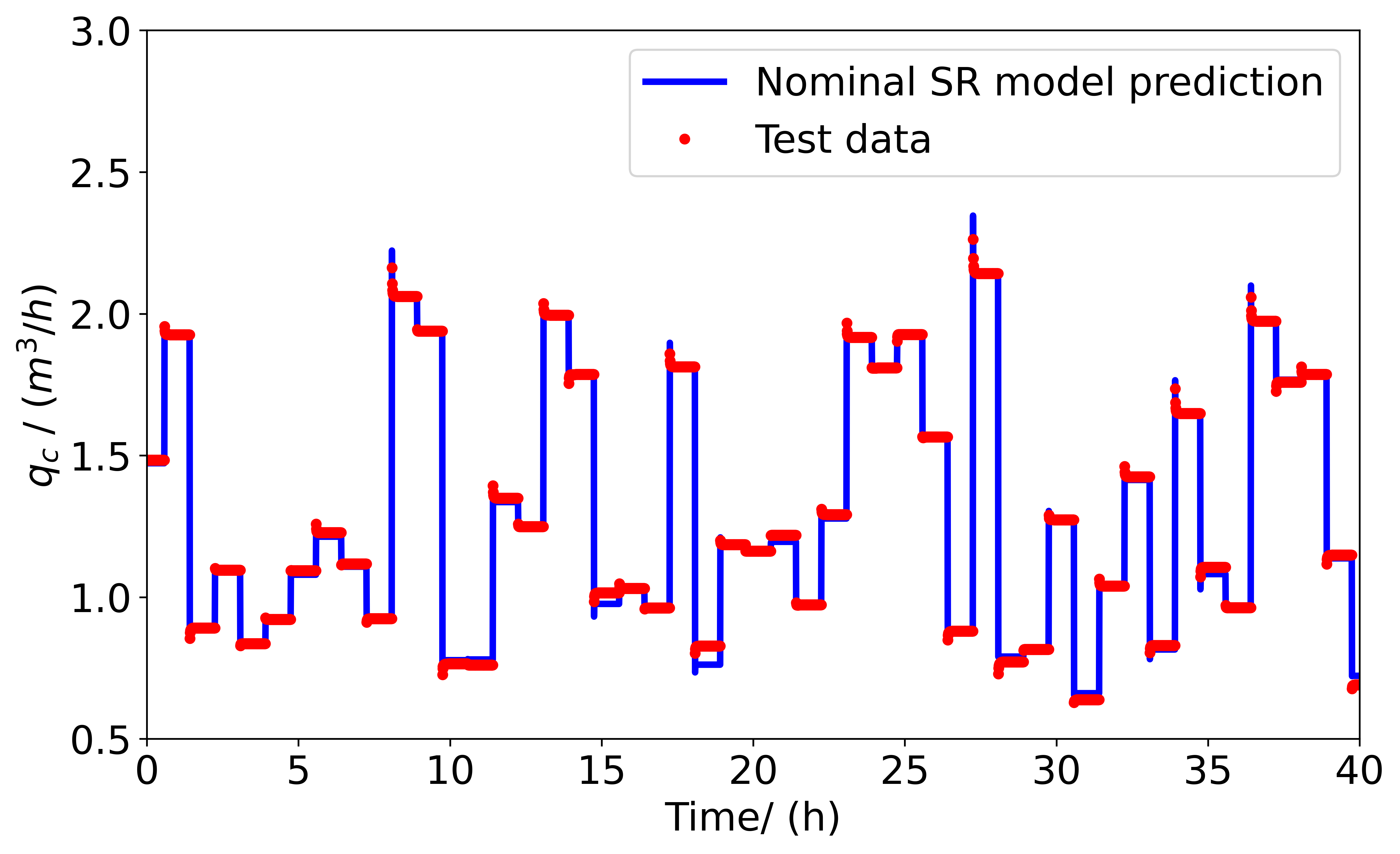}
        \caption{Zone 4 — $q_c$}
    \end{subfigure}

    \caption{Time-series dynamics of the nominal SR model predictions by zone. Each row corresponds to one operating region (zones 1–4). Left column: pump head $H$; right column: flow rate $q_c$.}
    \label{fig:comparacao_testes_2}
\end{figure}

Table~\ref{tab:metrics_zones} indicates that the multi-zone SR models achieve consistently low MAE and MSE across all regions for both $H(t)$ and $q_c(t)$. These small errors attest to the high fidelity of the identified surrogates in the original domain, supporting reliable one-step and multi-step predictions suitable for NMPC.

\begin{table}[h!]
\centering
\caption{MSE and MAE for SR models by operating zone (multi-zone).}
\label{tab:metrics_zones}
\renewcommand{\arraystretch}{1.3}
\begin{tabular}{l cc cc cc cc}
\toprule
 
& \multicolumn{2}{c}{\textbf{Zone 1}} 
& \multicolumn{2}{c}{\textbf{Zone 2}} 
& \multicolumn{2}{c}{\textbf{Zone 3}} 
& \multicolumn{2}{c}{\textbf{Zone 4}} \\
\cmidrule(lr){2-3}\cmidrule(lr){4-5}\cmidrule(lr){6-7}\cmidrule(lr){8-9}
& \textbf{MSE} & \textbf{MAE} & \textbf{MSE} & \textbf{MAE} & \textbf{MSE} & \textbf{MAE} & \textbf{MSE} & \textbf{MAE} \\
\midrule
$H(t)$      & $1.1\times 10^{-5}$ & $3.32\times 10^{-4}$ & $5.10\times 10^{-6}$ & $2.65\times 10^{-4}$ & $8.0\times 10^{-6}$ & $2.90\times 10^{-4}$ & $1.0\times 10^{-5}$ & $3.44\times 10^{-4}$ \\
$q_c(t)$    & $1.0\times 10^{-6}$ & $7.6\times 10^{-5}$  & $3.0\times 10^{-6}$ & $2.10\times 10^{-4}$   & $2.0\times 10^{-6}$ & $4.95\times 10^{-4}$ & $9.0\times 10^{-6}$ & $1.82\times 10^{-3}$ \\
\bottomrule
\end{tabular}
\end{table}

\subsection{Control scenarios and performance evaluation}

The control results are organised into three complementary scenarios that collectively expose the performance, robustness, and computational efficiency of the RNMPC strategies:

\begin{enumerate}
    \item Scenario~1 — Disturbance-rejection stress tests (with and without noise),  
    \item  Scenario~2 — Tight-constraint set-point tracking with controller comparison, and  
    \item Scenario~3 — Set-point–scheduled RNMPC\textsubscript{MZ} with multi-zone constraint across all operating regions.  
\end{enumerate}
 
The controllers compared in these scenarios are i) NMPC\textsubscript{MM}: a conventional nonlinear MPC based on the mechanistic model (described in Section \ref{ESP_model}), ii)  NMPC\textsubscript{SMR}: a nominal nonlinear MPC that employs a single symbolic regression model,  iii) RNMPC\textsubscript{SZ} and
iv) RNMPC\textsubscript{MZ}.

All controllers solve the same receding-horizon nonlinear program using MATLAB’s \texttt{fmincon} (interior-point, L-BFGS Hessian approximation, warm-start from the previous solution), with identical solver tolerances and a 30\,s per-iteration time limit to match the sampling time. 

The ability of all controllers to calculate the control action within a sampling time was evaluated, ensuring that the control task can be completed in real time under the given computational constraints.

It is important to emphasise that, in this work, the source of model uncertainty is not parameter mismatch but rather structural uncertainty of the symbolic regression models. Different symbolic equations, identified from the same dataset under distinct operating conditions or search paths, provide alternative but valid representations of the ESP dynamics. By embedding these structurally diverse models in the RNMPC formulation, the controller explicitly accounts for structural mismatch, ensuring regulation under model uncertainty.

\subsubsection{Scenario 1: Disturbance-rejection stress tests (with and without noise)}
A series of dynamic closed-loop simulations was conducted to evaluate the performance and robustness of the proposed NMPC controllers, based on symbolic regression models, across various operating scenarios. These scenarios were designed to assess the controller’s ability to (i) track reference trajectories for the controlled variables, (ii) respect operational and actuator constraints, and (iii) maintain robustness in the presence of model uncertainties.

The configuration of the RNMPC\textsubscript{MZ}, RNMPC\textsubscript{SZ}, NMPC\textsubscript{SRM}, and NMPC\textsubscript{MM} controllers was defined based on standard tuning practices, balancing performance, robustness, and constraint satisfaction. Table~\ref{tab:nmpc_tuning} summarises the tuning parameters adopted in this study, including prediction and control horizons, penalty weights for state deviations and control effort, and input bounds and rate limits. These values were selected to ensure the feasible, stable, and smooth operation of the closed-loop system, while enabling the real-time execution of the control algorithm.

\begin{table}[ht]
    \centering
    \caption{NMPC tuning and configuration parameters}
    \label{tab:nmpc_tuning}
    \begin{tabular}{@{}lll@{}}
        \toprule
        Parameter (Symbol)                   & Value       & Tuning Effect \\
        \midrule
        Prediction Horizon ($N_{p}$)         & 8           & ↑$N_{p}$: increase robustness, ↑computational load \\
        Control Horizon ($N_{c}$)            & 6           & ↑$N_{c}$: less aggressive action, ↑ solving time\\
        State Weight ($Q$)                   & 250, 800      & ↑$Q$: less overshoot, slower response \\
        Control Weight ($R$)                 & 1,1         & ↑$R$: smoother action, ↑tracking error \\
        Lower Control Limit ($u_{\min}$)     & 30, 0.1     & Defines safe operating range \\
        Upper Control Limit ($u_{\max}$)     & 60, 0.58    & Defines safe operating range \\
        Maximum Control Action Increment ($\Delta u_{\max}$) & 1, 0.03 & Protects actuators from abrupt commands \\
        \bottomrule
    \end{tabular}
\end{table}

\begin{figure}[h!]
  \centering
  \begin{subfigure}{0.49\textwidth}
    \centering
    \includegraphics[width=\linewidth]{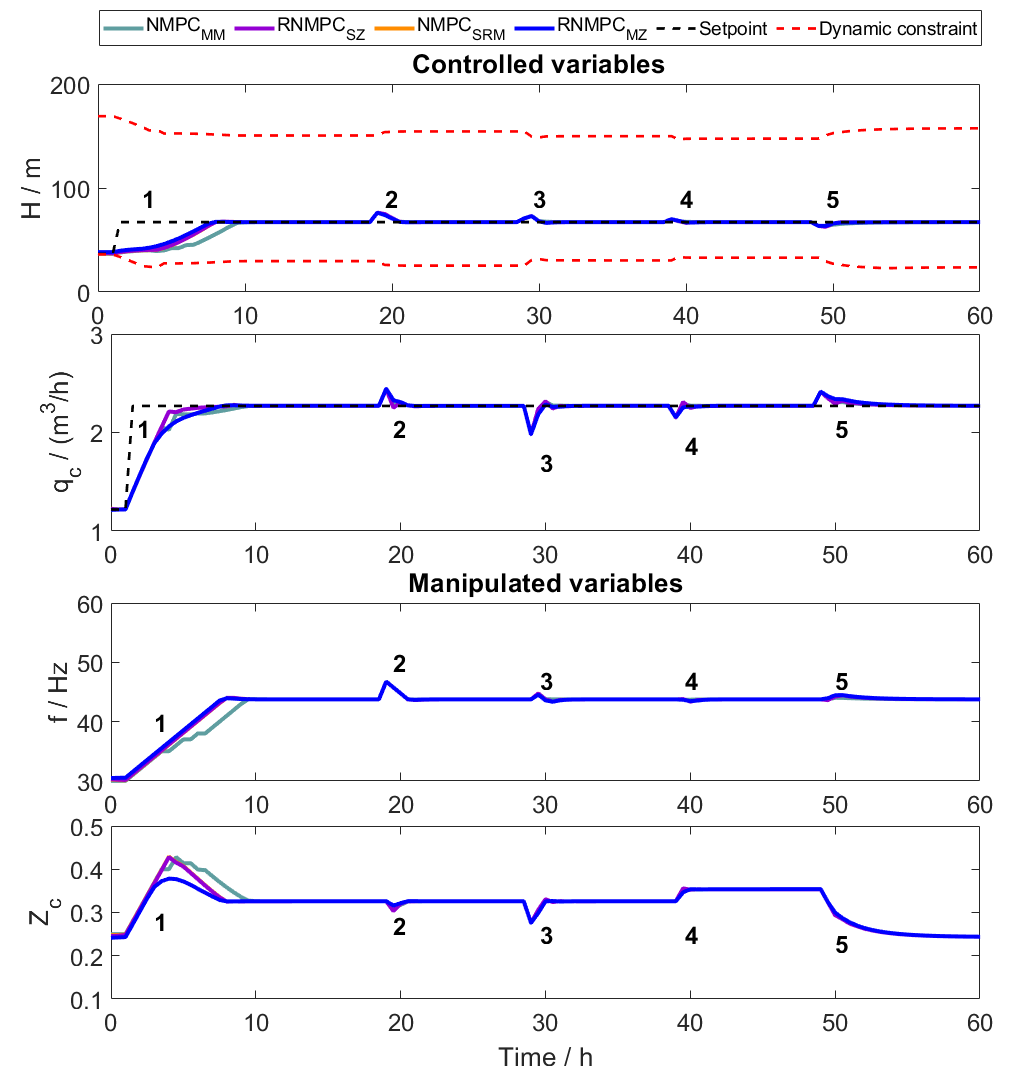}
    \caption{Without noise.}
    \label{fig:cen1}
  \end{subfigure}
  \hfill
  \begin{subfigure}{0.49\textwidth}
    \centering
    \includegraphics[width=0.99\linewidth]{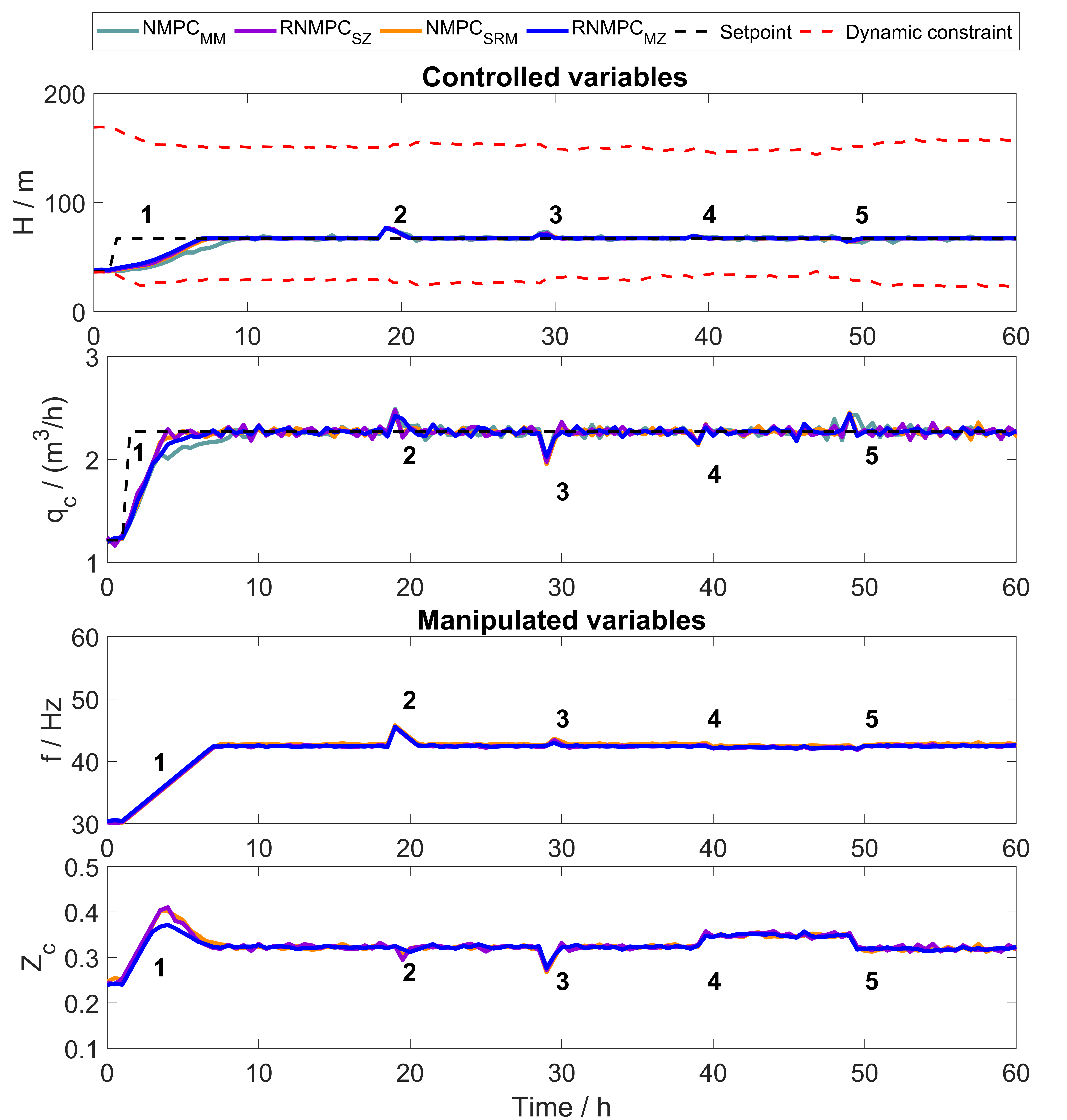} 
    \caption{With noise.}
    \label{fig:Cenario_1}
  \end{subfigure}

  \caption{Comparative trajectories of controller strategies within the operational envelope for scenario 1 (with and without noise).}
  \label{fig:cenarios_sidebyside}
\end{figure}

In the first scenario, illustrated in Figure \ref{fig:Cenario_1}, the disturbance handling was assessed for all controllers under a step in the operating point. The controlled variables are commanded initially to $H=36.6 m$ and $q_c=1.2 m^3/h$; later, the set-points are increased to $H=67.19 m$ and $q_c=2.27 m^3/h$. The objective is to evaluate how each controller rejects disturbances and regulates the manipulated variables (motor frequency and choke opening) to keep the outputs close to their targets while respecting the operational envelope.
 All disturbances applied in the regulatory cases are sustained over time to assess the controller’s ability to reject steady-state deviations and maintain constraint compliance. The same scenario was tested both with and without measurement noise to evaluate the controller’s performance under ideal and more realistic conditions.

\begin{itemize}
\item \textbf{Point 1 – Servo control with measured disturbance:} This scenario assesses the controller’s tracking capability in the presence of a known disturbance. Despite the perturbation, all controllers effectively track the reference trajectories for both $H$ and $q_c$, with smooth adjustments in the manipulated variables. The error correction mechanism reduces oscillations and speeds up convergence, particularly in the $q_c$.

\item \textbf{Point 2 – Regulatory control with measured disturbance (implementation error in frequency):} A persistent deviation is introduced in the motor frequency signal to emulate an implementation mismatch. The controller successfully compensates for the imposed bias, maintaining the outputs near their nominal values and ensuring operational feasibility. This test highlights the controller’s robustness against implementation errors in actuation.

\item \textbf{Point 3 – Regulatory control with measured disturbance (implementation error in choke opening):} Similar to the previous case, a steady disturbance is injected in the choke opening variable. The NMPC\textsubscript{MM}, NMPC\textsubscript{SMR}, RNMPC\textsubscript{SZ}, and RNMPC\textsubscript{MZ} adjust their control strategies accordingly, mitigating the offset introduced and maintaining the controlled variables within acceptable bounds. The symbolic prediction model, aided by the correction term, facilitates accurate compensation under nonlinear dynamics.

\item \textbf{Point 4 – Regulatory control with unmeasured disturbance (manifold pressure):} A disturbance in the downstream manifold pressure, unavailable to the controller, is introduced to evaluate the system’s resilience. Despite the unmeasured nature of the perturbation, the controller maintains stable operation and restricts the deviation of the controlled variables through adaptive adjustment of the control actions.

\item \textbf{Point 5 – Regulatory control with unmeasured disturbance (reservoir pressure):} In this final case, a deviation in the upstream reservoir pressure simulates changes in inlet conditions. All controllers once again demonstrate satisfactory regulation, compensating for the unknown disturbance and ensuring that both head and flow remain close to their desired values without violating constraints.
\end{itemize}

These scenarios collectively demonstrate the NMPC controller’s capacity to handle a range of disturbances, measured and unmeasured, in both manipulated and process variables, while maintaining performance, feasibility, and robustness. The sustained nature of the disturbances reinforces steady-state handling and the effectiveness of the symbolic-regression model with integrated error correction. Importantly, the SR surrogate models behaved well inside the NMPC optimiser: they did not introduce artificial local minima or ill-conditioning in the control objective, preserved solver convergence, and did not bias the computed moves in ways that would steer the plant away from the desired trajectory.

As shown in Figure~\ref{fig:Cenario_1}, all four controllers perform similarly in tracking setpoints and rejecting disturbances across the five evaluated scenarios. In both servo and regulator cases, the controlled variables $H$ and $q_c$ reach their setpoints effectively under all control strategies.

\begin{figure}[h!]
  \centering
  \begin{subfigure}{0.49\textwidth}
    \centering
    \includegraphics[width=\linewidth]{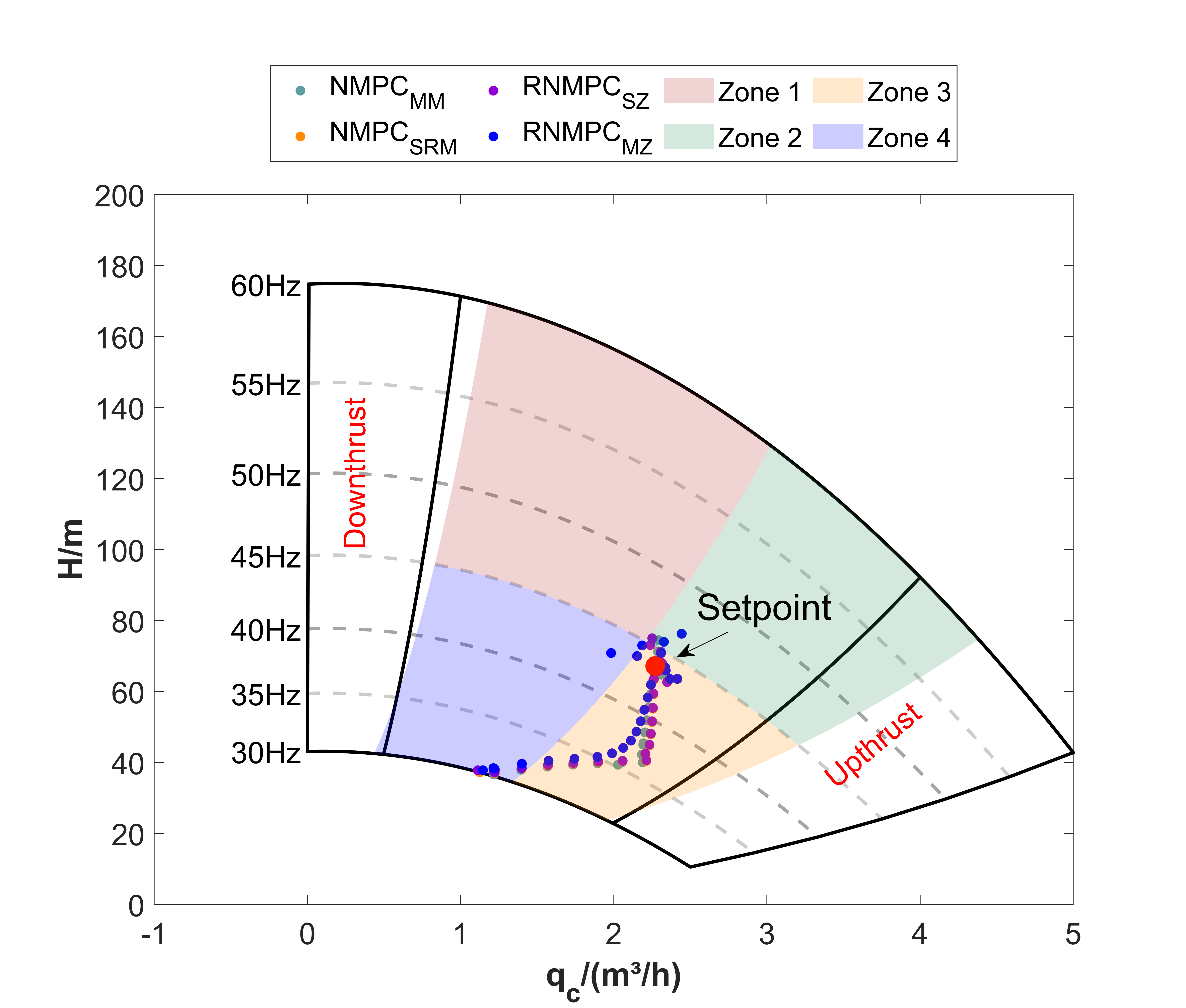}
    \caption{Without noise}
    \label{fig:envelope_p1}
  \end{subfigure}
  \hfill
  \begin{subfigure}{0.49\textwidth}
    \centering
    \includegraphics[width=\linewidth]{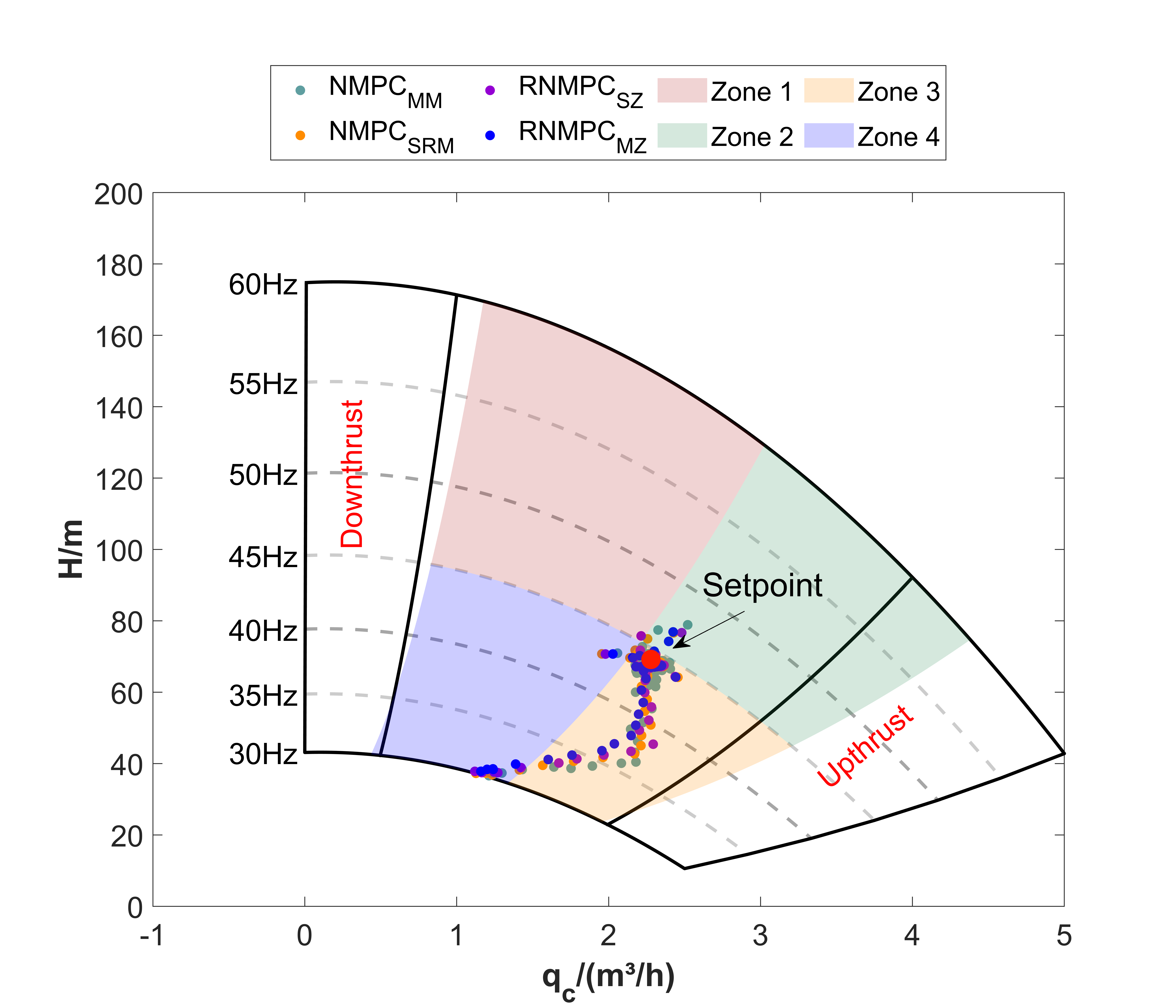} 
    \caption{With noise}
    \label{fig:envelope_p2}
  \end{subfigure}

  \caption{Operational envelope: nominal and noise conditions.}
  \label{fig:envelope}
\end{figure}

Figure \ref{fig:envelope} illustrates the operational envelope of the ESP system in the space of $q_c$ versus $H$. The shaded regions delineate areas of operational uncertainty: the downthrust zone on the left, associated with excessive pressure and low flow, and the upthrust zone on the right, corresponding to high flow and insufficient pressure. Superimposed on this map are the trajectories followed by each controller presented, as they regulate the system in closed-loop operation. The points trajectories indicate iso-frequency contours (30–60 Hz), which further contextualise the actuator behaviour within the dynamic limitations of the motor-pump system.


\begin{figure}[!ht]
  \centering

  \begin{subfigure}[t]{0.48\textwidth}
    \centering
    \includegraphics[width=\linewidth]{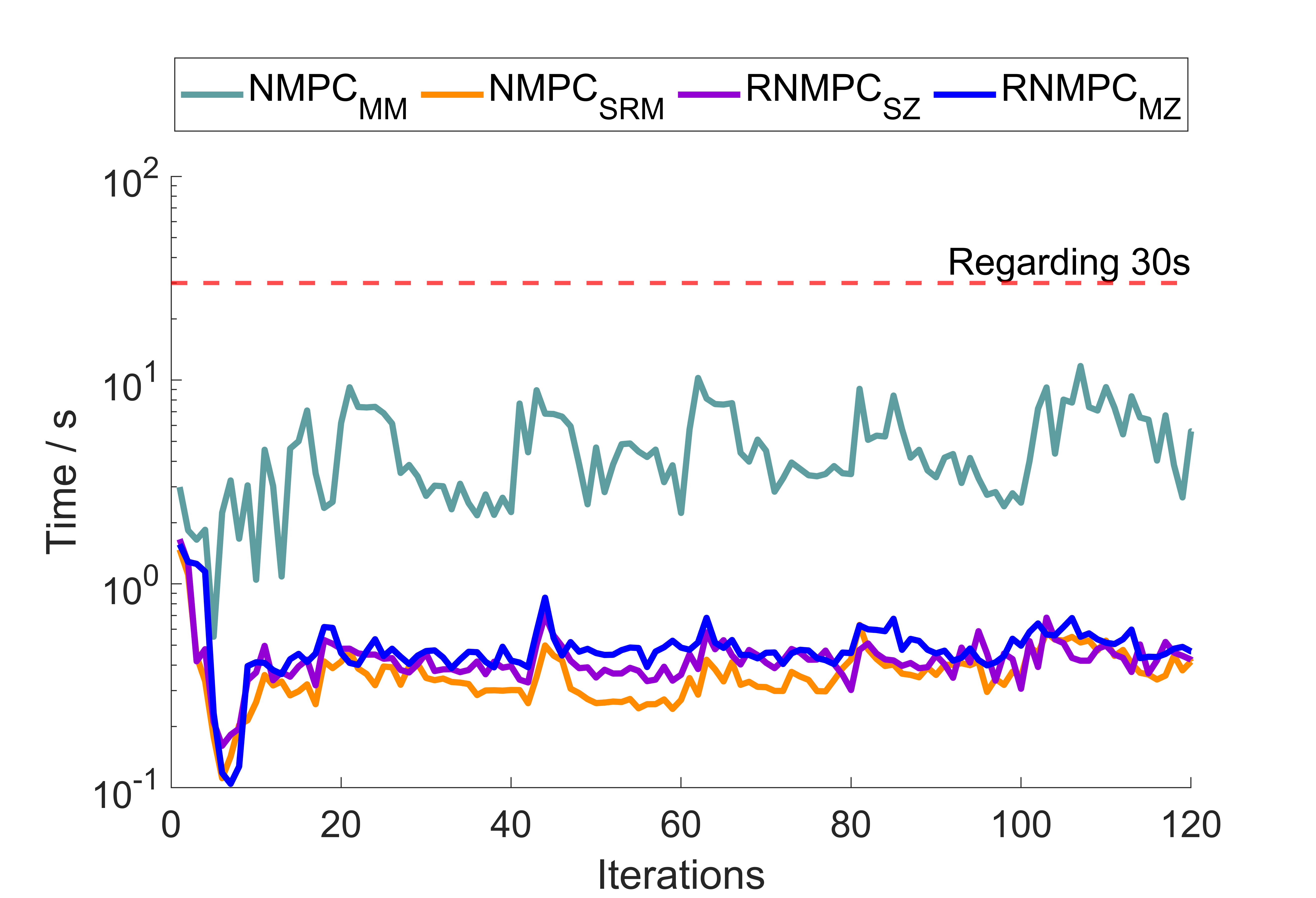}
    \caption{Computational time per iteration without noise.}
    \label{fig:tempo_sem_ruido}
  \end{subfigure}\hfill
  \begin{subfigure}[t]{0.48\textwidth}
    \centering
    \includegraphics[width=\linewidth]{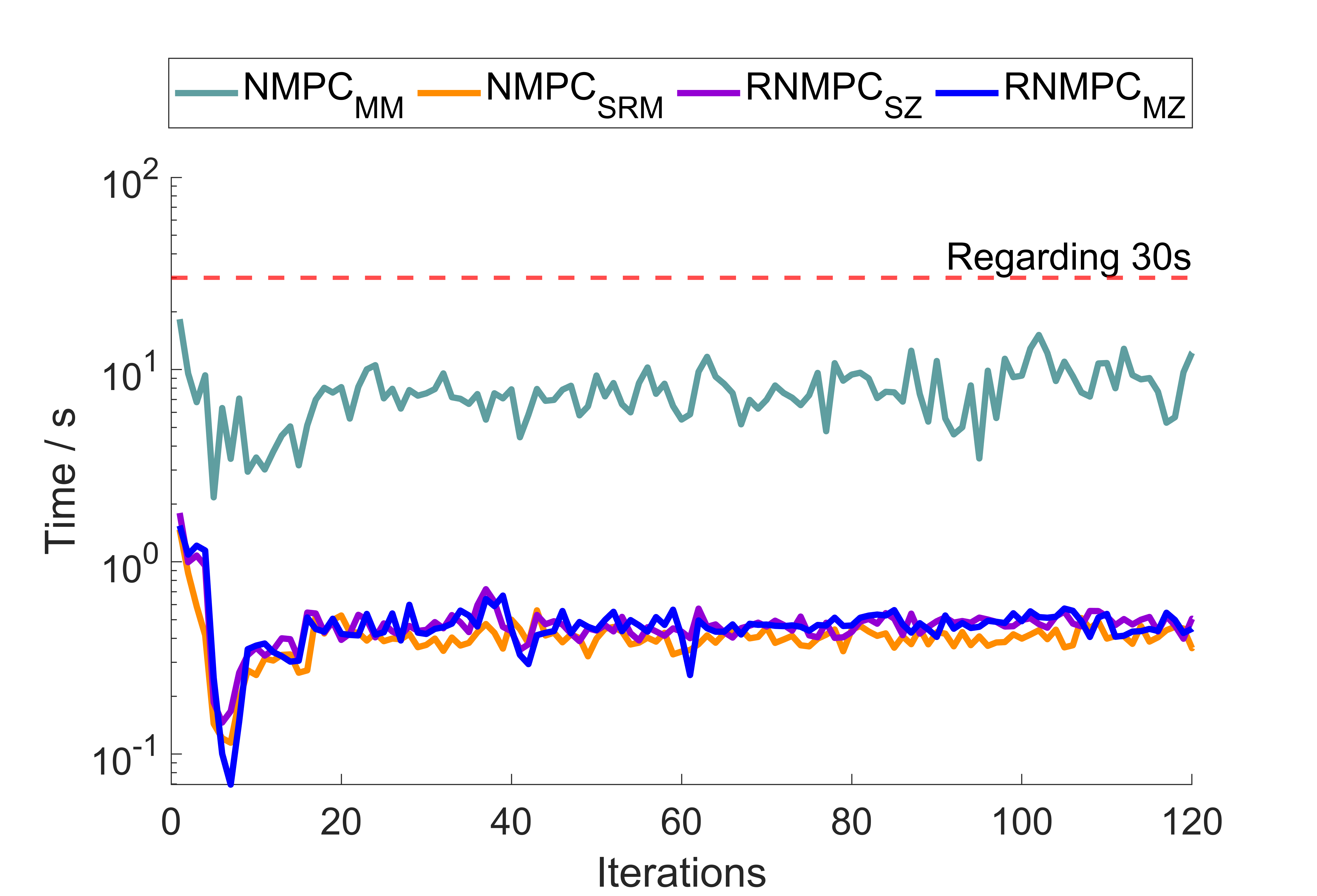}
    \caption{Computational time per iteration with noise.}
    \label{fig:tempo_com_ruido}
  \end{subfigure}

  \caption{Semi-log per-iteration computation time graph (with and without noise).}
  \label{fig:Tempo_1}
\end{figure}

To quantify runtime, the per-iteration solve time across controllers was measured (Figure~\ref {fig:Tempo_1}). 
The conventional NMPC\textsubscript{MM} averages 4.7 s/iteration. Replacing the plant model with symbolic-regression surrogates reduces the mean solve time markedly: 
NMPC\textsubscript{SRM} 0.37 s/it ($-92.1\%$, $12.7\times$ faster), 
RNMPC\textsubscript{SZ} 0.43 s/it ($-90.9\%$, $10.9\times$), and 
RNMPC\textsubscript{MZ} 0.50 s/it ($-89.4\%$, $9.4\times$). 
Relative to NMPC\textsubscript{SRM}, the robust variants add only modest overhead: 
+0.06 s for RNMPC\textsubscript{SZ} ($\sim$16\%) and +0.13 s for RNMPC\textsubscript{MZ} ($\sim$35\%).

The worst-case numbers reinforce this picture: NMPC\textsubscript{MM} peaks at 11.73 s/iteration, whereas SR-based variants remain below 2 s—1.48 s for NMPC\textsubscript{SRM} ($-87.4\%$), 
1.57 s for RNMPC\textsubscript{MZ} ($-86.7\%$), and 
1.66 s for RNMPC\textsubscript{SZ} ($-85.8\%$), confirming ample headroom for online implementation even with robustness and multi-model checks.

\subsubsection{Scenario 2: Tight-constraint set-point tracking with controller comparison}

A second scenario was performed in a reference-tracking regime. The controlled variables $H$ and $q_c$ are commanded through three successive set-points: initially $H=36.6 m$ with $q_c=1.2 m^3/h$; next, the targets are raised to $H=49.4 m$ with $q_c=2.3 m^3/h$; and later they are raised again to $H=108 m$ with $q_c=3.5 m^3/h$. The objective is to evaluate how each controller adjusts the manipulated variables to guide the outputs to these new targets while satisfying a tightened dynamic envelope (downthrust/upthrust) throughout the transitions.

\begin{figure}[h!]
	\centering
		\includegraphics[scale=.45]{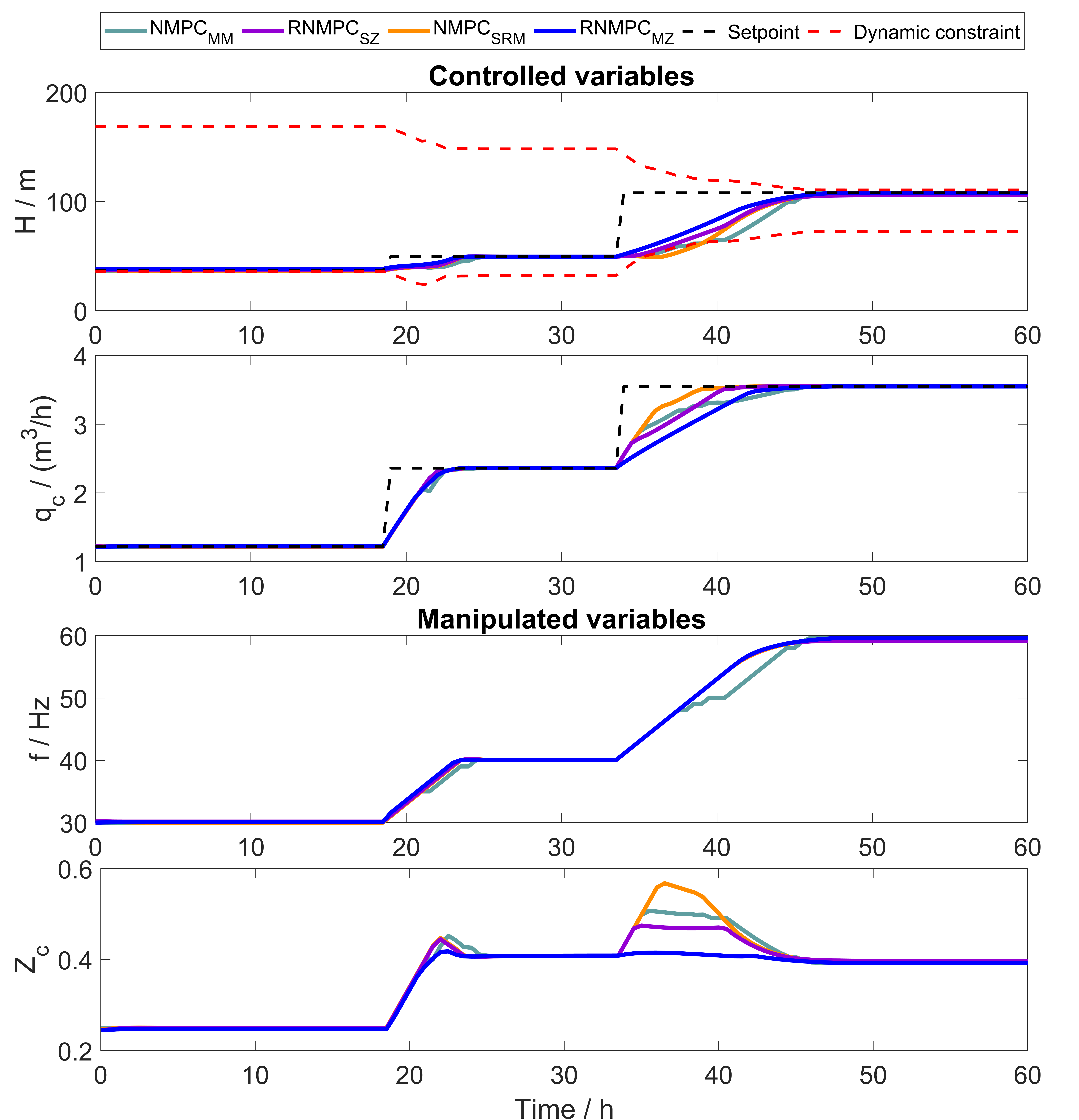}
	\caption{Trajectory comparison of control strategies within the operational envelope for scenario 2.}
	\label{fig:cenario2}
 \end{figure}

As shown in Fig.~\ref{fig:cenario2}, all controllers track the new references and drive the system to the desired operating point. The key difference lies in constraint enforcement. While NMPC\textsubscript{MM}, RNMPC\textsubscript{SZ}, and RNMPC\textsubscript{MZ} keep $H$ within the prescribed safety bounds throughout the transient, NMPC\textsubscript{SRM} exhibits a brief violation of the upper limit immediately after the setpoint step. This overshoot is consistent with the absence of robustness mechanisms in NMPC\textsubscript{SRM}: without uncertainty handling and constraint tightening, the optimiser selects more aggressive moves that do not hedge against model mismatch and sustained disturbances. 

Among the three strategies, RNMPC\textsubscript{MZ} clearly exhibits the most conservative behaviour. It maintains a greater distance from the constraint boundaries and adjusts the manipulated variables,$f$ and $Z_c$, more gradually. This results in smoother transitions and prevents the system from being pushed toward potentially unsafe regions. Although this conservative approach slows down the response slightly compared to the other controllers, it ensures strict compliance with constraints and operational safety. This scenario highlights the advantage of incorporating robustness in control design, even under nominal conditions, to promote feasibility.

\begin{figure}[h!]
	\centering
		\includegraphics[scale=.55]{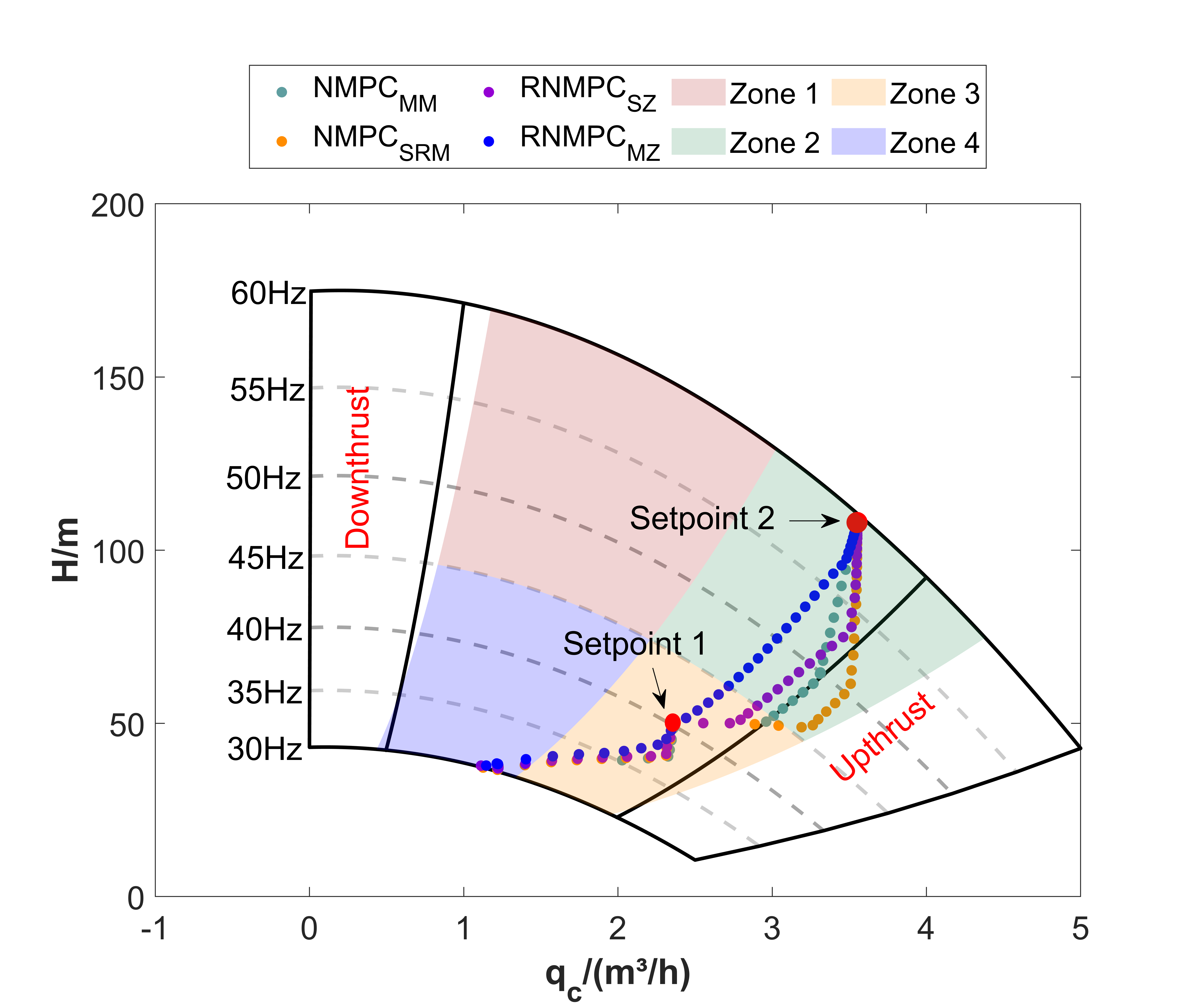}
	\caption{Operational envelope trajectories during the setpoint changes in scenario 2.}
	\label{fig:envelope_2}
 \end{figure}

The conservative behaviour of RNMPC\textsubscript{MZ}, and its larger clearance to constraint boundaries, is evident in the operational envelope of Fig.~\ref{fig:envelope_2}. The trajectories in the $(q_c,H)$ plane during the transition from setpoint~1 to setpoint~2 show that NMPC\textsubscript{MM} and NMPC\textsubscript{SRM} run closer to the feasible limits, especially near the upthrust region, whereas RNMPC\textsubscript{MZ} follows a more central route that deliberately avoids the envelope edges, prioritising constraint satisfaction and operational safety. Compared with its single-zone robust counterpart (RNMPC\textsubscript{SZ}), the multi-zone design keeps a consistently larger safety margin. RNMPC\textsubscript{SZ} remains feasible but tracks with tighter clearances, highlighting the additional conservatism gained by accounting for cross-zone variability in the robust optimisation.

To quantify runtime, we also analysed per-iteration solve times with the dataset of Fig.~\ref{fig:Tempo_1}. 
The conventional NMPC\textsubscript{MM} averaged 5.92 s/it, whereas SR-based controllers were much faster: 0.52 s/it (NMPC\textsubscript{SRM}), 0.56 s/it (RNMPC), and 0.60 s/it (multi-zone non-robust). 
Relative to NMPC\textsubscript{MM}, these means correspond to reductions of 91.3\% (11.44$\times$ faster), 90.6\% (10.62$\times$), and 89.8\% (9.84$\times$), respectively. 
Adding robustness to NMPC\textsubscript{SRM} increased the mean time by only +0.04 s ($\approx$7.7\% overhead), maintaining comfortable headroom for real-time operation.

At the worst-case per-iteration solve time over the run, NMPC\textsubscript{MM} reached 72.07 s, while SR variants stayed below 2.30 s: 2.09 s (NMPC\textsubscript{SRM}), 2.30 s (RNMPC\textsubscript{SZ}), and 1.7459 s (RNMPC\textsubscript{MZ}), i.e., worst-case cuts of 97.1\%, 96.8\%, and 97.6\% versus NMPC\textsubscript{MM}. 
Overall, the SR models keep the optimisation well within real-time bounds, and the RNMPC\textsubscript{MZ} checks add only a modest overhead without compromising online feasibility.

\begin{figure}[h!]
    \centering
    \begin{subfigure}[b]{0.45\textwidth}
        \centering
        \includegraphics[width=1.03\textwidth]{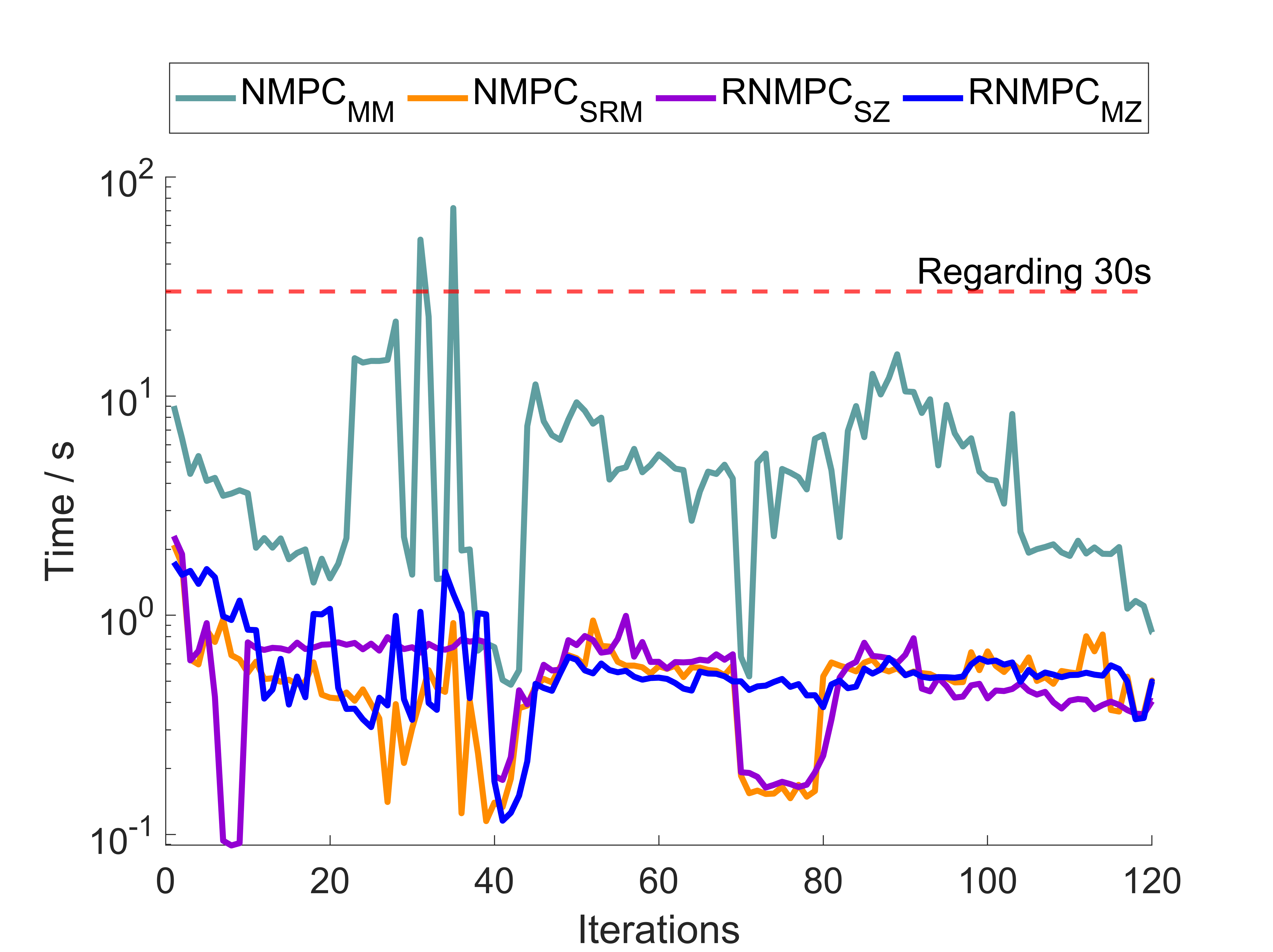}
        \caption{Semi-log computational time for each iteration graph.}
        \label{fig:Tempo_2}
    \end{subfigure}
    \hspace{-0.1cm} 
    \begin{subfigure}[b]{0.49\textwidth}
        \centering
        \includegraphics[width=\textwidth]{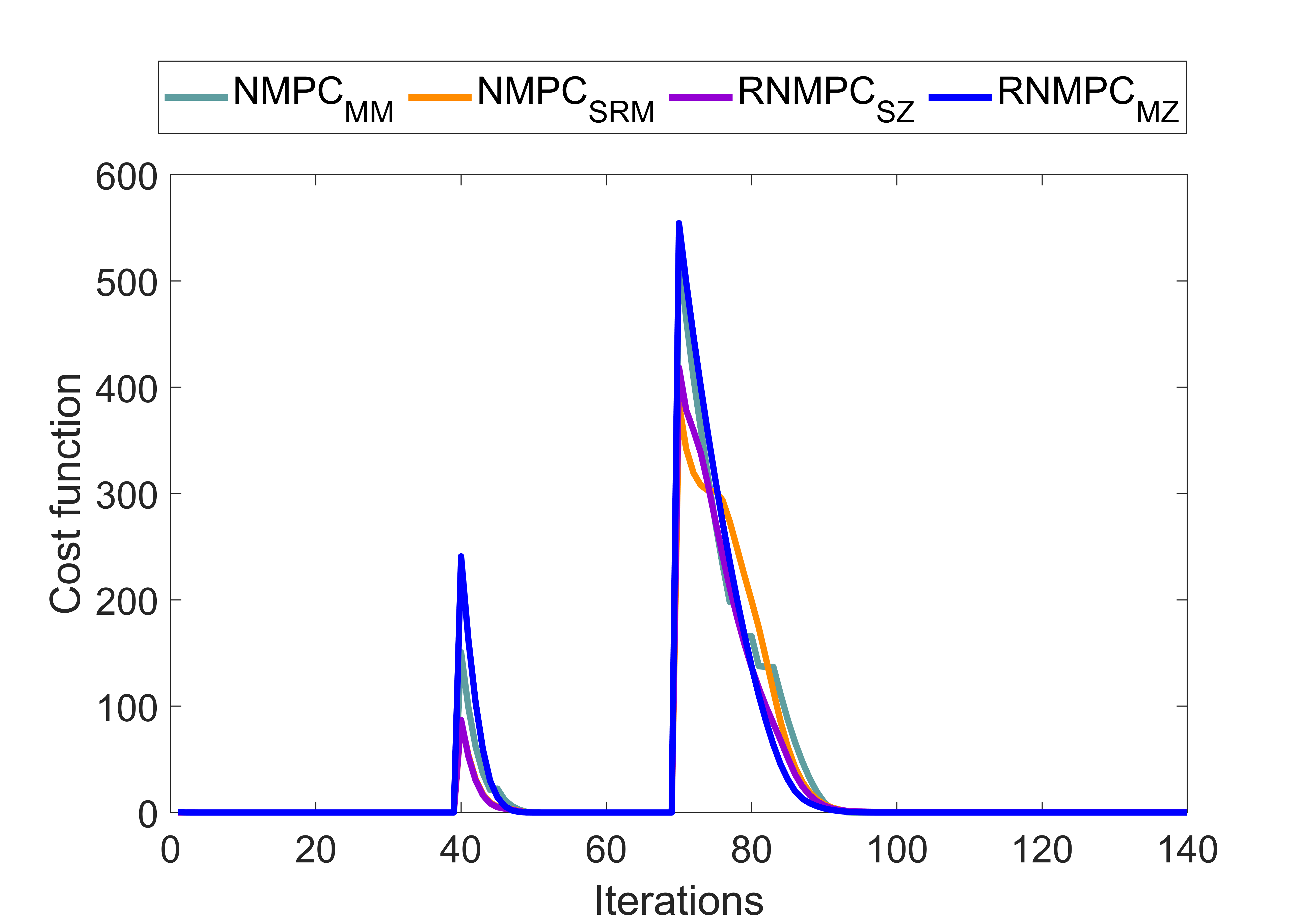}
        \caption{Cost function for each iteration.}
        \label{fig:cost_function_2}
    \end{subfigure}
    \caption{Controller performance: computational time and optimisation convergence for scenario 2.}
    \label{fig:cenario2_}
\end{figure}

Subfigure \ref{fig:cost_function_2} shows the evolution of the cost function across iterations during optimisation. All controllers converge toward near-optimal solutions. Despite introducing robustness into the prediction model, RNMPC\textsubscript{SZ} and RNMPC\textsubscript{MZ} maintain convergence speed and final cost values similar to those of the non-robust NMPC\textsubscript{SRM} and NMPC\textsubscript{MM}, indicating that the inclusion of robustness does not impose additional computational burden. These results confirm the practicality and scalability of symbolic regression models, especially when robustness is needed without sacrificing efficiency.
 
\subsubsection{Scenario 3: Set-point–scheduled RNMPC\textsubscript{MZ} with multi-zone constraint across all operating regions}

\begin{figure}[h!]
	\centering
		\includegraphics[scale=.35]{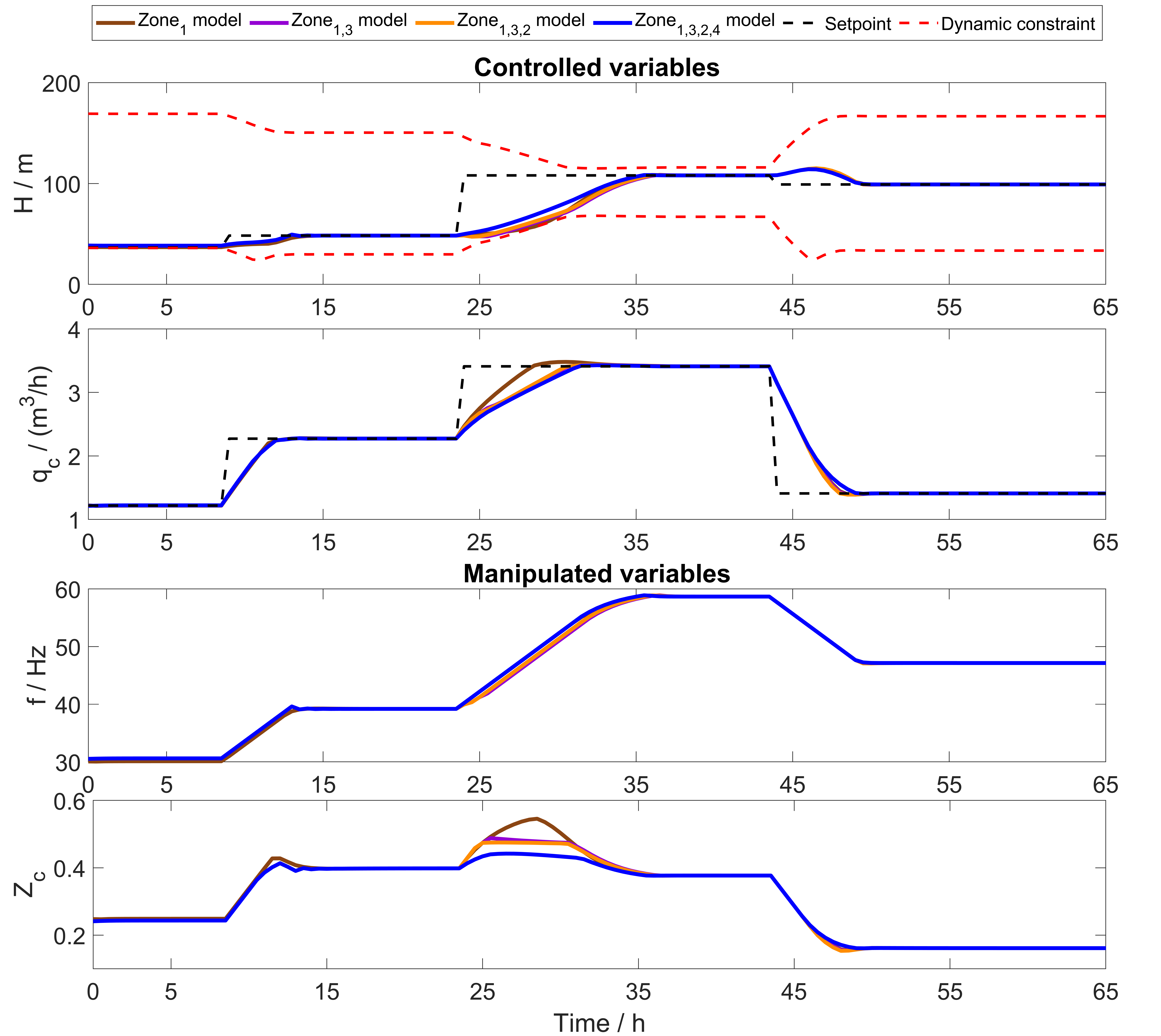}
	\caption{Trajectory comparison of control strategies within the operational envelope for scenario 3.}
	\label{fig:cenario_3}
 \end{figure}

Scenario 3 evaluates RNMPC\textsubscript{MZ} under a sequence of set-point changes that deliberately drives the process through all four operating regions. The set-point starts in zone\textsubscript{4} model  ($H$ = 36.6 m, $q_c$ = 1.2 $m^3/h$), then is stepped to zone\textsubscript{3} model ($H$ = 48.3 m, $q_c$ = 2.3 $m^3/h$), subsequently to zone\textsubscript{2} model ($H$ = 97.6 m, $q_c$ = 3.3 $m^3/h$), and finally to zone\textsubscript{1} model ($H$ = 121.7 m, $q_c$ = 1.4 $m^3/h$).
At each setpoint change, the nominal predictor is re-scheduled to the symbolic model of the newly active zone (set-point–scheduled nominal), while robustness is induced solely by embedding zone models as hard constraints. For each experiment, the constraint ensemble is held fixed and progressively expanded across runs, from zone\textsubscript{1} model to zone\textsubscript{1,2} model, zone\textsubscript{1,2,3} model, and zone\textsubscript{1,2,3,4} model, as shown in Figure~\ref {fig:cenario_3}. 

It is possible to notice in Figure \ref{fig:cenario_3} that, when only zone\textsubscript{1} model is enforced, the closed-loop trajectory violates operational boundaries, particularly around set-point transitions, from zone\textsubscript{3} $(H = 48.3 m,\; q_c = 2.2 m^3/h)$, to zone\textsubscript{2} $(H = 97.6 m,\; q_c = 3.3 m^3/h)$, at these moments, a slight overshoot in $q_c$ drives the controller past the envelope constraints, which can be seen in the dynamic-constraint traces of $H$ in Figure \ref{fig:cenario_3}.

Zone-specific SR models are progressively added to the constraint set, turning the feasible region into the intersection of the corresponding admissible sets. Practically, the drift toward the upthrust side arises from the control objective and tuning that prioritise higher production flow (weighting \(Q=250\) and \(R=500\) favours larger \(q_c\)). Hence, the optimiser naturally steers operation toward regions of higher throughput, which lie closer to upthrust in the envelope typically targeted in ESP operations. 

\begin{figure}[h!]
	\centering
		\includegraphics[scale=.45]{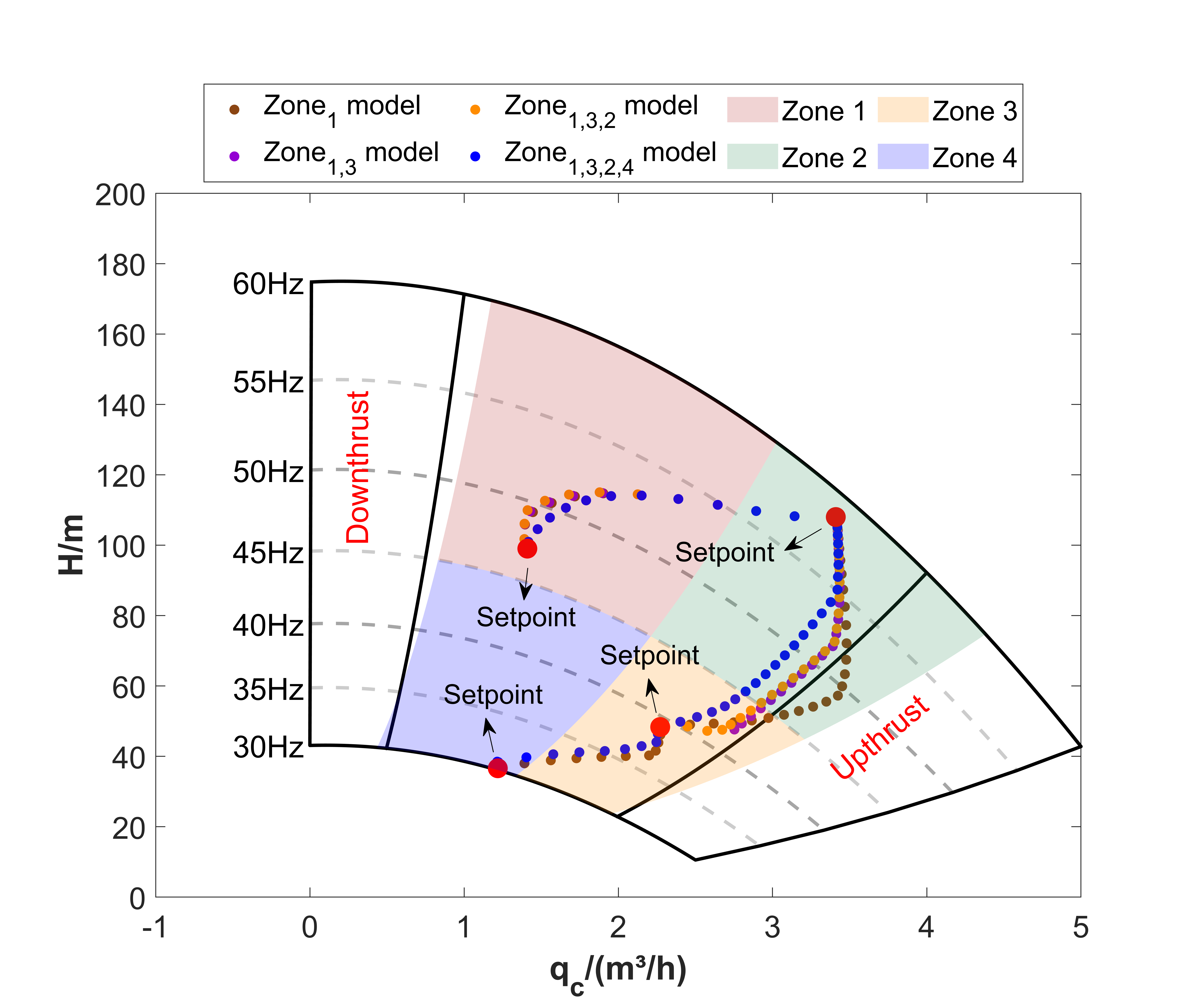}
	\caption{Operational envelope trajectories during the setpoint changes in scenario 3.}
	\label{fig:envelope_3}
 \end{figure}

Despite this upthrust-oriented incentive, the multi-zone constraint intersection tightens the admissible set and counteracts such drift: it consistently pushes the closed-loop trajectory away from the downthrust/upthrust limits throughout all set-point changes. With the zone\textsubscript{1,2,3,4} model enforced, the trajectory attains the most significant safety margin to the boundaries, the most conservative configuration, while still achieving set-point convergence in every region. This makes the trade-off explicit: expanding the enforced set improves coverage of nonlinearities. It mitigates structural/model bias during set-point moves, at the cost of increased conservatism and slightly milder control near constraints. This behaviour is visible in Fig.~\ref{fig:envelope_3}.

Near the final set-point in zone\textsubscript{1}, the controller with the full constraint ensemble zone\textsubscript{1,2,3,4} model again steers the plant farther from the envelope boundaries than any partial ensemble, reflecting its added conservatism; consequently, it exhibits a slower approach and longer settling time to the reference, as shown in Figure \ref{fig:cenario_3}.

\begin{figure}[h!]
        \centering
        \includegraphics[width=0.65\textwidth]{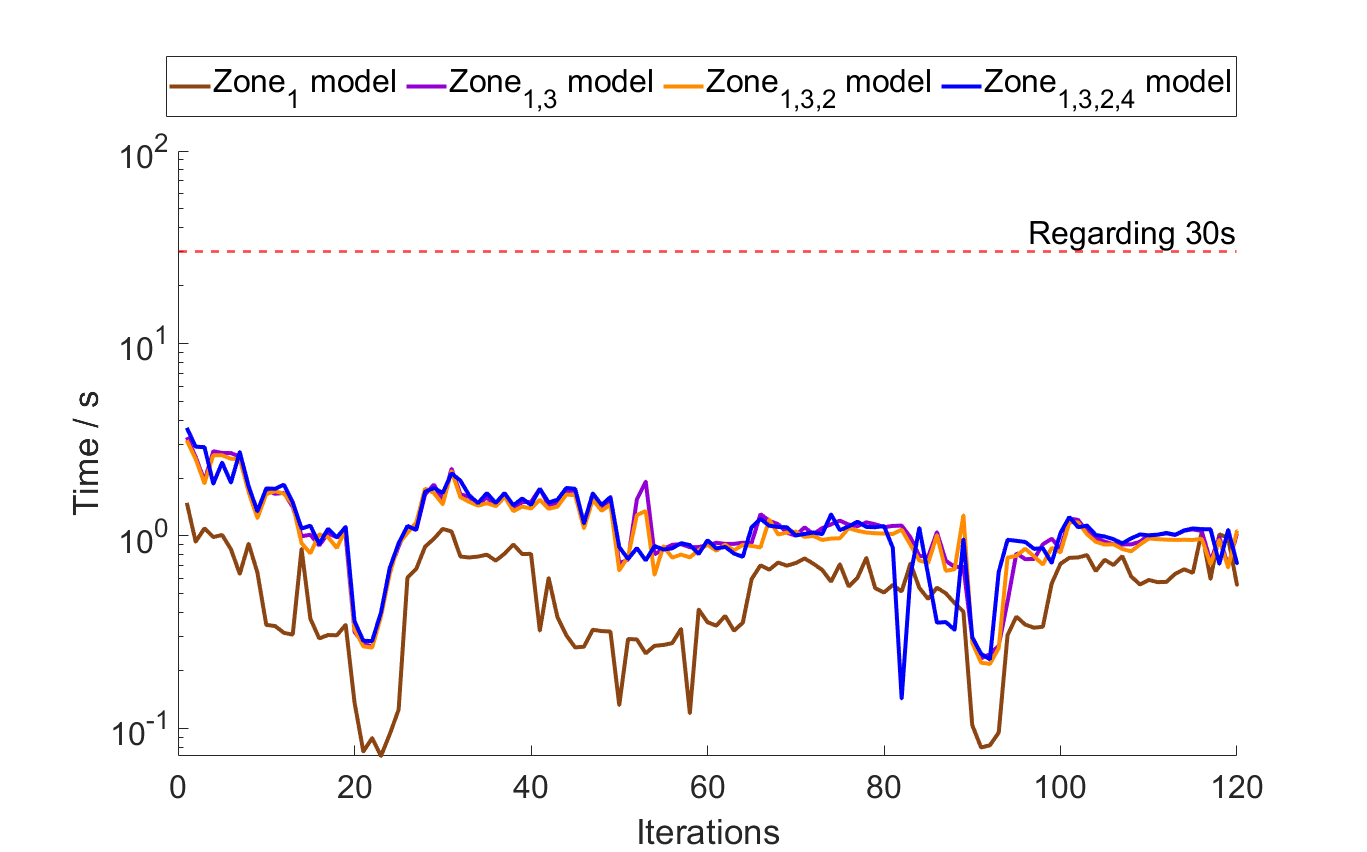}
        \label{fig:Tempo_3}
     \caption{Semi-log comparison of computational time graph for each iteration for scenario 3.}
    \label{fig:cenario_3_performance}
\end{figure}

Figure~\ref{fig:cenario_3_performance} reports the per-iteration time (log scale) for RNMPC\textsubscript{MZ} as the constraint ensemble grows from a single zone to four zones. Across the entire horizon, the four variants run within the same time band (sub-second to $\approx1 - 2 s$), with no systematic increase in runtime as more zone models are enforced. All curves lie well below the 30 s reference line, confirming real-time feasibility. In short, multi-zone robustness via constraint intersection adds negligible computational overhead.

\section{Conclusions}

This paper presents an offset-free robust nonlinear model predictive control scheme that integrates multiple data-driven representations derived via symbolic regression into the controller’s constraints, thereby guarding against structural uncertainty without imposing excessive computational burden. By assembling an ensemble of NARX-structure symbolic models and embedding them within a multi-model optimisation framework,  the controller for a simulated pilot-scale electric submersible pump system is developed. 

Two robust SR-based RNMPC configurations were evaluated: (i) a single-zone scheme, RNMPC\textsubscript{SZ}, in which synthetic data are generated within one operating zone and used to identify NARX-structure SR models embedded as hard constraints during optimisation; and (ii) a multi-zone scheme, RNMPC\textsubscript{MZ}, in which zone-specific SR models from multiple operating zones are jointly enforced, while the nominal predictor is re-scheduled to the zone associated with each new set-point.

 In Scenario 1, the resulting RNMPC\textsubscript{SZ} and RNMPC\textsubscript{MZ} achieved disturbance rejection and set-point tracking performance on par with conventional mechanistic NMPC\textsubscript{MM} and the NMPC\textsubscript{SRM}; however, the symbolic-regression controllers achieved shorter per-iteration solve times.

In Scenario 2, the RNMPC\textsubscript{MZ} and RNMPC\textsubscript{SZ} systematically delivered the most significant safety margins in relation to the other controllers and eliminated constraint violations across all set-point moves, especially near upthrust, at the cost of a slightly longer settling. Moreover, the RNMPC\textsubscript{MZ}, embedding up to four zone-specific SR models in the constraint set, consistently achieved the widest stand-off from the operational envelope, i.e., the most considerable minimum distance to the boundaries, and thus the most conservative closed-loop behaviour across all set-point changes. In scenario 3, we run the plant through a sequence of set-points that spans all four operating zones, exercising each nominal model in its domain. We evaluate increasing constraint sets  zone\textsubscript{1} model, to zone\textsubscript{1,2} model, to zone\textsubscript{1,2,3} model, and zone\textsubscript{1,2,3,4} model, which provides the most significant distance from the constraints (the most conservative configuration) and eliminates violations, including near the last set-point, at the cost of slightly slower settlement. From a computational perspective, including up to four zone models did not increase the solution time per iteration.
This outcome underscores the scalability of the multi-model approach: even many ensembles can be incorporated seamlessly to capture broader uncertainty without compromising speed.

Overall, the study demonstrates that symbolic regression models can generate lean, accurate surrogate models for high-performance robust control, paving the way for RNMPC deployment in fast, safety-critical processes where model uncertainty and computational constraints have historically been limiting factors. Future work will focus on automatic model selection from the Pareto front, extension to larger multivariable systems, and hardware-in-the-loop validation to confirm real-world applicability. Additionally, the parametric uncertainty of these models, as addressed by \citep{Rebello4c02104} using Bayesian inference, will be the subject of future investigations in the development of robust controllers.

\bibliographystyle{cas-model2-names}

\bibliography{cas-refs}

\appendix
\section{My Appendix} \label{appendix}
\setcounter{figure}{0}
\setcounter{table}{0}
\renewcommand{\thefigure}{A.\arabic{figure}}
\renewcommand{\thetable}{A.\arabic{table}}

\begin{figure}[h!]
    \centering
    \begin{subfigure}[b]{0.23\textwidth}
        \centering
        \includegraphics[width=\linewidth]{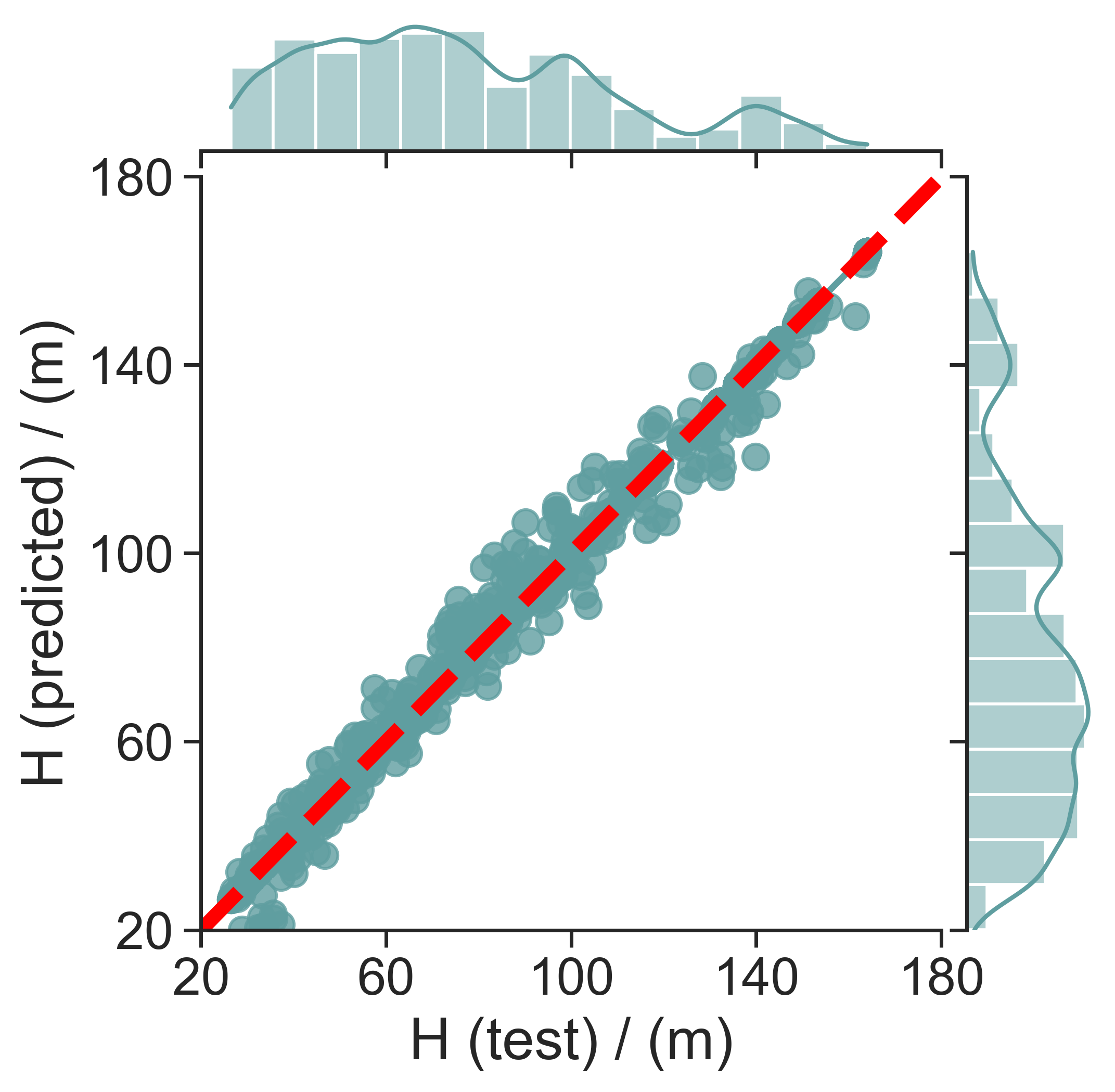}
        \caption{$H$ model I.}
    \end{subfigure}
    \hspace{0.01\textwidth}
    \begin{subfigure}[b]{0.23\textwidth}
        \centering
        \includegraphics[width=\linewidth]{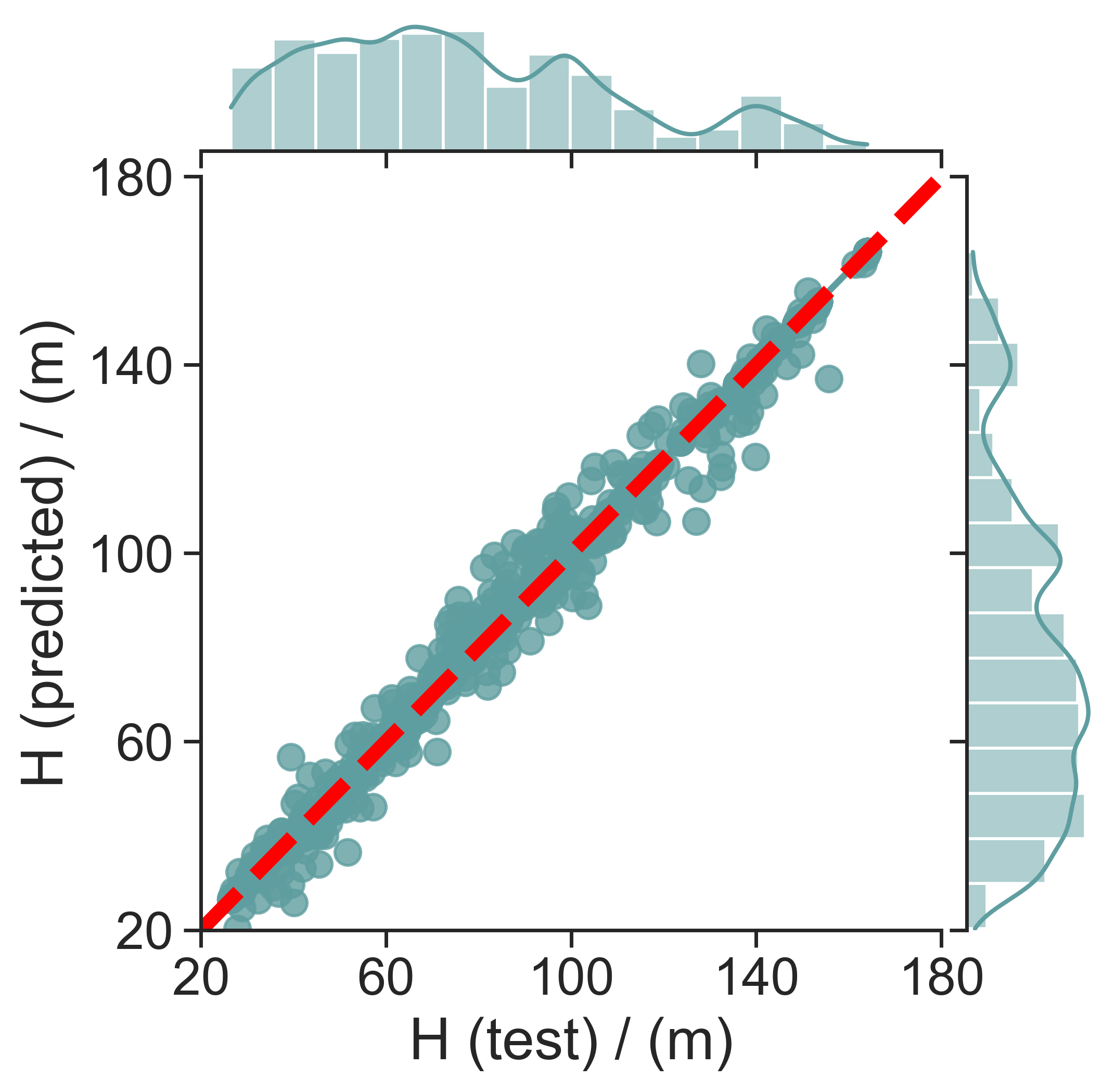}
        \caption{$H$ model II.}
    \end{subfigure}
    \hspace{0.01\textwidth}
    \begin{subfigure}[b]{0.23\textwidth}
        \centering
        \includegraphics[width=\linewidth]{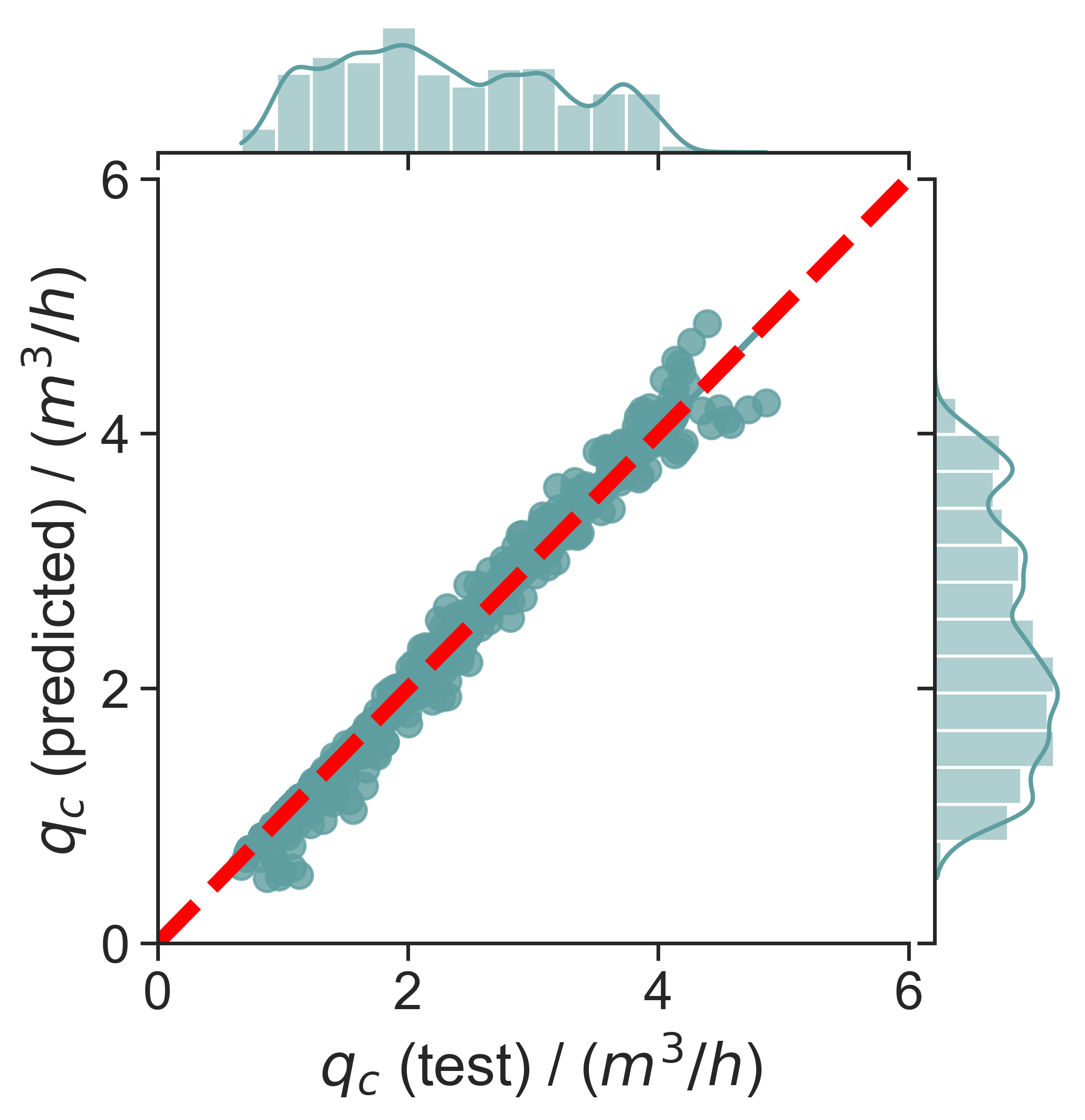}
        \caption{$q_c$ model I.}
    \end{subfigure}
    \hspace{0.01\textwidth}
    \begin{subfigure}[b]{0.23\textwidth}
        \centering
        \includegraphics[width=\linewidth]{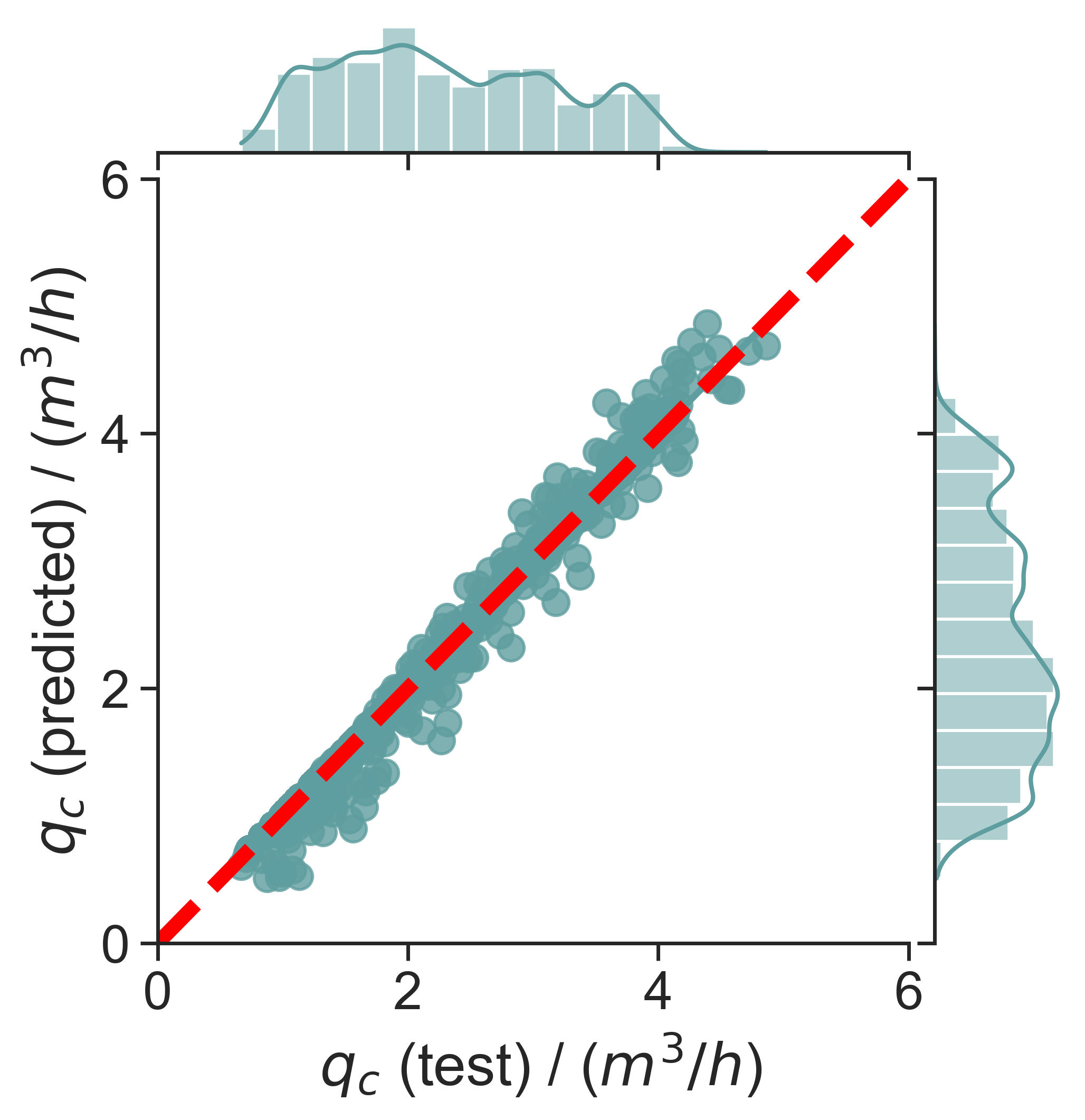}
        \caption{$q_c$ model II.}
    \end{subfigure}

    \caption{Parity plots of SR models on test data for model I and model II.}
    \label{fig:Parity_plots}
\end{figure}

\begin{figure}[h!]
    \centering
    \begin{subfigure}[b]{0.45\textwidth}
        \centering
        \includegraphics[width=\linewidth]{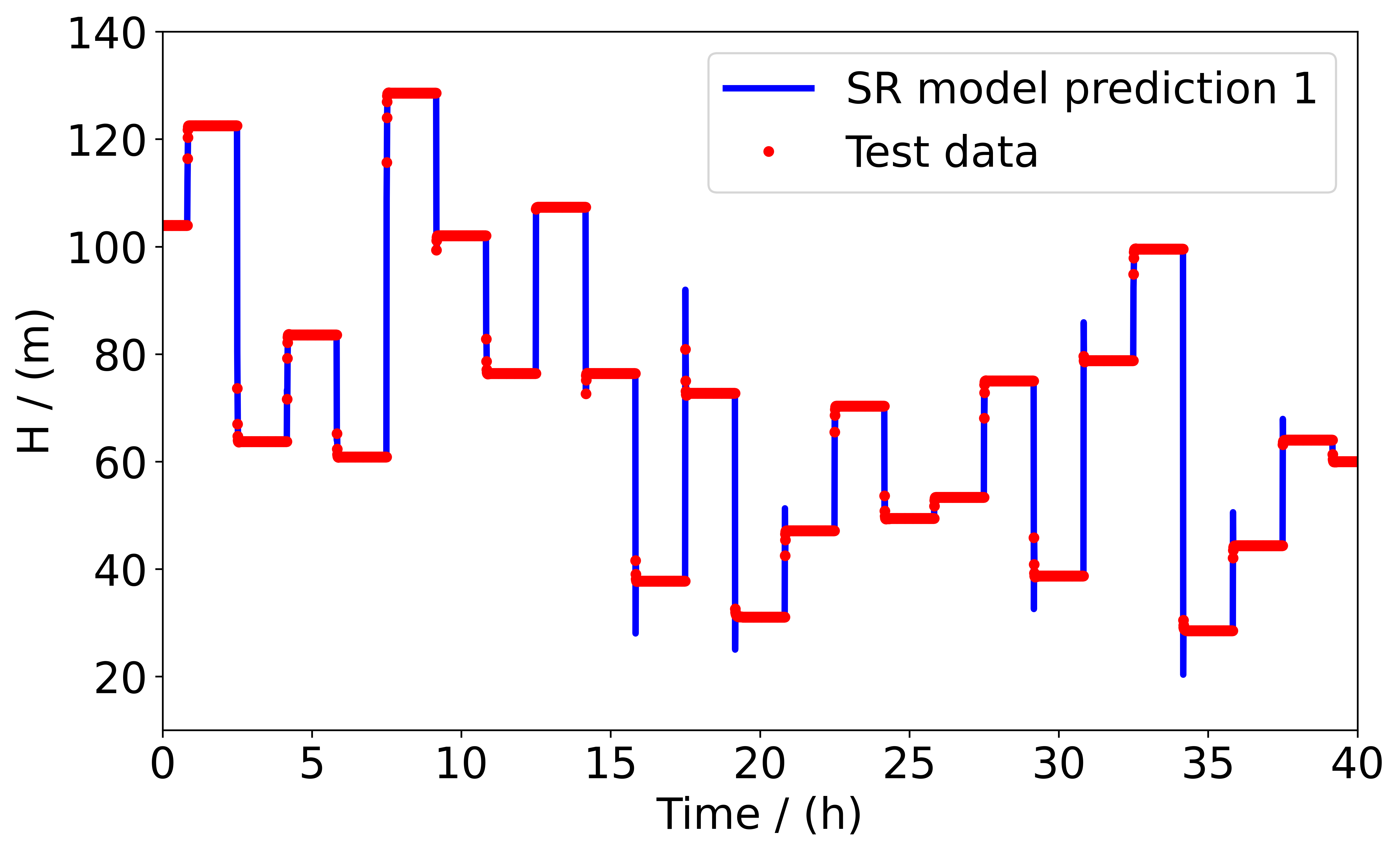}
        \caption{$H$ model I.}
    \end{subfigure}
    \hspace{0.03\textwidth}
    \begin{subfigure}[b]{0.45\textwidth}
        \centering
        \includegraphics[width=\linewidth]{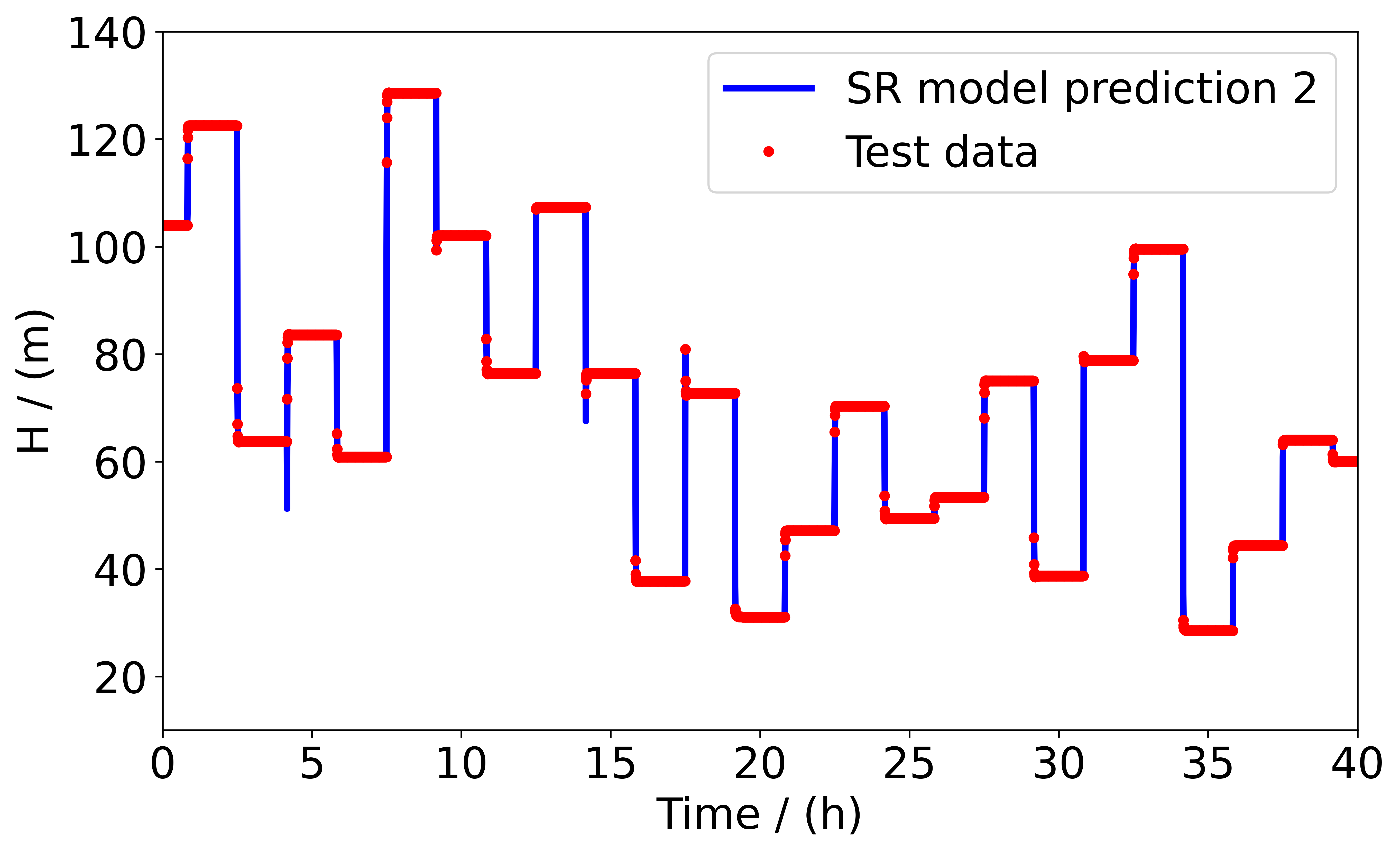}
        \caption{$H$ model II.}
    \end{subfigure}

    \vspace{0.3cm}

    \begin{subfigure}[b]{0.45\textwidth}
        \centering
        \includegraphics[width=\linewidth]{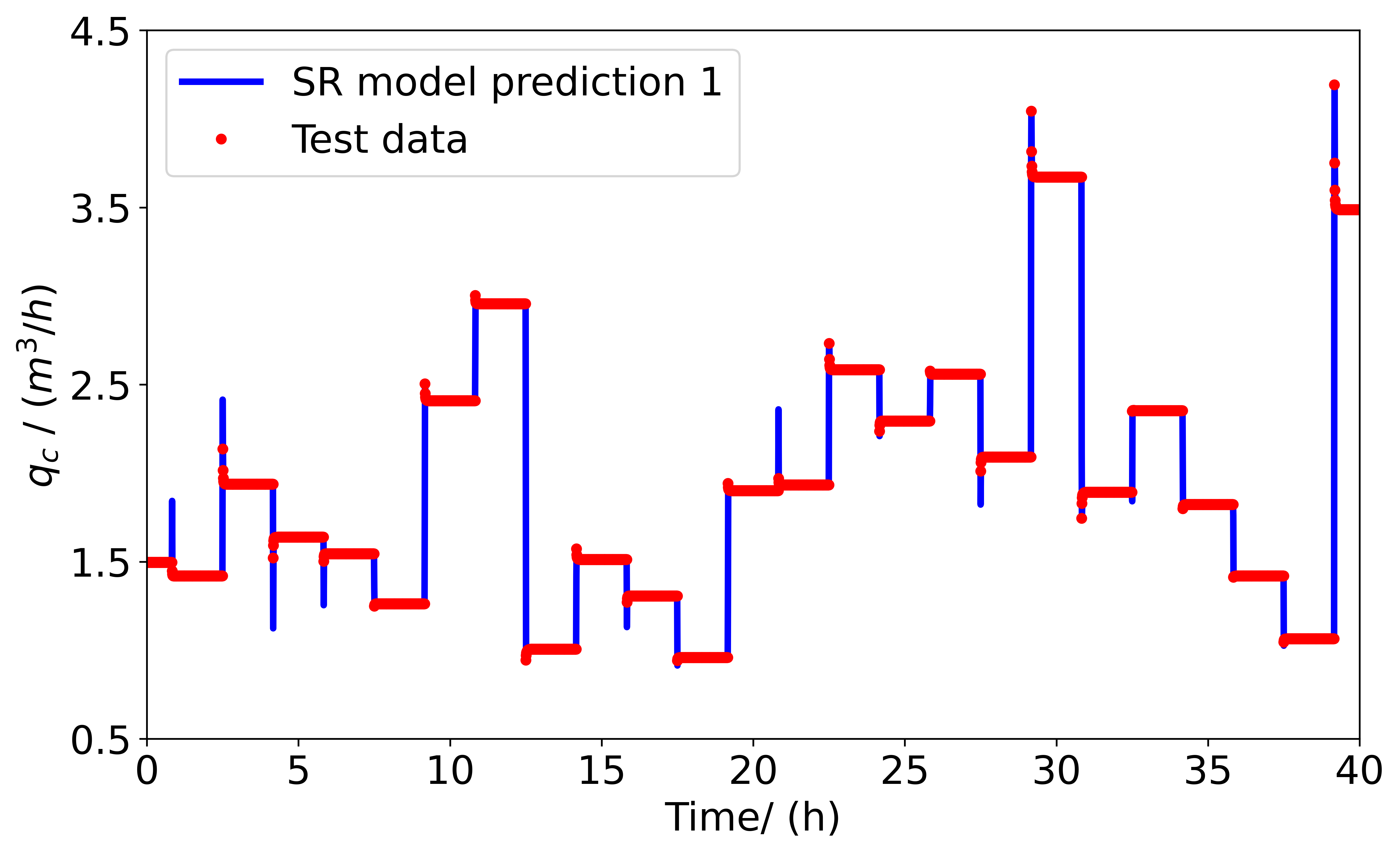}
        \caption{$q_c$ model I.}
    \end{subfigure}
    \hspace{0.03\textwidth}
    \begin{subfigure}[b]{0.45\textwidth}
        \centering
        \includegraphics[width=\linewidth]{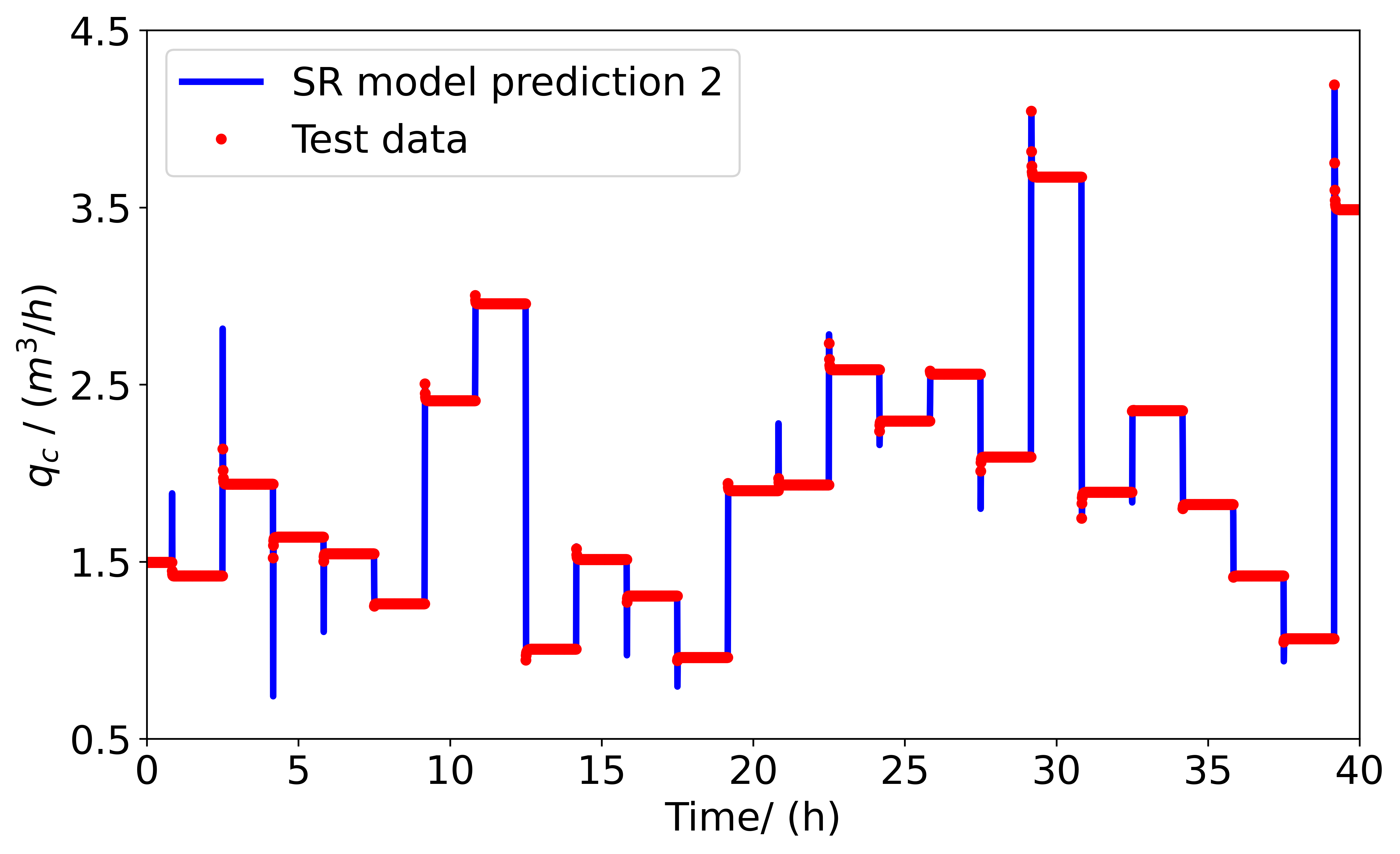}
        \caption{$q_c$ model II.}
    \end{subfigure}

    \caption{Time-series dynamics of the model I and II SR model predictions.}
    \label{fig:Time-series_dynamics}
\end{figure}

\begin{table}[H]
\centering
\caption{Default parameters of \texttt{PySRRegressor} as provided in the official documentation.}
\label{app:pysr_defaults}
\scriptsize
\begin{tabular}{lll}
\toprule
\textbf{Parameter} & \textbf{Description} & \textbf{Default value} \\
\midrule
\texttt{expression\_spec}         & Expression search space specification & \verb|ExpressionSpec()| \\
\texttt{maxsize}                  & Maximum equation size (nodes/operators) & 30 \\
\texttt{maxdepth}                 & Maximum expression tree depth & \verb|None| \\
\texttt{niterations}              & Number of evolutionary iterations & 100 \\
\texttt{populations}              & Number of parallel populations & 31 \\
\texttt{population\_size}         & Size of each population & 27 \\
\texttt{ncycles\_per\_iteration}  & Mutation cycles per 10 individuals per iteration & 380 \\
\texttt{elementwise\_loss}        & Elementwise loss function & \verb|"L2DistLoss()"| \\
\texttt{loss\_function}           & Custom loss function (Julia code) & \verb|None| \\
\texttt{loss\_function\_expression} & Loss function defined as symbolic expression & \verb|None| \\
\texttt{loss\_scale}              & Loss scaling in score computation & \verb|"log"| \\
\texttt{model\_selection}         & Criterion for model selection & \verb|"best"| \\
\texttt{tournament\_selection\_n} & Tournament pool size & 15 \\
\texttt{tournament\_selection\_p} & Probability of selecting best candidate in tournament & 0.982 \\
\texttt{optimizer\_algorithm}     & Constant optimizer algorithm & \verb|"BFGS"| \\
\texttt{optimizer\_nrestarts}     & Number of optimization restarts & 2 \\
\texttt{optimizer\_f\_calls\_limit} & Max function evaluations in optimization & 10,000 \\
\texttt{optimize\_probability}    & Probability of optimizing constants per iteration & 0.14 \\
\texttt{optimizer\_iterations}    & Iterations of constant optimizer per run & 8 \\
\texttt{should\_optimize\_constants} & Optimize constants after each iteration & \verb|True| \\
\texttt{fraction\_replaced}       & Fraction of population replaced by migration & 0.00036 \\
\texttt{fraction\_replaced\_hof}  & Fraction of population replaced from Hall of Fame & 0.0614 \\
\texttt{migration}                & Enable migration between populations & \verb|True| \\
\texttt{hof\_migration}           & Enable Hall of Fame migration & \verb|True| \\
\texttt{topn}                     & Number of top individuals migrating per population & 12 \\
\texttt{denoise}                  & Apply Gaussian process denoising before search & \verb|False| \\
\texttt{select\_k\_features}      & Feature pre-selection (e.g., Random Forest) & \verb|None| \\
\texttt{max\_evals}               & Maximum number of total evaluations & \verb|None| \\
\texttt{timeout\_in\_seconds}     & Maximum runtime before stopping & \verb|None| \\
\texttt{early\_stop\_condition}   & Early stopping rule (loss/complexity threshold) & \verb|None| \\
\bottomrule
\end{tabular}
\end{table}

\printcredits



\end{document}